\documentclass[prd,aps,nofootinbib,11pt]{revtex4}
\usepackage{amsmath,graphicx,color,epsfig}

\begin{document}
\title{ Radiative charmless  $B_{(s)}\to V \gamma$ and $B_{(s)}\to A \gamma$
 decays in pQCD approach}
\author{Wei Wang$^{a}$, Run-Hui Li$^{b,a}$ and Cai-Dian L\"u$^{a}$
} \affiliation{\small\it   $^a$
 Institute of High Energy Physics, P.O. Box 918(4) Beijing 100049, P.R.
 China\\
 \it  $^b$  Department of Physics, Shandong University, Jinan 250100, P.R.
 China \\
 \it  $^c$ Graduate University of Chinese Academy of Sciences,
Beijing 100049, P.R. China}

\begin{abstract}
We study the radiative charmless $B_{(s)}\to V(A)\gamma $ decays in
perturbative QCD (pQCD) approach to the leading order in $\alpha_s$
(here $V$ and $A$ denotes vector mesons and two kinds of
axial-vector mesons: $^3P_1$ and $^1P_1$ states, respectively.). Our
predictions of branching ratios are consistent with the current
available experimental data. We update all $B_{(s)}\to V$ form
factors and give the predictions for $B\to A$ form factors using the
recent hadronic inputs. In addition to the dominant factorizable
spectator diagrams, which is form factor like, we also calculate the
so-called ``power suppressed'' annihilation type diagrams, the
gluonic penguin, charming penguin, and two photon diagrams. These
diagrams give the main contributions to direct CP asymmetries,
mixing-induced CP asymmetry variables, the isospin asymmetry and
U-spin asymmetry variables. Unlike the branching ratios, these
ratios or observables possess higher theoretical precision in our
pQCD calculation, since they do not depend on the input hadronic
parameters too much. Most of the results still need experimental
tests in the on-going and forthcoming experiments.
\end{abstract}
\maketitle
\newpage
\section{Introduction}\label{section:introduction}

Present studies on $B$ decays are mainly concentrated on the precise
test of standard model (SM) and the search for signals of possible
new physics. Radiative processes $b\to s\gamma$ and $b\to d\gamma$
are among the most ideal probes and thus have received considerable
efforts  \cite{Hurth:2003vb}. In the standard model, these processes
are induced by the flavor-changing-neutral-current transitions which
are purely loop effects.  Such decays are rare and the measurement
of parameters in these channels, especially the CP asymmetry
variables, may shed light on detailed information on the flavor
structure of the electroweak interactions. This predictive power
relies on the accuracy of both the experimental side and the
theoretical side. Thanks to the technique of
operator-product-expansion, remarkable theoretical progress has been
made in the SM to the next-to-next-to-leading order accuracy
\cite{Misiak:2006zs}. Compared with experimental results
\cite{Barberio:2006bi}, the theoretical prediction is consistent
with but $1.2\sigma$ below the experimental data
\cite{Hurth:2007xa}. This consistence could certainly provide a
rather stringent constraint on non-standard model scenarios.

Compared with  inclusive processes, the exclusive processes $B\to
V\gamma$ is more tractable on the experimental side, but more
difficult on the theoretical side. Theoretical predictions are often
hampered by our ability to calculate the decay amplitude $\langle
V\gamma|O_i|B\rangle$, where $O_i$ is a magnetic moment or a
four-quark operator. We have to use some non-perturbative hadronic
quantities to describe the bound state effects. In the heavy-quark
limit, the non-perturbative contributions can be organized in a
universal and channel-independent manner. Factorization analysis,
the separation of the short-distance and long-distance dynamics,
can give many important predictions.

The dominant contribution to the radiative decay amplitudes is
proportional to the transition form factor. Different treatments on
the dynamics in form factor $F^{B\to V}$ result in different
explicit approaches. There are many approaches such as Lattice QCD
(LQCD) \cite{DelDebbio:1997kr}, light-front quark model (LFQM)
\cite{Cheng:2003sm,Cheng:2004yj} and light-cone sum rules (LCSR)
\cite{Ball:2004rg}. Recently three commonly-accepted approaches are
developed to study exclusive $B$ decays: QCD factorization (QCDF)
\cite{Beneke:1999br}, soft collinear effective theory (SCET)
\cite{Bauer:2001cu} and perturbative QCD approach
\cite{Keum:2000ph}. In pQCD approach, the recoiling meson moves on
the light-cone and a large momentum transfer is required. Keeping
quarks' transverse momentum, pQCD approach is free of endpoint
divergence and the Sudakov formalism makes it more self-consistent.
It has been successfully applied to various decay channels and quite
recently this approach is accessing to next-to-leading order
accuracy \cite{Nandi:2007qx}. A bigger advantage is that we can
really do the form factor calculation and the quantitative
annihilation type diagram calculation in this approach. The
importance of annihilation diagrams are already tested in the
predictions of direct CP asymmetries in $B^0 \to \pi^+\pi^-$,
$K^+\pi^-$ decays \cite{Keum:2000ph,direct} and in the explanation
of $B\to \phi K^*$ polarization problem \cite{kphi,kphi-e}. These
``power suppressed'' contributions are the main source of isospin
symmetry breaking (or SU(3) breaking) effects in radiative decays
$B\to V\gamma$ \cite{Keum:2004is,Lu:2005yz}.

Some of charmless $B\to V\gamma$ decay channels have been studied in
pQCD approach \cite{Keum:2004is,Lu:2005yz,Li:2006xe} separately.
According to the transition at the quark level, all the 10 decay
channels can be divided into three different groups: $b\to s$, $b\to
d$ and purely annihilation type.  The two $B\to K^* \gamma$ modes
($b\to s$ process) have been well measured experimentally. The
agreement of pQCD approach results and experimental data
\cite{Barberio:2006bi} is very encouraging. In this paper, we study
all those channels including the corresponding $B_s$ decays in a
comprehensive way. The input hadronic parameters will be chosen the
same as that in Ref. \cite{Ali:2007ff}. We also updated all the
$B\to V$ decay form factors in pQCD approach \cite{Lu:2002ny}.
Despite branching ratios and CP asymmetry parameters which heavily
depend on the input parameters, we also study some ratios
characterizing the isospin and SU(3) breaking effects, where most of
the uncertainties cancel. Experimental measurements of these
quantities will prove a good test of our theory, since different
method gives different prediction. With the ongoing $B$ factories
BaBar and Belle, the $B$-Physics program on CDF and the onset of the
LHC experiments, as well as the Super B-factories being contemplated
for the future, we expect a wealth of data involving these decays.

Although experimentalists have already measured one channel of the
$B\to A\gamma$ decays \cite{Abe:2004kr}, there are not many
discussions on the theoretical side. The pQCD study on $B\to
V\gamma$ can be straightforward extended to radiative processes
involving higher resonants such as $K_1(1270)$, $K_1(1400)$. In the
quark model, the quantum numbers $J^{PC}$ for the  orbitally excited
axial-vector mesons are $1^{++}$ or $1^{+-}$, depending on different
spins of the two quarks. In SU(3) limit, these mesons can not mix
with each other; but since the $s$ quark is heavier than $u,d$
quarks, $K_1(1270)$ and $K_1(1400)$ are not purely $1^3P_1$ or
$1^1P_1$ states. These two mesons are believed to be mixtures of
$K_{1A}$ and $K_{1B}$, where $K_{1A}$ and $K_{1B}$ are $^3P_1$ and
$^1P_1$ states, respectively. In general, the mixing angle can be
determined by experimental data. But unfortunately, there is not too
much data on the mixing of these mesons which leaves the mixing
angle much free. Analogous to $\eta$ and $\eta^\prime$, the
flavor-singlet and flavor-octet axial-vector meson can also mix with
each other. Using those  hadronic parameters determined in $B\to
V\gamma$ decays, the production of $A\gamma$ in $B$ decays could
provide a unique insight to these mysterious axial-vector mesons.

This paper is organized as follows: In section II, we briefly review
pQCD approach with the operator basis used subsequently. Some input
quantities which enter pQCD approach, wave function of the
$B$-meson, distribution amplitudes for light vector mesons, and for
light axial-vector mesons and input values of the various mesonic
decay constants, are also given here. In
section~\ref{section:formfactor}, we give the factorization formulae
and the numerical results  for $B\to V$ and $B\to A$ form factors.
Section~\ref{section:BtoV} contains the calculation of $B\to
V\gamma$ decays, making explicit the contributions from the
electromagnetic diploe operator, the chromo-magnetic moment
operator, some higher order (${\cal O}(\alpha_s)$) corrections from
tree operators, the contribution from two-photon diagrams, the tree
operator annihilation diagrams and penguin operator annihilation
diagrams. Numerical results for the charge-conjugated averages of
decay branching ratios are given in comparison  with the
corresponding numerical results obtained in QCDF approach and SCET,
as well as the available experimental data. We also give direct
CP-asymmetries, time-dependent CP asymmetries $S_f$ and observables
$H_f$ (for $B_s$ system) in the time-dependent decay rates in this
section. In  section \ref{section:BtoA}, we study $B\to A\gamma$
decays by predicting branching ratios and direct CP asymmetries. Our
summary is given in the last section. Some functions are relegated
to the appendix: appendix A contains the various functions that
enter the factorization formulae in the pQCD approach; appendix B
and C give the analytic formulae for the $B\to V\gamma$ and $B\to
A\gamma$ decays used in  numerical calculations, respectively.

\section{Formalism of pQCD approach}\label{section:pqcd}

\subsection{Notations and conventions}

We specify the weak effective Hamiltonian which describes $b\to D$
($D=d,s$) transitions \cite{Buchalla:1995vs}:
 \begin{eqnarray}
 {\cal H}_{eff} &=& \frac{G_{F}}{\sqrt{2}}
     \bigg\{ \sum\limits_{q=u,c} V_{qb} V_{qD}^{*} \big[
     C_{1}({\mu}) O^{q}_{1}({\mu})
  +  C_{2}({\mu}) O^{q}_{2}({\mu})\Big]\nonumber\\
  &&-V_{tb} V_{tD}^{*} \Big[{\sum\limits_{i=3}^{10,7\gamma,8g}} C_{i}({\mu}) O_{i}({\mu})
  \big ] \bigg\} + \mbox{H.c.} ,
 \label{eq:hamiltonian}
\end{eqnarray}
where $V_{qb(D)}$ and $V_{tb(D)}$ are Cabibbo-Kobayashi-Maskawa
(CKM) matrix elements. Functions $O_{i}$ ($i=1,...,10,7\gamma,8g$)
are local four-quark operators or the moment type operators:
 \begin{itemize}
 \item  current--current (tree) operators
    \begin{eqnarray}
  O^{q}_{1}=({\bar{q}}_{\alpha}b_{\beta} )_{V-A}
               ({\bar{D}}_{\beta} q_{\alpha})_{V-A},
    \ \ \ \ \ \ \ \ \
   O^{q}_{2}=({\bar{q}}_{\alpha}b_{\alpha})_{V-A}
               ({\bar{D}}_{\beta} q_{\beta} )_{V-A},
    \label{eq:operator12}
    \end{eqnarray}
     \item  QCD penguin operators
    \begin{eqnarray}
      O_{3}=({\bar{D}}_{\alpha}b_{\alpha})_{V-A}\sum\limits_{q^{\prime}}
           ({\bar{q}}^{\prime}_{\beta} q^{\prime}_{\beta} )_{V-A},
    \ \ \ \ \ \ \ \ \
    O_{4}=({\bar{D}}_{\beta} b_{\alpha})_{V-A}\sum\limits_{q^{\prime}}
           ({\bar{q}}^{\prime}_{\alpha}q^{\prime}_{\beta} )_{V-A},
    \label{eq:operator34} \\
     \!\!\!\! \!\!\!\! \!\!\!\! \!\!\!\! \!\!\!\! \!\!\!\!
    O_{5}=({\bar{D}}_{\alpha}b_{\alpha})_{V-A}\sum\limits_{q^{\prime}}
           ({\bar{q}}^{\prime}_{\beta} q^{\prime}_{\beta} )_{V+A},
    \ \ \ \ \ \ \ \ \
    O_{6}=({\bar{D}}_{\beta} b_{\alpha})_{V-A}\sum\limits_{q^{\prime}}
           ({\bar{q}}^{\prime}_{\alpha}q^{\prime}_{\beta} )_{V+A},
    \label{eq:operator56}
    \end{eqnarray}
 \item electro-weak penguin operators
    \begin{eqnarray}
     O_{7}=\frac{3}{2}({\bar{D}}_{\alpha}b_{\alpha})_{V-A}
           \sum\limits_{q^{\prime}}e_{q^{\prime}}
           ({\bar{q}}^{\prime}_{\beta} q^{\prime}_{\beta} )_{V+A},
    \ \ \ \
    O_{8}=\frac{3}{2}({\bar{D}}_{\beta} b_{\alpha})_{V-A}
           \sum\limits_{q^{\prime}}e_{q^{\prime}}
           ({\bar{q}}^{\prime}_{\alpha}q^{\prime}_{\beta} )_{V+A},
    \label{eq:operator78} \\
     O_{9}=\frac{3}{2}({\bar{D}}_{\alpha}b_{\alpha})_{V-A}
           \sum\limits_{q^{\prime}}e_{q^{\prime}}
           ({\bar{q}}^{\prime}_{\beta} q^{\prime}_{\beta} )_{V-A},
    \ \ \ \
    O_{10}=\frac{3}{2}({\bar{D}}_{\beta} b_{\alpha})_{V-A}
           \sum\limits_{q^{\prime}}e_{q^{\prime}}
           ({\bar{q}}^{\prime}_{\alpha}q^{\prime}_{\beta} )_{V-A},
    \label{eq:operator9x}
    \end{eqnarray}
     \item magnetic moment operators
    \begin{eqnarray}
     O_{7\gamma}=-\frac{e}{4\pi^2}{\bar{D}}_{\alpha}\sigma^{\mu\nu}(m_D P_L+m_b
     P_R)b_{\alpha}F_{\mu\nu},
    \ \ \ \
    O_{8g}=-\frac{g}{4\pi^2}{\bar{D}}_{\alpha}\sigma^{\mu\nu}(m_D P_L+m_b
     P_R)T^a_{\alpha\beta}b_{\beta}G^a_{\mu\nu},\label{eq:operator7gamma8g}
\end{eqnarray}
\end{itemize}
where $\alpha$ and $\beta$ are color indices and $q^\prime$ are the
active quarks at the scale $m_b$, i.e. $q^\prime=(u,d,s,c,b)$. The
left handed current is defined as $({\bar{q}}^{\prime}_{\alpha}
q^{\prime}_{\beta} )_{V-A}= {\bar{q}}^{\prime}_{\alpha} \gamma_\nu
(1-\gamma_5) q^{\prime}_{\beta}  $ and the right handed current
$({\bar{q}}^{\prime}_{\alpha} q^{\prime}_{\beta} )_{V+A}=
{\bar{q}}^{\prime}_{\alpha} \gamma_\nu (1+\gamma_5)
q^{\prime}_{\beta}$. The projection operators are defined as
$P_{L}=(1-\gamma_5)/2$ and $P_{R}=(1+\gamma_5)/2$.
 The combinations $a_i$ of Wilson coefficients are
defined as usual \cite{Ali:1998eb}:
\begin{eqnarray}
a_1= C_2+C_1/3, &~a_2= C_1+C_2/3, &~ a_3= C_3+C_4/3,  ~a_4=
C_4+C_3/3,~a_5= C_5+C_6/3,\nonumber \\
a_6= C_6+C_5/3, &~a_7= C_7+C_8/3, &~a_8= C_8+C_7/3,~a_9=
C_9+C_{10}/3,
 ~a_{10}= C_{10}+C_{9}/3.
\end{eqnarray}

For the explicit formulae, we will consider $\bar B$ meson decays
and use light-cone coordinates to describe the momentum:
\begin{eqnarray}
 p=(p^+,p^-,\vec{p}_{T})=\Big(\frac{p^0+p^3}{\sqrt{2}},\frac{p^0-p^3}{\sqrt{2}},\vec p_T\Big).
\end{eqnarray}
where $\vec p_T= (p^1,p^2)$. In the $\bar B$ meson rest frame,
momenta of $\bar B$ meson, vector (axial-vector) meson and the
photon are chosen as:
\begin{eqnarray}
P_1 &=&\frac{m_B}{\sqrt 2}(1,1,\vec 0_T),\;\;\;
 P_2=\frac{m_B}{\sqrt{2}}(0,1,\vec 0_T),\;\;\;
 P_\gamma=\frac{m_B}{\sqrt 2}(1,0,\vec 0_T),
\end{eqnarray}
where the vector (axial-vector) is mainly moving on the minus
direction $n_-$ and the photon is moving on the plus direction
$n_+$. Longitudinal momenta fractions of the spectator anti-quarks
in $\bar{B}$ and final state meson are chosen as ${x_1=k_1^+/P_1^+}$
and ${x_2=k_2^-/P_2^-}$. \footnote{One should be cautious that in
the discussion of light cone distribution amplitudes (LCDAs) for the
vectors and axial-vectors, $x$ is defined as the momentum fraction
of the positive quark which is different with our definition here.}
Including the transverse components, the momenta of these spectator
antiquarks are expressed by:
\begin{eqnarray}
k_1&=&(\frac{m_B}{\sqrt 2}x_1,0,\vec{k_{1T}}),\;\;\;
k_2=(0,\frac{m_B}{\sqrt 2} x_2,\vec{k_{2T}}),
\end{eqnarray}
then the ${b}$ and $D(=s,d)$ quark momenta are ${p_b=P_1-k_1}$ and
$p_D=P_2-k_2$. For convenience, we can define the following useful
ratio variables:
\begin{eqnarray}
r_b=\frac{m_b}{m_B},\;\;\; r_D=\frac{m_D}{m_B},\;\;\;
r_V=\frac{m_V}{m_B},\;\;\;r_A=\frac{m_A}{m_B},
\end{eqnarray}
where $m_D$ and $m_{V(A)}$ are masses for the $d(s)$ quark and the
vector (axial-vector) meson, respectively.  In the calculation of
decay amplitudes, we will only keep terms of leading order in $r_V$
but we will consider the corrections together with kinematic
corrections in the phase space as in Eq.~(\ref{eq:BRofBtoVgamma}).

According to the Lorentz structure, decay amplitudes from various
operators can be generally decomposed into scalar and pseudoscalar
components as:
\begin{eqnarray}
{\cal M}=(\epsilon^{\ast}_{\gamma}\cdot \epsilon^{*}_V){\cal M}^S
+i
 \epsilon_{\mu \nu \alpha\beta}\epsilon^{\ast \mu}_{\gamma}
\epsilon^{* \nu}_V n_+^\alpha n_-^\beta{\cal M}^P,
\end{eqnarray} and
where we adopt the convention $\epsilon^{0123}=+1$. In order to
study mixing-induced CP asymmetries, it is convenient to separate
different chiralities in the amplitudes. If the emitted photon is
left-handed, the relationship between the scalar ${\cal M}^S$ and
pseudoscalar component ${\cal M}^P$ is required as
\begin{eqnarray}
{\cal M}^S=-{\cal M}^P\equiv\frac{1}{2}{\cal M}^L,
\end{eqnarray}
while the condition
\begin{eqnarray}
{\cal M}^S={\cal M}^P\equiv\frac{1}{2}{\cal M}^R,
\end{eqnarray}
is required for the right-handed photon.

\subsection{A brief review of pQCD approach}

\begin{figure}[tb]
\vspace{-2.cm}
\begin{center}
\psfig{file=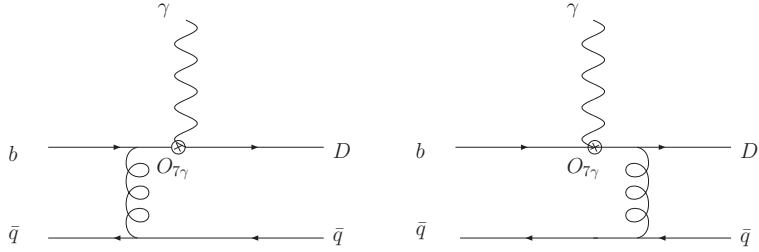,width=12.0cm,angle=0}
\end{center}
\vspace{-12.5cm} \caption{Feynman diagrams of the electromagnetic
penguin operator $O_{7\gamma}$ }\label{Feyn:O7gamma}
\end{figure}

The basic idea of pQCD approach is that it takes into account the
intrinsic transverse momentum of valence quarks. The decay
amplitude, taking the first diagram in Fig.~\ref{Feyn:O7gamma} as an
example, can be expressed as a convolution of wave functions
$\phi_B$, $\phi_V$ and hard scattering kernel $T_H$ by both
longitudinal and transverse momenta:
\begin{eqnarray}
{\cal M}=\int^1_0dx_1dx_2\int\frac{
d^2{\vec{k}}_{1T}}{(2\pi^2)}\frac{d^2{\vec{k}}_{2T}}{(2\pi^2)}\phi_B(x_1,{\vec{k}}_{1T},p_1,t)
T_H(x_1,x_2,{\vec{k}}_{1T},{\vec{k}}_{2T},t)
\phi_V(x_2,{\vec{k}}_{2T},p_2,t).
\end{eqnarray}
Usually it is convenient to compute the amplitude in coordinate
space. Through Fourier transformation, the above equation can be
expressed by:
\begin{eqnarray}
{\cal M}=\int^1_0dx_1dx_2\int
{d^2{\vec{b}}_{1}}{d^2{\vec{b}}_{2}}{\phi}_B(x_1,{\vec{b}}_{1},p_1,t)
T_H(x_1,x_2,{\vec{b}}_{1},{\vec{b}}_{2},t){\phi}_V(x_2,{\vec{b}}_{2},p_2,t).
\end{eqnarray}

\begin{figure}
\begin{center}
\includegraphics[width=3.5cm]{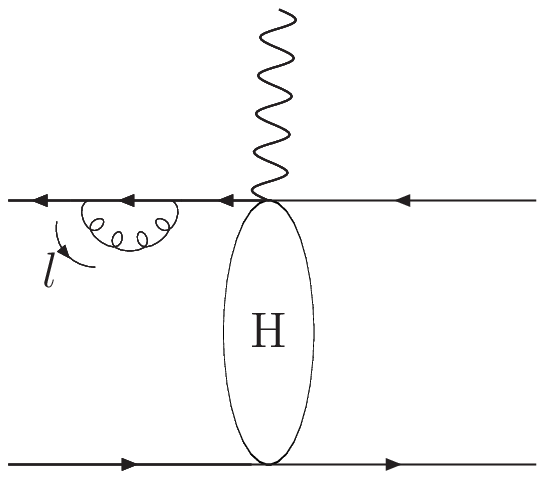}
\hspace{3mm}
\includegraphics[width=3.5cm]{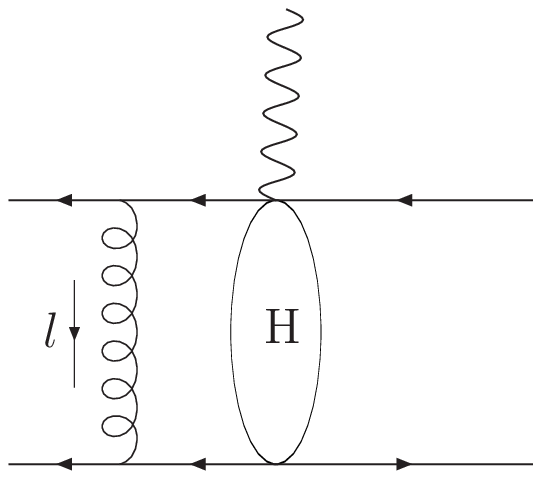}
\vspace{5mm}
\includegraphics[width=3.5cm]{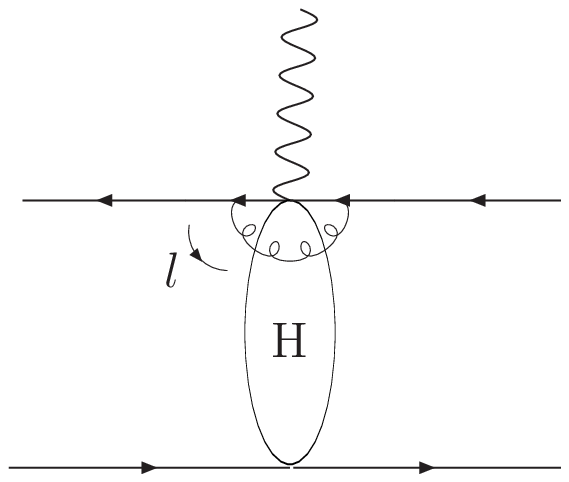}
\hspace{3mm}
\includegraphics[width=3.5cm]{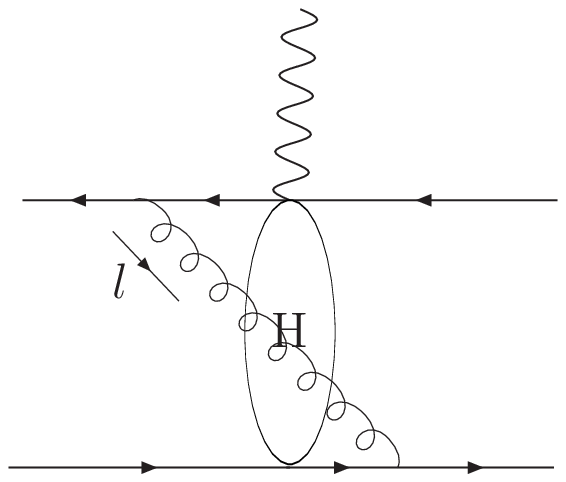}
\end{center}
\vspace{-.7cm} \caption{${O(\alpha_{s})}$ corrections to the hard
scattering kernel ${H}$.} \label{Feyn:alpha}
\end{figure}

This derivation is mainly concentrated on tree level diagrams, but
actually we have to take into account some loop effects which can
give sizable corrections. The ${\cal O}(\alpha_s)$ radiative
corrections to hard scattering process ${H}$ are depicted in
Fig.~\ref{Feyn:alpha}. In general, individual higher order diagrams
may suffer from two types of infrared divergences: soft and
collinear. Soft divergence comes from the region of a loop momentum
where all it's momentum components vanish:
\begin{eqnarray}
l^{\mu}=(l^+,l^-,\vec{l}_T)=(\Lambda,\Lambda,\vec{\Lambda}),
\end{eqnarray}
where $\Lambda$ is the typical scale for hadronization. Collinear
divergence originates from the gluon momentum region which is
parallel to the massless quark momentum,
\begin{eqnarray}
l^{\mu}=(l^+,l^-,\vec{l}_T) \sim
(m_B,{\Lambda}^2/m_B,\vec{\Lambda}).
\end{eqnarray}
In both cases, the loop integration corresponds to ${\int d^4 l/l^4
\sim \log{\Lambda}}$, thus logarithmic divergences are generated. It
has been shown order by order in perturbation theory that these
divergences can be separated from the hard kernel and absorbed into
meson wave functions using eikonal approximation \cite{Li:1994iu}.
But when soft and collinear momentum overlap, there will be double
logarithm divergences in the first two diagrams of
Fig.~\ref{Feyn:alpha}. These large double logarithm can be resummed
into the Sudakov factor whose explicit form is given in Appendix
\ref{PQCDfunctions}.

Furthermore, there are also another type of double logarithm which
comes from the loop correction for the weak decay vertex correction.
The left diagram in Fig.~\ref{Feyn:O7gamma}  gives an amplitude
proportional to $1/(x_2^2 x_1)$. In the threshold region with
$x_2\to 0$  [{(to be precise, $x_2\sim O(\Lambda_{QCD}/m_B)$)}],
additional soft divergences are associated with the internal quark
at higher orders. The QCD loop corrections to the electro-weak
vertex can produce the double logarithm $\alpha_s\ln^2 x_2$ and
resummation of this type of double logarithms lead to the Sudakov
factor $S_t(x_2)$. Similarly, resummation of $\alpha_s\ln^2 x_1$ due
to loop corrections in the other diagram leads to the Sudakov factor
$S_t(x_1)$. These double logarithm can also be factored out from the
hard part and grouped into the quark jet function. Resummation of
the double logarithms results in the threshold factor
\cite{Li:2001ay}. This factor decreases faster than any other power
of ${x}$ as ${x\rightarrow 0}$, which modifies the behavior in the
endpoint region to make pQCD approach more self-consistent. For
simplicity, this factor has been parameterized in a form which is
independent on channels, twists and flavors \cite{Li:2002mi}.

Combing all the elements together, we can get the typical
factorization formulae in pQCD approach:
\begin{eqnarray}
{\cal M}&=&\int^1_0dx_1dx_2\int
{d^2{\vec{b}}_{1}}{d^2{\vec{b}}_{2}}{(2\pi)^2}{\phi}_B(x_1,{\vec{b}}_{1},p_1,t)\nonumber\\
&&\times
T_H(x_1,x_2,Q,{\vec{b}}_{1},{\vec{b}}_{2},t){\phi}_V(x_2,{\vec{b}}_{2},p_2,t)
S_t(x_2)\exp[-S_B(t)-S_2(t)].
\end{eqnarray}

\subsection{Wave functions of $B$ mesons}

In order to calculate the analytic formulas of decay amplitudes, we
will use light cone wave functions $\Phi_{M,\alpha\beta}$ decomposed
in terms of the spin structure. In general, $\Phi_{M,\alpha\beta}$
with Dirac indices $\alpha,\beta$ can be decomposed into 16
independent components, $1_{\alpha\beta}$, $\gamma_{5\alpha\beta}$,
$\gamma^\mu_{\alpha\beta}$, $(\gamma^\mu\gamma_5)_{\alpha\beta}$,
$\sigma^{\mu\nu}_{\alpha\beta}$. If the considered meson is the $B$
meson, a heavy pseudo-scalar meson, the $B$ meson light-cone matrix
element can be
decomposed~\cite{Grozin:1996pq,Beneke:2000wa,Kawamura:2001jm} by:
\begin{eqnarray}
&&\int d^4ze^{ik_1\cdot z}
\langle 0|{b}_\beta(0)\bar D_\alpha(z)| {\bar B}(P_{B})\rangle \nonumber\\
&=&\frac{i}{\sqrt{2N_c}}\left\{(\not\! P_{B}+m_{B})\gamma_5
\left[\phi_{B} ({ k_1})-\frac{\not n -\not v }{\sqrt{2}}
\bar{\phi}_{B}({ k_1})\right]\right\}_{\beta\alpha},
\label{eq:BLCDA}
\end{eqnarray}
where $N_c=3$ is the color factor. $n$ and $v$ are two light-like
vectors: $n^2=v^2=0$. From equation (\ref{eq:BLCDA}), one can see
that there are two Lorentz structures in the $B$ meson distribution
amplitudes. They obey the following normalization conditions:
\begin{equation}
 \int\frac{d^4 k_1}{(2\pi)^4}\phi_{B}({
k_1})=\frac{f_{B}}{2\sqrt{2N_c}}, ~~~\int \frac{d^4
k_1}{(2\pi)^4}\bar{\phi}_{B}({ k_1})=0.
\end{equation}

In general, one should consider these two Lorentz structures in the
calculations of $B$ meson decays. However, it is found that the
contribution of $\bar{\phi}_{B}$ is numerically small
\cite{Lu:2002ny}, thus its contribution can be safely neglected.
With this approximation, we only retain the first term in the square
bracket from the full Lorentz structure in Eq.~(\ref{eq:BLCDA}):
\begin{equation}
 \Phi_{B}= \frac{i}{\sqrt{2N_c}} (\not \! P_{B} +m_{B}) \gamma_5
\phi_{B} ({ k_1}). \label{bmeson}
\end{equation}
In the following calculation, we will see that the hard part is
always independent of one of the $k_1^+$ and/or $k_1^-$. The ${B}$
meson wave function is then a function of the variables $k_1^-$ (or
$k_1^+$) and $k_{1}$ only,
\begin{eqnarray}
\phi_{B} (k_1^-, \vec k_{1T})=\int \frac{d k_1^+}{2\pi} \phi_{B}
(k_1^+, k_1^-, \vec k_{1T}). \label{int}
\end{eqnarray}
In the $b$-space, the $B$ meson's wave function can be expressed by
\begin{eqnarray}
 \Phi_{B}(x,b) = \frac{i}{\sqrt{2N_c}}
\left[ \not \! P_{B} \gamma_5 + m_{B} \gamma_{5} \right]
\phi_{B}(x,b),
\end{eqnarray}
where $b$ is the conjugate space coordinate of the transverse
momentum $k_T$.

In this study, we use the following phenomenological wave
function:
\begin{equation}
\phi_{B}(x,b) = N_{B} x^2(1-x)^2 \exp \left[ -\frac{m_{B}^2\
x^2}{2 \omega_b^2} -\frac{1}{2} (\omega_b b)^2
\right],\label{Eq:waveb}
\end{equation}
with $N_{B}$ the normalization factor. In recent years, a lot of
studies have been performed for $B_d^0$ and $B^\pm$ decays in pQCD
approach. The parameter $\omega_b=(0.40\pm 0.05)~\mathrm{GeV}$ has
been fixed  using the rich experimental data on $B_d^0$ and $B^\pm$
meson decays. In the SU(3) symmetry limit, this parameter should be
the same in $B_s$ decays. Considering a small SU(3) breaking, $s$
quark momentum fraction should be a little larger than that of $u$
or $d$ quark in the lighter $B$ mesons, since $s$ quark is heavier
than $u$ or $d$ quark. We will use $\omega_b=(0.50\pm
0.05)~\mathrm{GeV}$ in this paper for the $B_s$ decays
\cite{Ali:2007ff}.

\subsection{Light-cone distribution amplitudes of light vector mesons}

Decay constants for vector mesons are defined by:
\begin{equation}
\langle 0|\bar q_1\gamma_\mu
q_2|V(p,\epsilon)\rangle=f_Vm_V\epsilon_\mu,\;\;\; \langle 0|\bar
q_1\sigma_{\mu\nu}q_2|V(p,\epsilon)\rangle =if^T_V(\epsilon_\mu
p_\nu-\epsilon_\nu p_\mu).
\end{equation}
The longitudinal decay constants of charged vector mesons can be
extracted from the data on $\tau^- \to (\rho^-,K^{*-}) \nu_\tau$
\cite{Yao:2006px}. Neutral vector meson's longitudinal decay
constant can be determined by its electronic decay width through
$V^0\to e^+e^-$ \footnote{There is a recent study on extracting
vectors' decay constants from experimental data, which has taken
into account the effects of $\rho^0$-$\omega$ and $\omega$-$\phi$
mixing \cite{Ball:2006eu}.   Since we do not consider the mixing for
the decay amplitudes in our calculation, we will not use those
values for self-consistence.} and results are given in
Table~\ref{Table:Vdecayconstant}. Transverse decay constants are
mainly explored by QCD sum rules~\cite{Ball:2006eu} that are also
collected in Table~\ref{Table:Vdecayconstant}.

\begin{table}
\caption{Input values of the decay constants  for the vector mesons
(in MeV)}
\begin{tabular}{cccccccccc}
\hline\hline
   $f_\rho $   & $ f_\rho^T $   & $ f_\omega $ & $ f_\omega^T$
\ \ \
 & $ f_{K^*} $ & $ f_{K^*}^T $  & $f_\phi $    & $ f_\phi^T $  \\
\ \ \
   $ 209\pm 2$ & $ 165\pm 9$    & $ 195\pm 3$  & $ 151\pm 9$
\ \ \
 & $ 217\pm 5$ & $185\pm 10$    & $ 231\pm 4$  & $ 186\pm 9$\\
\hline \hline
\end{tabular}\label{Table:Vdecayconstant}
 \end{table}

We choose the vector meson momentum $P$ with $P^2=m_V^2$, which is
mainly on the plus direction. The polarization vectors $\epsilon$,
satisfying $P\cdot \epsilon=0$, include one longitudinal
polarization vector $\epsilon_L$ and two transverse polarization
vectors $\epsilon_T$. The vector meson distribution amplitudes up
to twist-3  are defined by:
\begin{eqnarray}
\langle V(P,\epsilon^*_L)|\bar q_{2\beta}(z) q_{1\alpha}
(0)|0\rangle &=&\frac{1}{\sqrt{2N_c}}\int_0^1 dx e^{ixP\cdot z}
\left[m_V\not\! \epsilon^*_L \phi_V(x) +\not\! \epsilon^*_L\not\!
P \phi_{V}^{t}(x) +m_V \phi_V^s(x)\right]_{\alpha\beta},
\nonumber\\
 \langle V(P,\epsilon^*_T)|\bar q_{2\beta}(z) q_{1\alpha}
(0)|0\rangle &=&\frac{1}{\sqrt{2N_c}}\int_0^1 dx e^{ixP\cdot z}
\left[ m_V\not\! \epsilon^*_T\phi_V^v(x)+ \not\!\epsilon^*_T\not\!
P\phi_V^T(x)\right.
\nonumber\\
& & \left.\;\;\;\;\;+m_V
i\epsilon_{\mu\nu\rho\sigma}\gamma_5\gamma^\mu\epsilon_T^{*\nu}
n^\rho v^\sigma \phi_V^a(x)\right ]_{\alpha\beta}\;, \label{spf}
\end{eqnarray}
for the longitudinal polarization and transverse polarization,
respectively. Here  $x$ is the momentum fraction associated with
the $q_2$ quark.  $n$ is the moving direction of the vector meson
and $v$ is the opposite direction. These distribution amplitudes
can be related to the ones used in QCD sum rules by:
\begin{eqnarray}
&&\phi_{V}(x)=\frac{f_{V}}{2\sqrt{2N_c}}\phi_{||}(x),\;\;\;
\phi_{V}^t(x)=\frac{f_{V}^T}{2\sqrt{2N_c}}h_{||}^{(t)}(x),\nonumber\\
&&\phi_{V}^s(x)=\frac{f_{V}^T}{4\sqrt{2N_c}}
\frac{d}{dx}h_{||}^{(s)}(x),\hspace{3mm}
\phi_{V}^T(x)=\frac{f_{V}^T}{2\sqrt{2N_c}}\phi_{\perp}(x)
,\nonumber\\
&&\phi_{V}^v(x)=\frac{f_{V}}{2\sqrt{2N_c}}g_{\perp}^{(v)}(x),
\hspace{3mm}\phi_{V}^a(x)=\frac{f_{V}}{8\sqrt{2N_c}}
\frac{d}{dx}g_{\perp}^{(a)}(x).
\end{eqnarray}

The twist-2 distribution amplitudes can be expanded in terms of
Gegenbauer polynomials $C_n^{3/2}$ with the coefficients called
Gegenbauer moments $a_n$:
\begin{eqnarray}
 \phi_{||,\perp} (x)&=&6x (1-x)
 \left[1+\sum_{n=1}^{\infty} a_{n}^{||,\perp}C_n^{3/2} (t) \right],\label{phiV}
\end{eqnarray}
where $t=2x-1$. The Gegenbauer moments $a_{n}^{||,\perp}$ are mainly
determined by the technique of QCD sum rules. Here we quote the
recent numerical results
\cite{Braun:2004vf,Ball:2005vx,Ball:2006nr,Ball:2007rt} as
\begin{eqnarray}
 a_1^\parallel(K^*)  & = & 0.03 \pm 0.02,\;\;\;\;\;    a_1^\perp(K^*)   =0.04 \pm 0.03,\\
 a_2^\parallel(\rho) & = & a_2^\parallel(\omega)= 0.15 \pm 0.07,\;\;\;
 a_2^\perp(\rho)   =a_2^\perp(\omega)   =0.14 \pm 0.06,\\
 a_2^\parallel(K^*)  & = & 0.11 \pm 0.09,\;\;\;\;\;    a_2^\perp(K^*)   =0.10 \pm 0.08,\\
 a_2^\parallel(\phi) & = & 0.18 \pm 0.08,\;\;\;\;\;    a_2^\perp(\phi)  =0.14 \pm 0.07,
\end{eqnarray}
where the values are taken at $\mu=1$ GeV.

Using equation of motion, two-particle twist-3 distribution
amplitudes are related to twist-2 LCDAs and three-particle twist-3
distribution amplitudes. But in some $B\to VV$ decays, there exists
the so-called polarization problem. It has been suggested that using
asymptotic LCDAs can resolve this problem in pQCD approach. Thus to
be self-consistent, we should also use the same form to calculate
radiative decays. As in Ref.~\cite{Ali:2007ff}, we use the
asymptotic form for twist-3 LCDAs:
\begin{eqnarray}
h_\parallel^{(t)}(x) & = & 3t^2 ,\;\;\;
h_{||}^{(s)}(x)   =  6 x(1-x),\\
g_\perp^{(a)}(x) & = & 6 x(1-x) ,\;\;\; g_\perp^{(v)}(x)  =
\frac{3}{4}\,(1+t^2).
\end{eqnarray}

\subsection{Light-cone distribution amplitudes of axial-vectors}

Longitudinal and transverse decay constants for axial-vectors are
defined by:
 \begin{eqnarray}
  \langle A(P,\epsilon)|\bar q_2 \gamma_\mu \gamma_5 q_1|0\rangle
   &= & if_{A}  m_{A} \,  \epsilon^{*}_\mu,\;\;\;
 \langle A(P,\epsilon)|
  \bar q_2 \sigma_{\mu\nu}\gamma_5 q_1
   |0\rangle
  =  f_{A}^{T}
 (\epsilon^{*}_{\mu} P_{\nu} - \epsilon_{\nu}^{*} P_{\mu}).
 \end{eqnarray}
In SU(2) limit, due to G-parity, the longitudinal (transverse) decay
constants vanish for the non-strange $^1P_1$[$^3P_1$] states. This
will affect the normalization for the corresponding distribution
amplitudes which will be discussed in the following. For
convenience, we take $f_{^3\! P_1}\equiv f$\, [$f_{^1\!
P_1}^T(\mu=1~{\rm GeV})\equiv f$] as the ``normalization constant".
The numbers of axial vector meson decay constants shown in
table~\ref{Table:Adecayconstant} are taken from Ref.
\cite{Yang:2005gk,Yang:2007zt}.

\begin{table}
\caption{Input values of the decay constants (absolute values) for
the axial-vector mesons (in MeV). The transverse decays constants
for $^1P_1$ are evaluated at $\mu =1$ GeV.}
\begin{tabular}{cccccccccc}
\hline\hline
   $f_{a_1(1260)} $   & $ f_{ f_1 (1^3P_1)} $   & $ f_{ f_8 (1^3P_1)} $ & $ f_{ K_{1A}}$
\ \ \
  & $ f^T_{b_1(1235)} $ & $ f^T_{h_1 (1^1P_1)} $  & $f^T_{h_8 (1^1P_1)} $    & $ f^T_{K_{1B}} $  \\
\ \ \
   $ 238\pm 10$ & $ 245\pm 13$    & $ 239\pm 13$  & $ 250\pm 13$
\ \ \
 & $  180\pm 8$ & $180\pm 12$    & $ 190\pm 10$  & $ 190\pm 10$\\
\hline \hline
\end{tabular}\label{Table:Adecayconstant}
 \end{table}

Distribution amplitudes for axial-vectors with quantum numbers
$J^{PC}=1^{++}$ or $1^{+-}$ are defined by:
\begin{eqnarray}
\langle A(P,\epsilon^*_L)|\bar q_{2\beta}(z) q_{1\alpha}
(0)|0\rangle &=&\frac{-i}{\sqrt{2N_c}}\int_0^1 dx e^{ixp\cdot z}
\left[m_A\not\! \epsilon^*_L \gamma_5\phi_A(x) -\not\!
\epsilon^*_L\not\! P \gamma_5\phi_{A}^{t}(x) -m_A\gamma_5
\phi_A^s(x)\right]_{\alpha\beta},
\nonumber\\
 \langle A(P,\epsilon^*_T)|\bar q_{2\beta}(z) q_{1\alpha}
(0)|0\rangle &=&\frac{-i}{\sqrt{2N_c}}\int_0^1 dx e^{ixp\cdot z}
\left[m_A\not\! \epsilon^*_T\gamma_5\phi_A^v(x) -
\not\!\epsilon^*_T\not\! P\gamma_5\phi_A^T(x)\right.
\nonumber\\
& & \left.\;\;\;\;\;-m_A
i\epsilon_{\mu\nu\rho\sigma}\gamma^\mu\epsilon_T^{*\nu} n^\rho
v^\sigma \phi_A^a(x)\right ]_{\alpha\beta}\;.
\label{eq:transverseWVA}
\end{eqnarray}
Besides the factor $-i\gamma_5$ from the right hand, axial-vector
mesons' distribution amplitudes can be related to the vector ones by
making the following replacement:
\begin{eqnarray}
\phi_V\to \phi_A,\;\;\phi_V^t\to -\phi_A^t,\;\;\phi_V^s\to
-\phi_A^s,\nonumber\\
\phi_V^T\to -\phi_A^T,\;\;\phi_V^v\to \phi_A^v,\;\;\phi_V^a\to
\phi_A^a.\label{eq:relationofAVDA}
\end{eqnarray}
These distribution amplitudes can be related to the ones
calculated in QCD sum rules by:
\begin{eqnarray}
 &&\phi_{A}(x)=\frac{f}{2\sqrt{2N_c}}\phi_{||}(x), \;\;\;\phi_{A}^t(x)=\frac{f}{2\sqrt{2N_c}}h_{||}^{(t)}(x),\nonumber\\
&&\hspace{-5mm}\phi_{A}^s(x)=\frac{f}{4\sqrt{2N_c}}
\frac{d}{dx}h_{\parallel}^{(s)}(x),\hspace{3mm}
\phi_{A}^T(x)=\frac{f}{2\sqrt{2N_c}}\phi_{\perp}(x)
,\nonumber\\
&&\hspace{-5mm}\phi_{A}^v(x)=\frac{f}{2\sqrt{2N_c}}g_{\perp}^{(v)}(x),
\hspace{3mm}\phi_{A}^a(x)=\frac{f}{8\sqrt{2N_c}}
\frac{d}{dx}g_{\perp}^{(a)}(x),
\end{eqnarray}
where we use $f$ as the ``normalization" constant for both
longitudinal polarized and transversely polarized mesons.

In SU(2) limit, due to G-parity, $\phi_\parallel$, $g_\perp^{(a)}$
and $g_\perp^{(v)}$ are symmetric [antisymmetric] under the
replacement $x\leftrightarrow 1-x$ for non-strange $1^3P_1$
[$1^1P_1$] states, whereas $\phi_\perp$, $h_{||}^{(t)}$, and
$h_{||}^{(s)}$ are antisymmetric [symmetric]. In the above, we have
taken $f_{^3\! P_1}^T=f_{^3\! P_1}=f$\, [$f_{^1\! P_1}=f_{^1\!
P_1}^T(\mu=1~{\rm GeV})=f$], thus we have
 \begin{eqnarray}
 \langle 1^3P_1(P,\epsilon)|
  \bar q_1 \sigma_{\mu\nu}\gamma_5 q_2
   |0\rangle
  &= & f_{^3P_1}^{T} a_0^{\perp,^3P_1} \,
(\epsilon^{*}_{\mu} P_{\nu} - \epsilon_{\nu}^{*} P_{\mu}),
 \\
  \langle 1^1P_1(P,\epsilon)|\bar q_1 \gamma_\mu \gamma_5 q_2|0\rangle
   &= & if_{^1P_1} a_0^{\parallel,^1P_1} \, m_{^1P_1} \,  \epsilon^{*}_\mu,
 \end{eqnarray}
where $a_0^{\perp,^3P_1}$ and  $a_0^{\parallel,^1P_1}$ are the
Gegenbauer zeroth moments. That can give the following
normalization for the distribution amplitudes:
\begin{eqnarray}
 \int_0^1 dx \phi_\perp (x) &=& a^\perp_0\Bigg[
 \int_0^1 dx \phi_{||}(x) =
 a^{||}_0\Bigg],\label{eq:normaizlationofA}
 \end{eqnarray}
for the $1^3P_1$ [$1^1P_1$] states. The zeroth Gegenbauer moments
$a_0^{\perp,^3P_1}$ and $a_0^{\parallel,^1P_1}$, characterizing the
degree of the flavor SU$(3)$ symmetry breaking, are non-zero for
only strange mesons. We normalize the distribution amplitude
$\phi_\parallel$ of the $1^3P_1$ states as
 \begin{equation}
 \int_0^1 dx\phi_{||}(x)=1.
 \end{equation}
For convenience, we formally define $a_0^{||}=1$ for the $1^3P_1$
states so that we can use Eq.~(\ref{eq:normaizlationofA}) as the
normalization condition. Similarly, we also define $a_0^{\perp}=1$
for $1^1P_1$ states so that $\phi_\perp (x)$ has a correct
normalization.

%
\begin{table}[ttb]
\caption{Gegenbauer moments of $\phi_\perp$ and $\phi_{||}$ for
$1^3P_1$ and $1^1P_1$ mesons evaluated in Ref. \cite{Yang:2007zt},
where the values are taken at $\mu=1$ GeV.}
\label{tab:AxialGegenbauer} $
\begin{array}{|c|c|c|c|c|c|}
 \hline
  a_2^{||, a_1(1260)}       & a_2^{||,f_1^{^3P_1}}        & a_2^{||,f_8^{^3P_1}}         & a_2^{||, K_{1A}}
                            & \multicolumn{2}{|c|}{a_1^{||, K_{1A}}}\\
  \hline
 -0.02\pm 0.02              &  -0.04\pm 0.03 & -0.07\pm 0.04 & -0.05\pm 0.03
                            & \multicolumn{2}{c|}{ {0.00\pm 0.26}}\\
 \hline
  a_1^{\perp, a_1(1260)}    & a_1^{\perp,f_1^{^3P_1}}     & a_1^{\perp,f_8^{^3P_1}}      & a_1^{\perp, K_{1A}}
                            & a_0^{\perp, K_{1A}}         & a_2^{\perp, K_{1A}}\\
  \hline
 -1.04\pm 0.34              & -1.06\pm 0.36               & -1.11\pm 0.31                & -1.08\pm 0.48
                            &  0.08\pm 0.09               & 0.02\pm 0.20\\
 \hline
 a_1^{||, b_1(1235)}        & a_1^{||,h_1^{^1P_1}}        &a_1^{||,h_8^{^1P_1}}          & a_1^{||, K_{1B}}
                            & a_0^{||, K_{1B}}            & a_2^{||, K_{1B}} \\
 \hline
    -1.95\pm 0.35           & -2.00\pm 0.35               & -1.95\pm 0.35                & -1.95\pm 0.45
                            &  0.14\pm 0.15               &  0.02\pm 0.10\\
 \hline
  a_2^{\perp, b_1(1235)}    & a_2^{\perp,h_1^{^1P_1}}     & a_2^{\perp,h_8^{^1P_1}}      & a_2^{\perp, K_{1B}}
                            & \multicolumn{2}{|c|}{a_1^{\perp, K_{1B}}} \\
 \hline
            0.03\pm 0.19    &  0.18\pm 0.22               &   0.14\pm 0.22               & -0.02\pm 0.22
                            &\multicolumn{2}{c|}{0.17\pm 0.22}\\
 \hline
\end{array}
$
\end{table}

Up to conformal spin 6, twist-2 distribution amplitudes for
axial-vector mesons can be expanded as:
\begin{eqnarray}
 \phi_\parallel(x) & = & 6 x \bar x \left[ a_0^\parallel + 3
a_1^\parallel\, t +
a_2^\parallel\, \frac{3}{2} ( 5t^2  - 1 ) \right], \label{eq:lcda-1p1-t2-1}\\
 \phi_\perp(x) & = & 6 x \bar x \left[ a_0^\perp + 3 a_1^\perp\, t +
a_2^\perp\, \frac{3}{2} ( 5t^2  - 1 ) \right],
\label{eq:lcda-3p1-t2-2}
\end{eqnarray}
wher the Gegenbauer moments are calculated in Refs.
\cite{Yang:2005gk,Yang:2007zt} shown in table
\ref{tab:AxialGegenbauer}. From the results in table
\ref{tab:AxialGegenbauer}, we can see that there are large
uncertainties in Gegenbauer moments which can inevitably induce
large uncertainties to branching ratios and CP asymmetries. We hope
the uncertainties could be reduced in future studies in order to
make more precise predictions.

As for twist-3 LCDAs, we use the following form:
\begin{eqnarray}
 g_\perp^{(v)}(x) & = & \frac{3}{4} a_0^\parallel (1+t^2)
 + \frac{3}{2}\, a_1^\parallel\, t^3,\;\;\;
 g_\perp^{(a)}(x)  =  6 x \bar x ( a_0^\parallel + a_1^\parallel t),
 \\
 h_\parallel^{(t)}(x) &= & 3a_0^\perp t^2+ \frac{3}{2}\,a_1^\perp\,t (3 t^2-1)
 ,\;\;\;
 h_\parallel^{(s)}(x)  =  6x\bar x( a_0^\perp + a_1^\perp t ).
 \end{eqnarray}

\section{$B\to V$ and $B\to A$  form factors}\label{section:formfactor}

$\bar B\to V$ form factors are defined under the conventional form
as follows:
 \begin{eqnarray}
  \langle V(P_2,\epsilon^*)|\bar q\gamma^{\mu}b|\bar B(P_1)\rangle
   &=&-\frac{2V(q^2)}{m_B+m_V}\epsilon^{\mu\nu\rho\sigma}
     \epsilon^*_{\nu}P_{1\rho}P_{2\sigma}, \nonumber\\
  \langle V(P_2,\epsilon^*)|\bar q\gamma^{\mu}\gamma_5 b|\bar
  B(P_1)\rangle
   &=&2im_V A_0(q^2)\frac{\epsilon^*\cdot q}{q^2}q^{\mu}
    +i(m_B+m_V)A_1(q^2)\left[\epsilon^*_{\mu}
    -\frac{\epsilon^*\cdot q}{q^2}q^{\mu} \right] \nonumber\\
    &&-iA_2(q^2)\frac{\epsilon^*\cdot q}{m_B+m_V}
     \left[ (P_1+P_2)^{\mu}-\frac{m_B^2-m_V^2}{q^2}q^{\mu} \right],\nonumber\\
  \langle V(P_2,\epsilon^*)|\bar q\sigma^{\mu\nu}q_{\nu}b|\bar
  B(P_1)\rangle
   &=&-2iT_1(q^2)\epsilon^{\mu\nu\rho\sigma}
     \epsilon^*_{\nu}P_{1\rho}P_{2\sigma}, \nonumber\\
  \langle V(P_2,\epsilon^*)|\bar q\sigma^{\mu\nu}\gamma_5q_{\nu}b|\bar
  B(P_1)\rangle
   &=&T_2(q^2)\left[(m_B^2-m_V^2)\epsilon^{*\mu}
       -(\epsilon^*\cdot q)(P_1+P_2)^{\mu} \right]\nonumber\\
   &&+T_3(q^2)(\epsilon^*\cdot q)\left[
       q^{\mu}-\frac{q^2}{m_B^2-m_V^2}(P_1+P_2)^{\mu}\right],
 \end{eqnarray}
where $q=P_1-P_2$ and the relation
$2m_VA_0(0)=(m_B+m_V)A_1(0)-(m_B-m_V)A_2(0)$ is obtained in order
to cancel the pole at $q^2=0$.

In pQCD approach, the factorization formulae for these form factors
at maximally recoiling ($q^2=0$) are expressed by:
\begin{eqnarray}
V&=&8\pi C_Fm_B(m_B+m_V)\int_0^1 dx_1dx_2\int_0^\infty
b_1db_1b_2db_2\phi_B(x_1,b_1)
 \nonumber\\
 &&\times \bigg\{E_e(t_a) h_e(x_1x_2,x_2,b_1,b_2)
 \Big[\phi_V^T(x_2)+(2+x_2)r_V\phi_V^a(x_2)-r_Vx_2\phi_V^v(x_2)\Big]\nonumber\\
 &&\;\;\;+ E_e'(t_a') h_e(x_1x_2,x_1,b_2,b_1)r_V\Big[\phi_V^a(x_2)+\phi_V^v(x_2)\Big]\bigg\},\label{eq:formfactorBtoV_V}\\
A_0&=&8\pi C_Fm_B^2\int_0^1 dx_1dx_2\int_0^\infty
b_1db_1b_2db_2\phi_B(x_1,b_1)
 \nonumber\\
 &&\times \bigg\{E_e(t_a) h_e(x_1x_2,x_2,b_1,b_2)
 \Big[(1+x_2)\phi_V(x_2)+(1-2x_2)r_V(\phi_V^s(x_2)+\phi_V^t(x_2))\Big]\nonumber\\
 &&\;\;\;+ 2r_VE_e'(t_a') h_e(x_1x_2,x_1,b_2,b_1)\phi_V^s(x_2)\bigg\},\\
A_1&=&8\pi C_Fm_B(m_B-m_V)\int_0^1 dx_1dx_2\int_0^\infty
b_1db_1b_2db_2\phi_B(x_1,b_1)
 \nonumber\\
 &&\times \bigg\{E_e(t_a) h_e(x_1x_2,x_2,b_1,b_2)
 \Big[\phi_V^T(x_2)+(2+x_2)r_V\phi_V^v(x_2)-r_Vx_2\phi_V^a(x_2)\Big]\nonumber\\
 &&\;\;\;+ E_e'(t_a') h_e(x_1x_2,x_1,b_2,b_1)r_V\Big[\phi_V^a(x_2)+\phi_V^v(x_2)\Big]\bigg\},\\
A_2&=&\frac{1}{m_B-m_V}\Big[(m_B+m_V)A_1-2m_VA_0\Big],
\end{eqnarray}
\begin{eqnarray}
T_1&=& T_2\nonumber\\
&=&8\pi C_Fm_B^2\int_0^1 dx_1dx_2\int_0^\infty
b_1db_1b_2db_2\phi_B(x_1,b_1)
 \nonumber\\
 &&\times \bigg\{E_e(t_a) h_e(x_1x_2,x_2,b_1,b_2)
 \Big[(1+x_2)\phi_V^T(x_2)+(1-2x_2)r_V(\phi_V^v(x_2)+\phi_V^a(x_2))\Big]\nonumber\\
 &&\;\;\;+ E_e'(t_a') h_e(x_1x_2,x_1,b_2,b_1)r_V(\phi_V^v(x_2)+\phi_V^a(x_2))\bigg\},
 \label{fft1}\\
T_3&=&T_2-16\pi C_Fm_B^2r_V\int_0^1 dx_1dx_2\int_0^\infty
b_1db_1b_2db_2\phi_B(x_1,b_1)
 \nonumber\\
 &&\times \bigg\{E_e(t_a) h_e(x_1x_2,x_2,b_1,b_2)
 \Big[\phi_V(x_2)+2r_V\phi_V^t(x_2)+r_Vx_2(\phi_V^t(x_2)-\phi_V^s(x_2))\Big]\nonumber\\
 &&\;\;\;+ 2r_VE_e'(t_a')
 h_e(x_1x_2,x_1,b_2,b_1)\phi_V^s(x_2)\bigg\},\label{ffv}
\end{eqnarray}
with $C_F=4/3$. The definitions of functions $E_i$, the
factorization scales $t_i$ and hard functions $h_i$, are given in
Appendix~\ref{PQCDfunctions}.

\begin{table}
\caption{$B\to V$ form factors at maximally recoil, i.e. $q^2=0$.
The first error comes from decay constants of $B$ mesons and shape
parameters $\omega_b$; while the second one is from hard scale $t$
and $\Lambda_{QCD}$.  }
\begin{center}
\begin{tabular}{cc|c|c|c|c|c}
\hline \hline
         &       & $B\to\rho$ &$B\to K^*$     &$B\to\omega$ &$B_s\to K^*$  &$B_s\to\phi$ \\
 \hline
LFQM\cite{Cheng:2003sm}&$V$ &$0.27$      &$ 0.31$        &             &              &        \\
      & $A_0$  &$0.28$      &$0.31$         &             &              &        \\
     & $A_1$  &$0.22$      &$0.26$         &             &              &        \\
      & $A_2$  &$0.20$      &$ 0.24$        &             &              &        \\
\hline
LCSR\cite{Ball:2004rg}&$V$&$0.323$  &$0.411$        &$0.293$      &$0.311$       &$0.434$ \\
      & $A_0$  &$0.303$     &$0.374$        &$0.281$      &$0.360$       &$0.474$ \\
      & $A_1$  &$0.242$     &$0.292$        & $0.219$     &$0.233$       &$0.311$ \\
      & $A_2$  &$0.221$     &$0.259$        &$0.198$      &$0.181$       &$0.234$ \\
      & $T_2$  &$0.267$     &$0.333$        &$0.242$      &$0.260$       &$0.349$ \\
\hline
LQCD\cite{DelDebbio:1997kr} &$V$&$0.35 $     &               &             &              &        \\
     & $A_0$  &$0.30 $     &               &             &              &        \\
     & $A_1$  &$0.27$      &               &             &              &        \\
      & $A_2$  &$0.26$      &               &             &              &        \\
  \cite{Becirevic:2006nm}  & $T_1$  &         &$0.24$  &             &              &        \\
\hline SCET LCQM\cite{Lu:2007sg}
     & $V$    &$0.298$     &$0.339$        &$0.275$      &$0.323$       &$0.329$ \\
     & $A_0$  &$0.260$     &$0.283$        &$0.240$      &$0.279$       &$0.279$ \\
     & $A_1$  &$0.227$     &$0.248$        &$0.209$      &$0.228$       &$0.232$ \\
      & $A_2$  &$0.215$     &$0.233$        &$0.198$      &$0.204$       &$0.210$ \\
     & $T_1=T_2$  &$0.260$     &$0.290$        &$0.239$      &$0.271$       &$0.276$ \\
     & $T_3$  &$0.184$     &$0.194$        &$0.168$      &$0.165$       &$0.170$\\
\hline
This work  & $V$    &$0.21_{-0.04-0.00}^{+0.05+0.00}$     &$0.25_{-0.05-0.00}^{+0.05+0.00}$        &$0.20_{-0.04-0.00}^{+0.04+0.00}$      &$0.21_{-0.03-0.01}^{+0.04+0.00}$   &$0.25_{-0.04-0.01}^{+0.05+0.00}$\\
       & $A_0$  &$0.25_{-0.05-0.01}^{+0.05+0.00}$     &$0.30_{-0.05-0.01}^{+0.06+0.00}$       &$0.24_{-0.04-0.01}^{+0.05+0.00}$      &$0.26_{-0.04-0.01}^{+0.05+0.00}$   &$0.30_{-0.05-0.01}^{+0.05+0.00}$\\
       & $A_1$  &$0.17_{-0.03-0.00}^{+0.04+0.00}$     &$0.19_{-0.03-0.00}^{+0.04+0.00}$       &$0.15_{-0.03-0.00}^{+0.03+0.00}$      &$0.16_{-0.02-0.01}^{+0.03+0.00}$   &$0.18_{-0.03-0.01}^{+0.03+0.00}$\\
       & $A_2$  &$0.13_{-0.02-0.00}^{+0.03+0.00}$     &$0.14_{-0.03-0.00}^{+0.03+0.00}$       &$0.13_{-0.02-0.00}^{+0.03+0.00}$      &$0.12_{-0.02-0.01}^{+0.02+0.00}$   &$0.13_{-0.02-0.01}^{+0.02+0.00}$\\
      & $T_1=T_2$  &$0.20_{-0.04-0.00}^{+0.04+0.00}$     &$0.23_{-0.04-0.00}^{+0.05+0.00}$       &$0.18_{-0.03-0.00}^{+0.04+0.00}$      &$0.19_{-0.03-0.01}^{+0.04+0.00}$   &$0.22_{-0.04-0.01}^{+0.04+0.00}$\\
      & $T_3$  &$0.13_{-0.02-0.00}^{+0.03+0.00}$     &$0.13_{-0.02-0.00}^{+0.03+0.00}$       &$0.12_{-0.02-0.00}^{+0.03+0.00}$      &$0.11_{-0.02-0.01}^{+0.02+0.00}$   &$0.12_{-0.02-0.01}^{+0.02+0.00}$\\
 \hline \hline
\end{tabular}\label{Tab:BtoVformfactor}
\end{center}
\end{table}

With the above expressions for form factors, we obtain the numerical
results that are collected in table~\ref{Tab:BtoVformfactor}. When
evaluating the form factor $A_0$, we used
$2m_VA_0(0)=(m_B+m_V)A_1(0)-(m_B-m_V)A_2(0)$ and assume $A_1$ and
$A_2$ are linearly correlated to estimate the uncertainties. The
first error comes from decay constants of $B_{(s)}$ meson and shape
parameters $\omega_b$; while the second one is from hard scale $t$
and $\Lambda_{QCD}$. In the calculation, $f_B=(0.19 \pm 0.02)\mbox{
GeV}$, $\omega_B=(0.40\pm 0.05)\mbox{ GeV}$ (for $B^\pm$ and $B_d^0$
mesons) and $f_{B_s}=(0.23 \pm 0.02)$ GeV, $\omega_{B_s}=(0.50 \pm
0.05)$ GeV (for $B_s^0$ meson) have been used. It is clear that
these hadronic parameters give the dominant theoretical
uncertainties. They quantify the SU(3)-symmetry breaking effects in
the form factors in pQCD approach. To make a comparison, we also
collect the results using other approaches
\cite{Cheng:2003sm,Ball:2004rg,DelDebbio:1997kr,Becirevic:2006nm,Lu:2007sg}.
From table~\ref{Tab:BtoVformfactor}, we can see that most of our
results are consistent with others within theoretical errors.

Likewise, $\bar B\to A$ form factors are defined by:
 \begin{eqnarray}
  \langle A(P_2,\epsilon^*)|\bar q\gamma^{\mu}\gamma_5 b|\bar B(P_1)\rangle
   &=&-\frac{2iA(q^2)}{m_B-m_A}\epsilon^{\mu\nu\rho\sigma}
     \epsilon^*_{\nu}P_{1\rho}P_{2\sigma}, \nonumber\\
  \langle A(P_2,\epsilon^*)|\bar q\gamma^{\mu}b|\bar
  B(P_1)\rangle
   &=&-2m_V V_0(q^2)\frac{\epsilon^*\cdot q}{q^2}q^{\mu}
    -(m_B-m_A)V_1(q^2)\left[\epsilon^*_{\mu}
    -\frac{\epsilon^*\cdot q}{q^2}q^{\mu} \right] \nonumber\\
    &&+V_2(q^2)\frac{\epsilon^*\cdot q}{m_B-m_A}
     \left[ (P_1+P_2)^{\mu}-\frac{m_B^2-m_A^2}{q^2}q^{\mu} \right],\nonumber\\
  \langle A(P_2,\epsilon^*)|\bar q\sigma^{\mu\nu}\gamma_5q_{\nu}b|\bar
  B(P_1)\rangle
   &=&-2T_1(q^2)\epsilon^{\mu\nu\rho\sigma}
     \epsilon^*_{\nu}P_{1\rho}P_{2\sigma}, \nonumber\\
  \langle A(P_2,\epsilon^*)|\bar q\sigma^{\mu\nu}q_{\nu}b|\bar
  B(P_1)\rangle
   &=&-iT_2(q^2)\left[(m_B^2-m_A^2)\epsilon^{*\mu}
       -(\epsilon^*\cdot q)(P_1+P_2)^{\mu} \right]\nonumber\\
   &&-iT_3(q^2)(\epsilon^*\cdot q)\left[
       q^{\mu}-\frac{q^2}{m_B^2-m_A^2}(P_1+P_2)^{\mu}\right],
 \end{eqnarray}
with a factor $-i$ different from $B\to V$ and the factor $m_B+m_V$
($m_B-m_V$) is replaced by $m_B-m_A$ ($m_B+m_A$). Similar with $B\to
V$ form factors, the relation $2m_AV_0=(m_B-m_A)V_1-(m_B+m_A)V_2$ is
obtained at $q^2=0$. In pQCD approach, $B\to A$ form factors'
formulas can be derived from the corresponding $B\to V$ form factor
formulas in eq.(\ref{eq:formfactorBtoV_V}-\ref{ffv}) using the
replacement in Eq.~(\ref{eq:relationofAVDA}) with the proper change
of the momentum fraction.

{\small
\begin{table}
\caption{$B\to A$ form factors at maximally recoil, i.e. $q^2=0$.
Results in the first line of each form factor are calculated using
$\theta_K=45^\circ$, $\theta_{^1P_1}=10^\circ$ or
$\theta_{^3P_1}=38^\circ$, while the second line corresponds to the
angle $\theta_K=-45^\circ$, $\theta_{^1P_1}=45^\circ$ or
$\theta_{^3P_1}=50^\circ$. The errors are from: decay constants of
$B_{(s)}$ meson and shape parameter $\omega_b$; Gegenbauer moments
in axial-vectors' LCDAs. }
\begin{center}
\begin{tabular}{cc|c|c|c|c|c|c}
\hline \hline
         &        &$B\to K_1(1270)$  &$B\to h_1(1170)$ &$B_s\to K_1(1270)$  &$B\to h_1(1380)$   & $B_s\to h_1(1170)$   & $B_s\to h_1(1380)$          \\
 \hline
\hline
\ \ \   & $A$       &$-0.05_{-0.01-0.05}^{+0.01+0.05}$       &$0.13_{-0.02-0.02}^{+0.03+0.02}$      &$-0.05_{-0.01-0.04}^{+0.01+0.04}$    &$0.07_{-0.01-0.01}^{+0.01+0.01}$          &$0.07_{-0.01-0.03}^{+0.01+0.00}$    &$-0.17_{-0.03-0.02}^{+0.03+0.02}$     \\
\ \ \   &           &$0.33_{-0.06-0.05}^{+0.07+0.05}$        &$0.14_{-0.03-0.02}^{+0.03+0.02}$      &$-0.11_{-0.02-0.04}^{+0.02+0.04}$    &$-0.02_{-0.01-0.00}^{+0.00+0.00}$         &$-0.03_{-0.01-0.04}^{+0.01+0.04}$   &$-0.19_{-0.03-0.02}^{+0.03+0.02}$\\
\hline   & $V_0$     &$0.11_{-0.02-0.10}^{+0.02+1.10}$        &$0.29_{-0.05-0.02}^{+0.06+0.02}$     &$-0.12_{-0.02-0.09}^{+0.02+0.09}$    &$0.16_{-0.03-0.02}^{+0.03+0.01}$          &$0.16_{-0.03-0.01}^{+0.03+0.01}$    &$-0.40_{-0.07-0.03}^{+0.06+0.03}$     \\
\ \ \   &           &$0.60_{-0.11-0.10}^{+0.12+0.10}$        &$0.32_{-0.06-0.03}^{+0.06+0.03}$      &$0.11_{-0.02-0.09}^{+0.03+0.09}$   &$-0.05_{-0.01-0.00}^{+0.01+0.00}$         &$-0.08_{-0.01-0.01}^{+0.01+0.01}$   &$-0.43_{-0.08-0.03}^{+0.07+0.03}$\\
\hline   & $V_1$     &$-0.09_{-0.02-0.08}^{+0.02+0.08}$       &$0.22_{-0.04-0.03}^{+0.04+0.03}$     &$-0.09_{-0.01-0.07}^{+0.01+0.07}$    &$0.12_{-0.02-0.02}^{+0.03+0.02}$          &$0.12_{-0.02-0.02}^{+0.02+0.01}$    &$-0.31_{-0.06-0.04}^{+0.05+0.04}$        \\
\ \ \   &           &$0.57_{-0.10-0.08}^{+0.12+0.08}$        &$0.24_{-0.04-0.03}^{+0.05+0.03}$      &$-0.20_{-0.04-0.07}^{+0.04+0.07}$    &$-0.04_{-0.01-0.01}^{+0.01+0.01}$         &$-0.05_{-0.01-0.01}^{+0.01+0.01}$   &$-0.34_{-0.06-0.04}^{+0.05+0.04}$\\
\hline   & $V_2$     &$-0.10_{-0.02-0.02}^{+0.01+0.02}$       &$0.03_{-0.01-0.01}^{+0.01+0.01}$     &$-0.01_{-0.00-0.02}^{+0.00+0.02}$    &$0.01_{-0.00-0.01}^{+0.00+0.01}$          &$0.02_{-0.00-0.00}^{+0.00+0.00}$    &$-0.02_{-0.00-0.01}^{+0.00+0.01}$        \\
\ \ \   &           &$0.11_{-0.02-0.02}^{+0.02+0.02}$        &$0.04_{-0.01-0.01}^{+0.01+0.01}$      &$-0.17_{-0.04-0.02}^{+0.03+0.02}$    &$-0.004_{-0.001-0.001}^{+0.001+0.001}$  &$-0.01_{-0.00-0.00}^{+0.00+0.00}$   &$-0.02_{-0.00-0.01}^{+0.00+0.01}$\\
\hline  &$T_1(T_2)$ &$-0.06_{-0.01-0.07}^{+0.01+0.07}$        &$0.18_{-0.03-0.02}^{+0.04+0.02}$      &$-0.08_{-0.01-0.06}^{+0.01+0.06}$   &$0.10_{-0.02-0.01}^{+0.02+0.01}$          &$0.10_{-0.02-0.01}^{+0.02+0.01}$    &$-0.25_{-0.05-0.03}^{+0.04+0.03}$     \\
\ \ \   &           &$0.45_{-0.08-0.07}^{+0.10+0.07}$       &$0.20_{-0.03-0.02}^{+0.04+0.02}$       &$-0.12_{-0.03-0.06}^{+0.02+0.06}$   &$-0.04_{-0.01-0.00}^{+0.06+0.04}$         &$-0.04_{-0.00-0.01}^{+0.00+0.01}$   &$-0.27_{-0.05-0.03}^{+0.04+0.03}$\\
\hline  & $T_3$     &$-0.11_{-0.03-0.03}^{+0.02+0.03}$        &$0.05_{-0.01-0.01}^{+0.01+0.01}$     &$-0.02_{-0.00-0.02}^{+0.00+0.02}$   &$0.02_{-0.00-0.01}^{+0.00+0.01}$          &$0.03_{-0.01-0.01}^{+0.01+0.01}$    &$-0.05_{-0.01-0.02}^{+0.01+0.02}$  \\
\ \ \   &           &$0.17_{-0.03-0.03}^{+0.04+0.03}$       &$0.06_{-0.01-0.01}^{+0.01+0.01}$       &$-0.19_{-0.05-0.02}^{+0.04+0.02}$   &$-0.01_{-0.00-0.00}^{+0.00+0.00}$         &$-0.01_{-0.00-0.00}^{+0.00+0.00}$   &$-0.05_{-0.01-0.02}^{+0.01+0.02}$\\
 \hline \hline
         &          &$B\to K_1(1400)$     &$B\to f_1(1285)$  &$B_s\to K_1(1400)$  &$B\to f_1(1420)$      &$B_s\to f_1(1420)$   &$B_s\to f_1(1285)$\\
 \hline
\ \ \   & $A$       &$0.05_{-0.01-0.05}^{+0.01+0.05}$        &$0.19_{-0.03-0.02}^{+0.04+0.02}$     &$0.12_{-0.02-0.04}^{+0.03+0.04}$     &$-0.01_{-0.00-0.00}^{+0.00+0.00}$    &$-0.24_{-0.04-0.03}^{+0.04+0.03}$   &$-0.01_{-0.00-0.00}^{+0.00+0.00}$ \\
\ \ \   &           &$0.34_{-0.06-0.05}^{+0.07+0.05}$        &$0.18_{-0.03-0.02}^{+0.04+0.02}$     &$-0.05_{-0.01-0.04}^{+0.01+0.04}$      &$-0.05_{-0.01-0.01}^{+0.01+0.01}$    &$-0.23_{-0.04-0.03}^{+0.04+0.03}$   &$-0.06_{-0.01-0.00}^{+0.01+0.00}$\\
\hline  & $V_0$     &$-0.12_{-0.03-0.11}^{+0.02+0.11}$         &$0.26_{-0.05-0.06}^{+0.05+0.06}$   &$-0.14_{-0.04-0.10}^{+0.03+0.10}$      &$-0.01_{-0.00-0.01}^{+0.00+0.01}$    &$-0.34_{-0.06-0.08}^{+0.05+0.08}$   &$-0.01_{-0.00-0.00}^{+0.00+0.00}$\\
\ \ \   &           &$0.64_{-0.11-0.11}^{+0.13+0.11}$        &$0.25_{-0.04-0.06}^{+0.05+0.06}$     &$-0.12_{-0.02-0.10}^{+0.02+0.10}$      &$-0.07_{-0.01-0.02}^{+0.01+0.02}$    &$-0.33_{-0.06-0.08}^{+0.05+0.08}$   &$-0.08_{-0.01-0.01}^{+0.01+0.01}$\\
\hline   & $V_1$     &$0.09_{-0.02-0.09}^{+0.02+0.09}$       &$0.32_{-0.06-0.04}^{+0.07+0.04}$     &$0.24_{-0.04-0.08}^{+0.05+0.08}$     &$-0.02_{-0.00-0.01}^{+0.00+0.00}$    &$-0.44_{-0.08-0.05}^{+0.07+0.05}$   &$-0.02_{-0.00-0.00}^{+0.00+0.00}$\\
\ \ \   &           &$0.62_{-0.11-0.09}^{+0.13+0.09}$        &$0.32_{-0.06-0.04}^{+0.07+0.04}$     &$-0.09_{-0.01-0.08}^{+0.01+0.08}$      &$-0.09_{-0.02-0.01}^{+0.02+0.01}$    &$-0.43_{-0.08-0.05}^{+0.07+0.05}$   &$-0.10_{-0.02-0.01}^{+0.02+0.01}$\\
\hline   & $V_2$     &$0.10_{-0.02-0.02}^{+0.02+0.02}$       &$0.10_{-0.02-0.00}^{+0.02+0.00}$     &$0.20_{-0.04-0.02}^{+0.05+0.02}$     &$-0.01_{-0.00-0.00}^{+0.00+0.00}$    &$-0.11_{-0.02-0.04}^{+0.02+0.04}$   &$-0.01_{-0.00-0.00}^{+0.00+0.00}$\\
\ \ \   &           &$0.10_{-0.02-0.02}^{+0.02+0.02}$        &$0.10_{-0.02-0.00}^{+0.02+0.00}$     &$-0.00_{-0.00-0.02}^{+0.00+0.02}$     &$-0.02_{-0.01-0.02}^{+0.00+0.02}$    &$-0.11_{-0.02-0.04}^{+0.02+0.04}$   &$-0.03_{-0.01-0.00}^{+0.00+0.00}$\\
\hline  &$T_1(T_2)$ &$0.06_{-0.01-0.07}^{+0.02+0.07}$          &$0.26_{-0.05-0.03}^{+0.05+0.03}$    &$0.14_{-0.03-0.06}^{+0.03+0.06}$      &$-0.01_{-0.00-0.00}^{+0.00+0.00}$    &$-0.33_{-0.06-0.04}^{+0.05+0.04}$   &$-0.01_{-0.00-0.00}^{+0.00+0.00}$\\
\ \ \   &           &$0.48_{-0.09-0.07}^{+0.10+0.07}$         &$0.25_{-0.05-0.03}^{+0.05+0.03}$    &$-0.08_{-0.01-0.06}^{+0.01+0.06}$     &$-0.07_{-0.02-0.01}^{+0.01+0.01}$    &$-0.32_{-0.06-0.04}^{+0.05+0.04}$   &$-0.08_{-0.01-0.01}^{+0.01+0.01}$\\
\hline   &$T_3$      &$0.13_{-0.02-0.03}^{+0.03+0.03}$        &$0.14_{-0.03-0.04}^{+0.03+0.04}$    &$0.24_{-0.05-0.02}^{+0.06+0.02}$      &$-0.01_{-0.00-0.00}^{+0.00+0.00}$    &$-0.16_{-0.03-0.00}^{+0.03+0.00}$   &$-0.01_{-0.00-0.00}^{+0.00+0.00}$\\
\ \ \   &           &$0.15_{-0.03-0.03}^{+0.03+0.03}$         &$0.13_{-0.02-0.04}^{+0.03+0.04}$    &$-0.01_{-0.00-0.02}^{+0.01+0.02}$     &$-0.03_{-0.01-0.00}^{+0.01+0.00}$    &$-0.15_{-0.03-0.00}^{+0.02+0.00}$   &$-0.04_{-0.01-0.00}^{+0.01+0.00}$\\
\hline\hline
        &   This work         &$B\to b_1$    & $B\to a_1$  &LFQM\cite{Cheng:2003sm,Cheng:2004yj}    &$B\to b_1(K_{1B})$    & $B\to a_1(K_{1A})$\\
\hline
\ \ \   &$A$         &$0.19_{-0.03-0.02}^{+0.04+0.03}$     &$0.26_{-0.05-0.03}^{+0.06+0.03}$  &$A$  &$0.10$ (0.11)         & $0.25$(0.26) \\
\ \ \   &$V_0$       &$0.45_{-0.08-0.04}^{+0.09+0.04}$     &$0.34_{-0.06-0.08}^{+0.07+0.08}$  &$V_0$&$0.39$(0.41)          & $0.13$(0.14) \\
\ \ \   &$V_1$       &$0.33_{-0.06-0.04}^{+0.07+0.04}$     &$0.43_{-0.08-0.05}^{+0.09+0.05}$  &$V_1$&$0.18$(0.19)          & $0.37$(0.39) \\
\ \ \   &$V_2$       &$0.03_{-0.00-0.01}^{+0.01+0.01}$     &$0.14_{-0.03-0.00}^{+0.03+0.00}$  &$V_2$&$-0.03(-0.05)$        & $0.18$(0.17) \\
\ \ \   &$T_1(T_2)$   &$0.27_{-0.05-0.03}^{+0.06+0.03}$    &$0.34_{-0.06-0.05}^{+0.07+0.05}$  &$T_1(T_2)$& --($0.13$)       &--($0.11$) \\
\ \ \   &$T_3$       &$0.06_{-0.01-0.02}^{+0.01+0.02}$    &$0.19_{-0.03-0.01}^{+0.04+0.01}$   &$T_3$& -- ($-0.07$)         &--($0.19$) \\
\hline\hline
\end{tabular}\label{Tab:formfactorBA}
\end{center}
\end{table}}

In the following, we will use $a_1$ to denote $a_1(1260)$, $b_1$ to
denote $b_1(1235)$. 
In Table~\ref{Tab:formfactorBA}, we give the numerical results for
$B\to A$ form factors, in which we have used minus values for decays
constants of $^1P_1$ mesons \footnote{Decay constants given in QCD
sum rules \cite{Yang:2005gk,Yang:2007zt} are both positive for two
kinds of axial-vectors and we find that this will give negative
values for $B\to ^1P_1$ form factors, like in \cite{Cheng:2007mx}.
For non-strange $^1P_1$ mesons, this minus sign will not give any
differences as it can not be observed experimentally. But we should
point out the minus sign will affect the mixing between $K_{1A}$ and
$K_{1B}$ by changing the mixing angle $\theta$ to $-\theta$.}. The
errors are from: decay constants of  $B_{(s)}$ mesons and shape
parameters $\omega_b$; Gegenbauer moments in axial-vectors' LCDAs.
In the calculation, we use the mass of the two physical states
$K_1(1270)$, $K_1(1400)$ as that of two spin states $K_{1A}$,
$K_{1B}$ for simplicity and similar for the branching ratios which
are given in the following. That only involves a slight difference
to the form factors. As the quark contents (to be precise the mixing
angles) of the axial-vectors $K_1(f_1,h_1)$ have not been uniquely
determined, we give two different kinds of results for form factors
as in Ref. \cite{Yang:2007zt}: the results in the first line are
calculated using $\theta_K=45^\circ$ while the second line
corresponds to the angle $\theta_K=-45^\circ$. This is also done for
the results involving the flavor-singlet and flavor-octet mesons:
the results in the first line are calculated using
$\theta_{^1P_1}=10^\circ$, $\theta_{^3P_1}=38^\circ$;  while the
second line corresponds to the angle $\theta_{^1P_1}=45^\circ$,
$\theta_{^3P_1}=50^\circ$.

\begin{table}[tb]
\caption{Distinct contributions to form factor $T_1$ from various
distribution amplitudes.}
 \label{Tab:formfactorcomparison}
\begin{center}
 \begin{tabular}{c|c|c|c}
  \hline\hline
            &  $B\to\rho$ & $B\to a_1(1260)$    & $B\to b_1(1235)$
          \\  \hline
   $\phi^T$
                                                          &$0.086$
                                                          &$0.142$
                                                          &$0.080$
                                                          \\\hline
   $\phi^a$
                                                          &$0.047$
                                                          &$0.086$
                                                          &$0.086$
                                                          \\ \hline
   $\phi^v$
                                                          &$0.063$
                                                          &$0.115$
                                                          &$0.100$
                                                           \\
 \hline
 total                                                    &$0.196$
                                                          &$0.343$
                                                          &$0.266$
                                                          \\
 \hline\hline
\end{tabular}
\end{center}
\end{table}

A number of remarks on $B\to A$ form factors are in order.
\begin{enumerate}
\item Form factors are strongly dependent on mixing angles.
Many of them even are different by order of magnitude, because the
mixing angles describe directly the inner quark contents of the
meson. The large difference of form factors surely will induce large
differences in branching ratios, which we will see later.

\item We give a comparison of the $B\to\rho$, $B\to a_1(1260)$ and $B\to
b_1(1235)$ form factors. Form factors $V_0$, $V_1$, $T_1$ for $B\to
A$ transition are larger than the corresponding $B\to V$ ones. It
seems that the form factor $A^{B\to (a_1,b_1)}$ is somewhat equal to
or even smaller than $V^{B\to \rho}$. But actually that is
artificial: as in Eq.~(\ref{eq:formfactorBtoV_V}), the pre-factor is
$m_B+m_V$ while for $B\to A$ form factor $A$, the factor becomes
$m_B-m_A$.  We take $T_1$ as an example to explain the reason for
the large $B\to A$ form factors. In
table~\ref{Tab:formfactorcomparison}, we give contributions from
three kinds of LCDAs: $\phi^T$, $\phi^v$ and $\phi^a$. The
contribution from $\phi^T$ is larger for $B\to a_1$, than the other
two transitions only because the axial-vector $a_1$ decay constant
is larger. Furthermore, larger axial vector meson mass implies
larger contribution from twist-3 distribution amplitudes $\phi^v$,
$\phi^a$ for both of $T_1^{B\to b_1}$ and $T_1^{B\to a_1}$.

\item
In our calculation for form factors involving $f_1$ mesons, we have
used the mixing angle between the octet and singlet:
$\theta=38^\circ(50^\circ)$ which is very close to the ideal mixing
angle $\theta=35.3^\circ$. That implies that the lighter meson
$f_1(1285)$ is almost made up of $\frac{\bar uu+\bar dd}{\sqrt 2}$
while the heavier meson $f_1(1420)$ is dominated by the $\bar ss$
component. Thus $B\to f_1(1420)$ and $B_s\to f_1(1285)$ form factors
are suppressed by the flavor structure and are numerically small.
The form factors involving $h_1$ are similar if the mixing angle is
taken as $45^\circ$.

\item From the table~\ref{Tab:formfactorBA}, we can see that the form
factor $T_1^{B\to a_1}$ is almost equal to $T_1^{B\to b_1}$. In the
flavor SU(3) symmetry limit, $B\to K_{1A}$ and $B\to K_{1B}$ form
factors are also almost equal with each other. But the physical
states  $K_1(1270)$ and $K_1(1400)$ are mixtures of $B\to
K_{1A,1B}$. With the mixing angle $\theta_K=\pm45^\circ$, the
$B_{d,s}\to K_1(1270)(K_1(1400))$ form factors are either enhanced
by a factor $\sqrt 2$ or highly suppressed. This feature will
definitely affect branching ratios which will be discussed later.
\end{enumerate}


Up to now, there are not too many experimental constraints on $B\to
A$ form factors. However there are lots of studies using some
non-perturbative methods: quark meson model \cite{Deandrea:1998ww},
ISGW \cite{Isgur:1988gb,Scora:1995ty}, QCD sum rules and light-cone
sum rules \cite{Lee:2006qj,Aliev:1999mx,Yang:} and light-front quark
model \cite{Cheng:2003sm,Cheng:2004yj}. Results in LFQM are also
given in table \ref{Tab:formfactorBA} to make a comparison. These
two approaches are very different in the treatment of dynamics of
transition form factors, but at first we will analyze differences
caused by  non-perturbative inputs. For $B\to a_1$ and $B\to K_{1A}$
form factors, most of our results (except $V_0$ and $T_{1,2}$) are
slightly larger than or almost equal with these of evaluated in
LFQM, as slightly larger decay constants for $a_1$ and $K_{1A}$ are
used: $f_{a_1}=203$ MeV and $f_{K_{1A}}=186$ MeV.  The form factor
$V_0$ is calculated by the relation $2m_AV_0=(m_B-m_A)V_1
-(m_B+m_A)V_2$. Small differences in $V_1$ and $V_2$ have induced a
large difference in $V_0$, which could be reduced in future studies
using more precise hadronic inputs. We have found there are large
differences in $B\to ^1P_1$ transition form factors. As the decay
constant of $b_1$ is zero in isospin limit, thus in LFQM the shape
parameter $\omega$ can not be directly determined and Cheng and Chua
used the same value with that of $a_1$ \cite{Cheng:2003sm}. It is
also similar for $K_{1B}$: they used the same shape parameter with
that of $K_{1A}$ which predicts $f_{K_{1B}}=11$ MeV. Compared with
the QCD sum rule results $f_{K_{1B}}=f_{K_{1B}}^T\times a_0^{||}$
given in table~\ref{Table:Adecayconstant} and
\ref{tab:AxialGegenbauer}, we can see: although they are consistent
within large theoretical errors, there are still large differences
in the central value. Thus our predictions for $B\to {^1P_1}$ form
factors (central values) are larger than those in LFQM. We have to
confess that the differences in decays constants are not responsible
for all differences in form factors. That may arise from further
differences in the dynamics. Compared with the recent light-cone sum
rules results \cite{Yang:}:
\begin{eqnarray}
V^{Ba_1}_0(0)&=0.303^{+0.022}_{-0.035},\;\;\;\;\;\;\; &V^{BK_{1A}}_0(0)=0.316^{+0.048}_{-0.042},\\
 V^{Bb_1}_0(0)&=-0.356^{+0.039}_{-0.033},\;\;\;\ &
 V^{BK_{1B}}_0(0)=-0.360^{+0.030}_{-0.028},
\end{eqnarray}
where uncertainties are from Borel window and input parameters, we
can see that they are well consistent with our calculations in pQCD
approach. As mentioned in Ref.~\cite{Cheng:2007mx}, the Babar
measurement of $\bar B^0\to a_1^+\pi^-$ \cite{Aubert:2006gb} favors
$V_0^{B\to a_1}(0)\simeq 0.30$ and this is very close to our result:
$V_0^{B\to a_1}(0)= 0.34$. It is also noted that there are large
uncertainties in our numerical results, especially due to large
uncertainties of hadronic parameters, such as large Gegenbauer
moments uncertainties shown in table~\ref{tab:AxialGegenbauer}.

\section{Calculation of radiative decay $B\to V\gamma$} \label{section:BtoV}
\subsection{The factorization formulae for decay amplitude}

For convenience, we define a common factor $F$ which appears in
many diagrams by:
\begin{eqnarray}
F=\frac{em_B^5C_F}{\pi}.
\end{eqnarray}
As we have mentioned in the above section, we  have to use the
amplitudes with distinct chiralities. The explicit factorization
formulae for the left-handed and right-handed photon from  operator
$O_{7\gamma}$ depicted in Fig.~\ref{Feyn:O7gamma} are given by:
\begin{eqnarray}
{\cal M}_{7\gamma}^L&=& 4r_bF\int _0^1 dx_1dx_2\int_0^\infty
    b_1db_1 b_2db_2\phi_{B}(x_1,b_1)
    \bigg\{
      C_{7\gamma}(t_a) E_e(t_a)\nonumber\\ &&\times
     \Big[(1+x_2)\phi_V^T(x_2)+r_V(1-2x_2)(\phi_V^v(x_2)+\phi_V^a(x_2))\Big]
      h_e(x_1x_2,x_2,b_1,b_2)
     \nonumber\\
     && \;\;+r_V\Big[\phi_V^v(x_2)+\phi_V^a(x_2)\Big]C_{7\gamma}(t_a^\prime)
         E_e'(t_a^\prime)h_e(x_1x_2,x_1,b_2,b_1)
    \bigg\},
    \label{eqs:O7gammaleft}\\
 {\cal M}_{7\gamma}^R&=& -\frac{r_D}{r_b}{\cal M}_{7\gamma}^L,
  \label{eqs:O7gammaright}
\end{eqnarray}
where the left-helicity amplitude is from the $m_b$ term in the
effective Hamiltonian and the right-helicity amplitude is from the
$m_D$ term which is obviously highly suppressed. This $O_{7\gamma}$
contribution is the dominant one characterizing by the form factor
$T_1(T_2)$. The formulas in
eq.(\ref{eqs:O7gammaleft},\ref{eqs:O7gammaright}) are the same as
that in eq.(\ref{fft1}) times the Wilson coefficient $C_7$.

\begin{figure}[tb]
\vspace{-1.5cm}
\begin{center}
\psfig{file=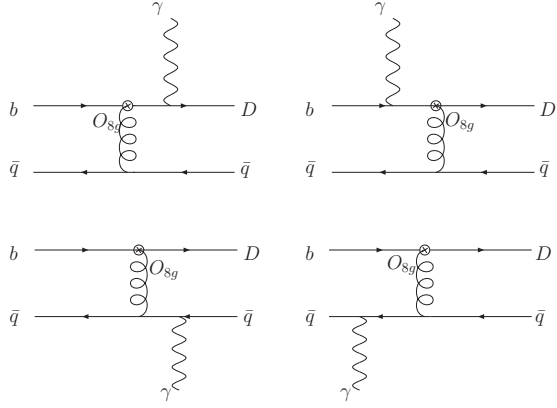,width=11.0cm,angle=0}
\end{center}
 \vspace{-9.5cm}
\caption{Feynman diagrams of the chromo-magnetic penguin operator
$O_{8g}$ }\label{Feyn:O8g}
\end{figure}

In pQCD approach, a hard gluon is required to kick the soft
spectator in the $B$ meson to turn into an energetic collinear
anti-quark. This gluon could be generated from QCD interaction
Hamiltonian  or from the $O_{8g}$ operator. In Fig.~\ref{Feyn:O8g},
we give the four diagrams from $O_{8g}$ operator given in the
effective Hamiltonian. The factorization formulae for the first two
diagrams in Fig.~\ref{Feyn:O8g} are
\begin{eqnarray}
{\cal M}_{8g}^{L(a)}&=& 2r_bF\int _0^1 dx_1dx_2\int_0^\infty b_1db_1
b_2db_2\phi_{B}(x_1,b_1)
 \bigg\{Q_DC_{8g}(t_b) E_e(t_b)\nonumber\\ &&\times
\Big[(2x_2-x_1)\phi_V^T(x_2)-3x_2r_V(\phi_V^v(x_2)+\phi_V^a(x_2))\Big]h_e(x_1x_2,x_2-1,b_1,b_2)
\label{eqs:O8glefta}\\ &&
\;\;+\Big[x_1\phi_V^T(x_2)+x_2r_V(\phi_V^v(x_2)+\phi_V^a(x_2))\Big]Q_b
 C_{8g}(t_b^\prime)
 E_e'(t_b^\prime)h_e(x_1x_2,1+x_1,b_2,b_1)
 \bigg\}, \nonumber\\
 {\cal M}_{8g}^{R(a)}&=& -\frac{r_D}{r_b}{\cal M}_{8g}^{L(a)}.\label{eqs:O8grighta}
\end{eqnarray}
If we consider the last two diagrams in Fig.~\ref{Feyn:O8g}, there
will be more sources to generate the right-handed photon in addition
to the $m_D$ term in the effective electro-weak Hamiltonian. The
third diagram can give a small contribution which is from the higher
twist component:
\begin{eqnarray}
 {\cal M}_{8g}^{L(b)}(Q_q)&=&
       2r_bQ_qF\int _0^1dx_1dx_2\int_0^\infty b_1db_1 b_2db_2\phi_{B}(x_1,b_1)
  \bigg \{
      C_{8g}(t_c) E_e(t_c)\nonumber\\
      &&\;\times
     \Big[(2+x_2-x_1)\phi_V^T(x_2)+3x_2r_V(\phi_V^v(x_2)+\phi_V^a(x_2))\Big]h_e(x_1-x_2,-x_2,b_1,b_2)
     \nonumber\\
  && \;\;\;+\Big[x_2r_2(\phi_V^v(x_2)+\phi_V^a(x_2))-x_1\phi_V^T(x_2)\Big]
       C_{8g}(t_c^\prime) E_e'(t_c')h_e(x_1-x_2,x_1,b_2,b_1)
  \bigg \}\nonumber\\
  &&   -2r_DQ_qF\int _0^1 dx_1dx_2\int_0^\infty b_1db_1
       b_2db_2\phi_{B}(x_1,b_1) C_{8g}(t_c) E_e(t_c)\nonumber\\
  &&\times \Big[3x_2r_V(\phi_V^v(x_2)-\phi_V^a(x_2))\Big]h_e(x_1-x_2,-x_2,b_1,b_2),
  \label{eqs:O8gleftb}
 \end{eqnarray}
 \begin{eqnarray}{\cal M}_{8g}^{R(b)}(Q_q)&=&
   -2r_DQ_qF\int _0^1 dx_1dx_2\int_0^\infty b_1db_1 b_2db_2\phi_{B}(x_1,b_1)
 \bigg\{C_{8g}(t_c) E_e(t_c)\nonumber\\
  &&\;\times
    \Big[(2+x_2-x_1)\phi_V^T(x_2)+3x_2r_V(\phi_V^v(x_2)+\phi_V^a(x_2))\Big]h_e(x_1-x_2,-x_2,b_1,b_2)
  \nonumber\\
 && \;\;+\Big[x_2r_V(\phi_V^v(x_2)+\phi_V^a(x_2))-x_1\phi_V^T(x_2)\Big]
     C_{8g}(t_c') E_e'(t_c')h_e(x_1-x_2,x_1,b_2,b_1)
  \bigg\}\nonumber\\
 &&    +2r_bQ_qF\int _0^1 dx_1dx_2\int_0^\infty b_1db_1
      b_2db_2\phi_{B}(x_1,b_1) C_{8g}(t_c) E_e(t_c)\nonumber\\
  &&\;\times
    \Big[3x_2r_V(\phi_V^v(x_2)-\phi_V^a(x_2))\Big]h_e(x_1-x_2,-x_2,b_1,b_2).
    \label{eqs:O8grightb}
\end{eqnarray}

\begin{figure}[tb]
\vspace{-2.0cm}
\begin{center}
\psfig{file=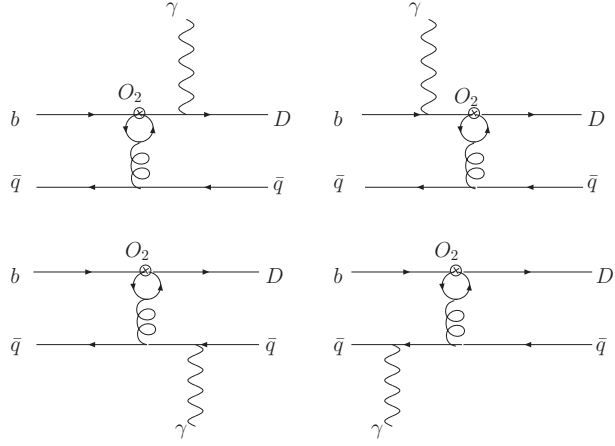,width=12.0cm,angle=0}
\end{center}
 \vspace{-10cm}
\caption{Feynman diagrams in which the operator $O_2$ is inserted in
the loop with a photon   emitted from the external quark line
}\label{Feyn:O2}
\end{figure}

\begin{figure}[tb]
\vspace{-5.cm}
\begin{center}
\psfig{file=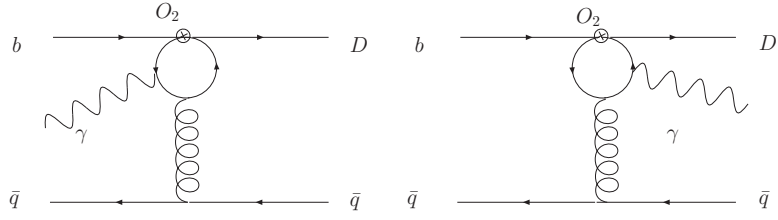,width=12.0cm,angle=0}
\end{center}
 \vspace{-10.cm}
\caption{Feynman diagrams in which the operator $O_2$ is inserted in
the loop with both of the photon and the virtual gluon  emitted from
the internal quark line. }\label{Feyn:O22}
\end{figure}

\begin{figure}[tb]
\vspace{-1.5cm}
\begin{center}
\psfig{file=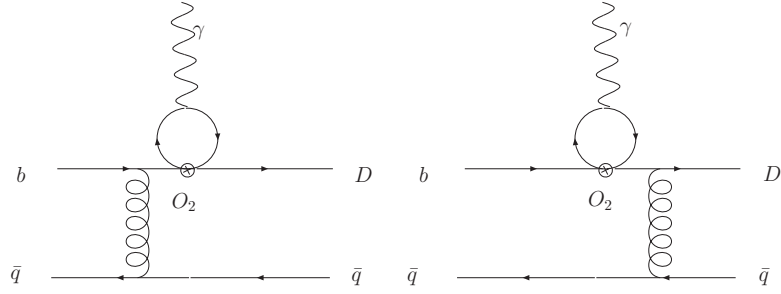,width=12.0cm,angle=0}
\end{center}
 \vspace{-12cm}
\caption{Feynman diagrams in which the operator $O_2$ is inserted in
the loop with a photon    emitted from the internal quark line}
\label{Feyn:O23}
\end{figure}

Next we want to mention some higher order corrections as usual which
may give important contributions: charm and up quark loop (${\cal
O}(\alpha_s)$) contributions in Fig.~\ref{Feyn:O2},
Fig.~\ref{Feyn:O22} and Fig.~\ref{Feyn:O23}. It should be pointed
out that these contributions are not related to next-to-leading
order corrections in pQCD approach, while next-to-leading order
corrections to the exclusive processes $\pi\gamma^*\to\gamma$ in
pQCD approach have been investigated in Ref.~\cite{Nandi:2007qx}.

We use the subtitle ``quark line photon emission'' to denote that a
photon is emitted through the external quark lines as in
Fig.\ref{Feyn:O2}. We define the ${c}$ and ${u}$ loop function in
order that the ${b\to Dg}$ vertex can be expressed as
${\bar{D}\gamma^{\mu}(1-\gamma^5)I^a_{\mu\nu}A^{a\nu}b}$. It has the
gauge invariant form \cite{Bander:1979px}  as follows:
\begin{eqnarray}
I_{\mu\nu}^a&=&\frac{gT^a}{2\pi^2}(k^2 g_{\mu\nu}-k_{\mu}k_{\nu})
\int^1_0 dx x(1-x)
\left[1+\log\left(\frac{m_i^2-x(1-x)k^2}{t^2}\right)\right]\nonumber\\
&=&-\frac{gT^a}{8\pi^2}(k^2
g_{\mu\nu}-k_{\mu}k_{\nu})\left[G(m_i^2,k^2,t)-\frac{2}{3}\right],
\end{eqnarray}
where ${k}$ is the gluon momentum and ${m_i}$ is the loop internal
quark mass. G is the function from the loop integration:{\small
\begin{eqnarray}
G(m_i^2,k^2,t) &=& \theta(-k^2)\frac{2}{3}\left[
\frac{5}{3}+\frac{4m_i^2}{k^2}-\ln{\frac{m_i^2}{t^2}}
+\left(1+\frac{2m_i^2}{k^2}\right) \sqrt{1-\frac{4m_i^2}{k^2}}
\ln{\frac{\sqrt{1-4m_i^2/k^2}-1}{\sqrt{1-4m_i^2/k^2}+1}}
\right]\nonumber\\
&&+\theta(k^2)\theta(4m_i^2-k^2) \frac{2}{3}\Bigg[
\frac{5}{3}+\frac{4m_i^2}{k^2}-\ln{\frac{m_i^2}{t^2}}\nonumber\\
&& -2\left(1+\frac{2m_i^2}{k^2}\right)
\sqrt{\frac{4m_i^2}{k^2}-1}\arctan{\left(\frac{1}{\sqrt{4m_i^2/k^2-1}}\right)}
\Bigg]\nonumber\\
&&+\theta(k^2-4m_i^2) \frac{2}{3}\Bigg[
\frac{5}{3}+\frac{4m_i^2}{k^2}-\ln{\frac{m_i^2}{t^2}}
\nonumber\\
&& +\left(1+\frac{2m_i^2}{k^2}\right) \sqrt{1-\frac{4m_i^2}{k^2}}
\left( \ln{\frac{1-\sqrt{1-4m_i^2/k^2}}{1+\sqrt{1-4m_i^2/k^2}}}
+i\pi \right) \Bigg].
\end{eqnarray}}
The loop function ${G}$ has the dependence of gluon momentum square
of ${k^2}$. But there is no singularity when we take the limit of
${k\rightarrow 0}$, so we can neglect ${k_T}$ components of ${k^2}$
in the loop function ${G}$. Using this effective vertex, the
factorization formulae for the first two diagrams of
Fig.\ref{Feyn:O2} is calculated as:
\begin{eqnarray}
 M^{L(a)}_{1i}&=&-Q_DF\int^1_0 dx_1 dx_2 \int b_1 db_1 b_2 db_2
   \phi_B(x_1,b_1)
   C_{2}(t_b) E_e(t_b)\Big[G(m_i^2,-x_1x_2m_B^2,t_b)-\frac{2}{3}\Big]\nonumber\\
&&\times  \Big[ x_2^2r_V(\phi_{V}^{v}(x_2)+\phi_{V}^{a}(x_2))
   +3x_1x_2\phi_{V}^T(x_2)
   \Big]h_e(x_1x_2,x_2-1,b_1,b_2),\\
 M^{R(a)}_{1i}&=&-Q_bF \int^1_0 dx_1 dx_2 \int b_1 db_1 b_2 db_2
    \phi_B(x_1,b_1)C_{2}(t_b')
    E_e'(t_b')\Big[G(m_i^2,-x_1x_2m_B^2,t_b')-\frac{2}{3}\Big]\nonumber\\
   &&\times x_1x_2r_V(\phi_{V}^a(x_2)-\phi_{V}^v(x_2))
    h_e(x_1x_2,1+x_1,b_2,b_1),
\end{eqnarray}
where $Q_D=-\frac{1}{3}$.   For the other two diagrams of
Fig.\ref{Feyn:O2}, the factorization formulas are
\begin{eqnarray}
{\cal M}_{1i}^{L(b)}(Q_q)&=& Q_qF\int _0^1 dx_1dx_2\int_0^\infty
  b_1db_1 b_2db_2\phi_{B}(x_1,b_1)
 \bigg\{
   C_{2}(t_c) E_e(t_c) \nonumber\\
   &&\;\;\;\times
  \Big[x_2r_{V}(1+2x_2)[\phi_{V}^v(x_2)
 +\phi_{V}^a(x_2)] +3(x_2-x_1)\phi_{V}^T(x_2)
\Big]\nonumber\\
   &&\;\;\;\times
    \Big[G(m_i^2,-(x_1-x_2)m_B^2,t_c)-\frac{2}{3}\Big]h_e(x_1-x_2,-x_2,b_1,b_2)
  \nonumber\\
  && +\Big[
  x_2r_{V}(\phi_{V}^v(x_2)+\phi_{V}^a(x_2))-x_1\phi_{V}^T(x_2) \Big]
  \Big[G(m_i^2,-(x_1-x_2)m_B^2,t_c')-\frac{2}{3}\Big]\nonumber\\
  &&\;\;\;\times  C_{2}(t_c') E_e'(t_c') h_e(x_1-x_2,x_1,b_2,b_1)
 \bigg\},
 \label{eqs:O8gleftb2}\\
 {\cal M}_{1i}^{R(b)}(Q_q)&=& Q_qF\int _0^1 dx_1dx_2\int_0^\infty
 b_1db_1 b_2db_2\phi_{B}(x_1,b_1)  C_{2}(t_c) E_e(t_c)h_e(x_1-x_2,-x_2,b_1,b_2)
 \nonumber\\
 &&\;\;\;\times \Big[G(m_i^2,-(x_1-x_2)m_B^2,t_c)-\frac{2}{3}\Big]\times
 (2+x_2)x_2r_{V}\Big[\phi_{V}^v(x_2)-\phi_{V}^a(x_2)\Big].
\label{eqs:O8grightb2}
\end{eqnarray}

In Fig.~\ref{Feyn:O22}, we give the diagrams in which a photon
emitted from the internal loop quark line. The sum of the effective
vertex $b\to D \gamma g^*$ in Fig.\ref{Feyn:O22}  has been derived
by \cite{Liu:1990yb,Simma:1990nr}:
\begin{equation}
I=\bar{D}\gamma^{\rho}(1-\gamma_5)T^a
bI_{\mu\nu\rho}A^{\mu}A^{a\nu},
\end{equation}
with the tensor structure given by
\begin{eqnarray}
I_{\mu\nu\rho} &=&A_4\left[(q\cdot
k)\epsilon_{\mu\nu\rho\sigma}(q-k)^{\sigma}
+\epsilon_{\nu\rho\sigma\tau}q^{\sigma}k^{\tau}k_{\mu}
-\epsilon_{\mu\rho\sigma\tau}q^{\sigma}k^{\tau}q_{\nu}\right]\nonumber\\
&&\hspace{-3mm}+A_5\left[\epsilon_{\mu\rho\sigma\tau}q^{\sigma}k^{\tau}k_{\nu}
     -k^2\epsilon_{\mu\nu\rho\sigma}q^{\sigma}\right],
\end{eqnarray}
and
\begin{eqnarray}
 A_4 &=&\frac{4ieg}{3\pi^2}\int_0^1dx\int_0^{1-x}dy \frac{x
 y}{x(1-x)k^2+ 2xyq\cdot k-m_i^2+i\varepsilon},\\
 A_5 &=&-\frac{4ieg}{3\pi^2}\int_0^1dx\int_0^{1-x}dy
 \frac{x(1-x)}{x(1-x)k^2+ 2xyq\cdot k-m_i^2+i\varepsilon},
\end{eqnarray}
where $q$ is the momentum of the photon $q=P_B-P_V$, and $k$ is the
momentum of the gluon. Then the amplitudes can be expressed as
follows:
\begin{eqnarray}
 {\cal M}_{2i}^{L}&=&-\frac{8}{3}F \int^1_0 dx \int^{1-x}_0 dy
 \int^1_0 dx_1 dx_2 \int b_1 db_1
 \phi_B(x_1,b_1)C_2(t_d)\alpha_s(t_d)\exp[-S_B(t_d)]
 \nonumber\\
 &&\times \frac{h_e'}{xyx_2 m_B^2-m_i^2}\times
 \bigg\{xyx_2\Big[(1-x_2)r_{V}(\phi_{V}^v(x_2)+\phi_{V}^a(x_2))
 -(1+2x_1) \phi_{V}^T(x_2)
 \Big]\nonumber\\
 &&\;\;+x(1-x) \Big[ x_2^2r_{V}(\phi_{V}^v(x_2)
 +\phi_{V}^a(x_2))+3x_1x_2\phi_{V}^T(x_2) \Big]\bigg\},\\
 {\cal M}_{2i}^R&=&\frac{8}{3}F \int^1_0 dx \int^{1-x}_0 dy
 \int^1_0 dx_1 dx_2 \int b_1 db_1
 \phi_B(x_1,b_1)C_2(t_d)\alpha_s(t_d)\exp[-S_B(t_d)]
 \nonumber\\
 &&\times \frac{h_e'}{xyx_2 m_B^2-m_i^2}\times
 xyx_2^2r_{V}\Big[\phi_{V}^v(x_2)-\phi_{V}^a(x_2) \Big],
\end{eqnarray}
where the function $h_e'$  is defined by:
\begin{eqnarray}
h_e' &\equiv & K_0(\sqrt {x_1x_2} m_Bb_1)-\Big[\theta(B^2)K_0(b_1
\sqrt{|B^2|})+\theta(-B^2)i\frac{\pi}{2}H_0(b_1
\sqrt{|B^2|})\Big],\end{eqnarray} with
 \begin{eqnarray}
 B^2&=&x_1x_2m_B^2
-\frac{y}{1-x}x_2m_B^2+\frac{m_i^2}{x(1-x)}.
\end{eqnarray}

Diagrams in Fig.~\ref{Feyn:O23} in which the photon is emitted from
the external loop and the gluon is emitted from QCD interaction
Hamiltonian  do not give any contribution to $B\to V\gamma$. The
reason is as follows. Similar with $b\to Dg$, the vertex for $b\to
D\gamma$ can also be expressed as
${\bar{D}\gamma^{\mu}A^\nu(1-\gamma^5)(k^2g_{\mu\nu}-k_\mu k_\nu)b}$
only with a different coefficient. For an on-shell photon, the
following conditions are required: $k^2=0$ and $\epsilon\cdot k=0$,
thus the contribution from diagrams in Fig.~\ref{Feyn:O23} vanishes
in $b\to s(d)\gamma$ decays. But it should be noted that these
diagrams can give a non-zero contribution to $b\to s(d)\gamma^*\to
s(d)l^+l^-$.

\begin{figure}[tb]
\vspace{-1.2cm}
\begin{center}
\psfig{file=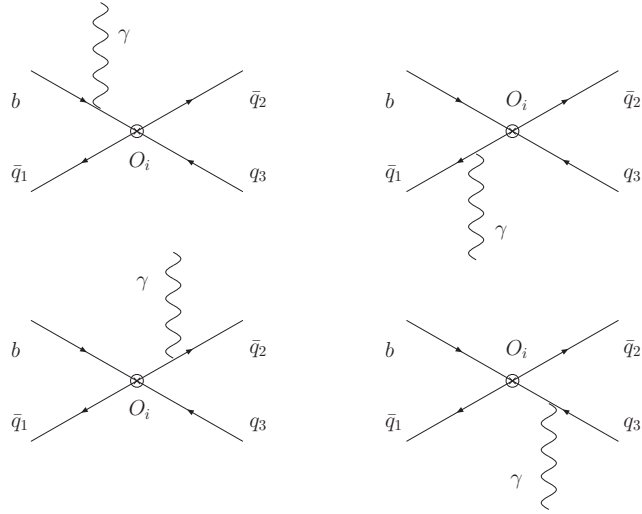,width=12.0cm,angle=0}
\end{center}
 \vspace{-9.5cm}
\caption{Feynman diagrams for annihilation topologies}
\label{Feyn:annihilation}
\end{figure}

In annihilation diagrams, there are three different kinds of
operators in the $\otimes$ depicted in Fig.~\ref{Feyn:annihilation}.
In the following, we use $LL$ to denote the left-handed current
between $b$ and $\bar q$ quark and the left-handed current between
the final state two quarks; $LR$ denotes the left-handed current
between $b$ and $\bar q$ quark and the right-handed current between
the final state two quarks; we use $SP$ to denote the $(S-P)(S+P)$
current which is from the Fierz transformation of $(V-A)(V+A)$
operators. The factorization formulae for these diagrams are given
by:
\begin{eqnarray}
 {\cal M}_{ann}^{L(a,LL)}(a_i,Q_{q_1})&=& {\cal M}_{ann}^{L(a,LR)}(a_i,Q_{q_1})\nonumber\\
 &=&F\frac{3\sqrt{6}Q_{q_1}r_{V}f_{V}\pi}{2m_B^2}
\int^1_0 dx_1\int_0^\infty b_1
db_1a_i(t_e')E_a(t_e')\phi_B(x_1,b_1)K_{0}(\sqrt x_1m_Bb_1),\\
 {\cal M}_{ann}^{R(a,LL)}(a_i,Q_{q_1})&=& {\cal M}_{ann}^{R(a,LR)}(a_i,Q_{q_1})\\
 &=&-F\frac{3\sqrt{6}Q_br_{V}f_{V}\pi}{2m_B^2}
\int^1_0 dx_1\int_0^\infty b_1 db_1 a_i(t_e) E_a(t_e)\phi_B(x_1,b_1)
K_0(\sqrt {1+x_1}m_Bb_1),\nonumber
\end{eqnarray}
\begin{eqnarray}
 {\cal M}_{ann}^{L(b,LL)}(a_i,Q_{q_2},Q_{q_3})&=& {\cal M}_{ann}^{R(b,LR)}(a_i,Q_{q_2},Q_{q_3})\nonumber\\
 &=&-F\frac{3\sqrt{6}r_{V}f_B\pi}{2m_B^2}
 \int^1_0 dx_2 \int b_2 db_2\Big[\phi_{V}^{v}(x_2)+\phi_{V}^a(x_2)\Big]\nonumber\\
&\times& \bigg\{Q_{q_2}a_i(t_f)
  E'_a(t_f)i\frac{\pi}{2}H_0^{(1)}(\sqrt{1-x_2}m_Bb_2) \nonumber\\
&&
  -x_2Q_{q_3}a_i(t_f')  E'_a(t_f')i\frac{\pi}{2}H_0^{(1)}(\sqrt{x_2}m_Bb_2)\bigg\},
 \\
  {\cal M}_{ann}^{R(b,LL)}(a_i,Q_{q_2},Q_{q_3})&=&    {\cal M}_{ann}^{L(b,LR)}(a_i,Q_{q_2},Q_{q_3})\nonumber\\
   &=&-F\frac{3\sqrt{6}r_{V}f_B\pi}{2m_B^2}
 \int^1_0 dx_2 \int b_2 db_2\nonumber\\
&\times&
\bigg\{(1-x_2)\Big[\phi_{V}^{v}(x_2)-\phi_{V}^a(x_2)\Big]Q_{q_2}a_i(t_f)
  E'_a(t_f)i\frac{\pi}{2}H_0^{(1)}(\sqrt{1-x_2}m_Bb_2)\nonumber\\
&&\;\;\;
  -\Big[\phi_{V}^{v}(x_2)-\phi_{V}^a(x_2)\Big]Q_{q_3}a_i(t_f')  E'_a(t_f')
  i\frac{\pi}{2}H_0^{(1)}(\sqrt{x_2}m_Bb_2)\bigg\},
\end{eqnarray}
\begin{eqnarray}
 {\cal M}_{ann}^{L(SP)}(a_i)&=&F\frac{3\sqrt{6} f_B\pi}{m_B^2}
\int^1_0 dx_2 \int b_2 db_2
\phi_{V}^T(x_2)\\
 &\times& \bigg\{Q_{q_2}a_i(t_f) E'_a(t_f) i\frac{\pi}{2}H_0^{(1)}(\sqrt{1-x_2}m_Bb_2)
 + Q_{q_3}a_i(t_f') E'_a(t_f')
 i\frac{\pi}{2}H_0^{(1)}(\sqrt{x_2}m_Bb_2)\bigg\}.\nonumber
\end{eqnarray}

\begin{figure}
\vspace{-2.cm}
\begin{center}
\psfig{file=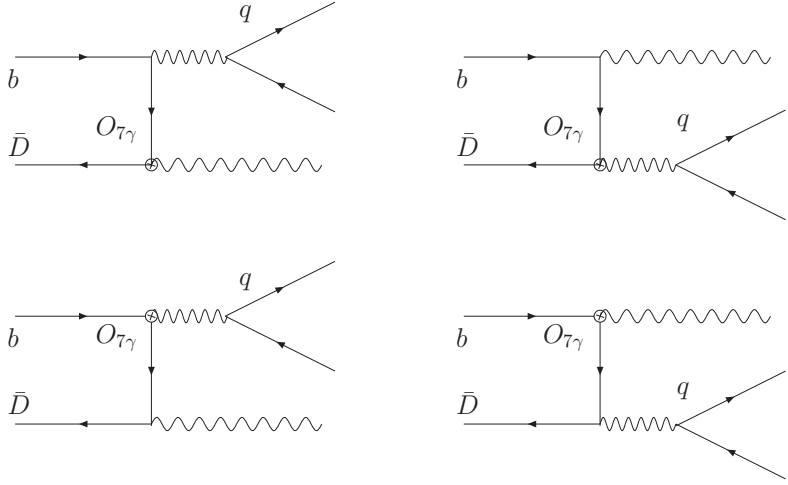,width=15.0cm,angle=0}
\end{center}
 \vspace{-13.5cm}
\caption{Feynman diagrams with double-photon contributions}
\label{Feyn:twophoton}
\end{figure}

Finally, there is another kind of contribution from $O_{7\gamma}$:
the neutral vector meson $\bar qq$ is generated by a photon as
depicted in Fig.~\ref{Feyn:twophoton}. Although these diagrams are
suppressed by the electromagnetic coupling constant, the enhancement
factor $m_B/\Lambda_{QCD}$ can make it important in some cases
\cite{Lu:2006nz}. We include these diagrams in our calculation.  The
first two diagrams of Fig.~\ref{Feyn:twophoton} are equal to each
other, so are the last two diagrams. The factorization formulae are
given by:
\begin{eqnarray}
 {\cal M}_{en}^{L}(Q_q)&=& -F\frac{3\sqrt{6}\alpha_{em}
Q_{q}Q_b r_bf_V}{m_Bm_V} \int^1_0 dx_1\int_0^\infty b_1 db_1\phi_B(x_1,b_1)\nonumber\\
 &&\times \bigg\{C_{7\gamma}(t_e)
 E_a(t_e) K_0(\sqrt {1+x_1}m_Bb_1)+C_{7\gamma}(t_e')
 E_a(t_e') K_0(\sqrt {x_1}m_Bb_1)\bigg\},\\
{\cal M}_{en}^{R}(Q_q)&=&-\frac{r_D}{r_b}{\cal M}_{en}^{L}(Q_q).
\end{eqnarray}

\subsection{Numerical results of Branching ratios}

With those decay amplitude formulas for different Feynman diagrams
in the last subsection, it is easy to get the total decay amplitude
for each channel of $B\to V\gamma$: $B^-\to \rho^-\gamma$, $B^0\to
\rho^0\gamma$, $B^0\to \omega\gamma$, $B^-\to K^{*-}\gamma$, $B^0\to
\phi\gamma$, $B^0\to K^{*0}\gamma$, $B_s^0\to K^{*0}\gamma$,
$B_s^0\to \rho^{0}\gamma$,  $B_s^0\to \omega\gamma$, and $B_s^0\to
\phi\gamma$. The explicit expressions are shown in
Appendix~\ref{section:BtoVgammaformulae}. The CP-averaged decay
width is then
\begin{eqnarray}
\Gamma(B\to V\gamma)=\frac{|{\cal A}(\bar B\to V\gamma)|^2+| {\cal
A}( B\to \bar V\gamma)|^2}{32\pi
m_B}(1-r_V^2)^3,\label{eq:BRofBtoVgamma}
\end{eqnarray}
where the summation on polarizations is implemented.

For CKM matrix elements, we use the same values as in Ref.
\cite{Ali:2007ff}:
\begin{equation}
 \begin{array}{lll}
 |V_{ud}|=0.974,   &|V_{us}|=0.226,    &|V_{ub}|=(3.68^{+0.11}_{-0.08})\times 10^{-3},
\\
  |V_{td}|=(8.20^{+0.59}_{-0.27})\times 10^{-3}, &|V_{ts}|=40.96\times 10^{-3},   &|V_{tb}|=1.0,\\
 \alpha={(99^{+4}_{-9.4})}^\circ,&   \gamma=(59.0^{+9.7}_{-3.7})^\circ,&
  \mbox{arg}[-V_{ts}V^*_{tb}]={1.0}^\circ,
\end{array}
\end{equation}
where we have adopted the updated results from \cite{Charles:2004jd}
and drop the (small) errors on $V_{ud}$, $V_{us}$, $V_{ts}$ and
$V_{tb}$. The CKM factors mostly give an overall factor to branching
ratios. However, the CKM angles do give large uncertainties to
branching ratios of some decay modes and to all the non-zero CP
asymmetries which will be discussed in the following subsection.

\begin{table}[tb]
\caption{$CP$-averaged branching ratios ($\times
 10^{-6}$) of $B\to V\gamma$ decays obtained in pQCD approach
(This work); the errors for these entries correspond to
uncertainties in input hadronic quantities, from the
scale-dependence, and CKM matrix elements, respectively. For
comparison, we also listed the current experimental measurements
\cite{Barberio:2006bi,Drutskoy:2007aj,Aubert:2006pu,Abe:2005rj} and
theoretical estimates of branching ratios recently given in Ref.
\cite{Ball:2006eu} (QCDF) and in Ref.~\cite{Ali:2007sj} (SCET).}
 \label{Tab:BRVgamma}
\begin{center}
 \begin{tabular}{c|c|c|c|cc}
  \hline\hline
        {Modes}     &  QCDF &SCET&   This work & \multicolumn{2}{c}{ Exp.}
          \\  \hline
   $B^-\to K^{*-}\gamma$
                                                          & $53.3\pm13.5\pm5.8$
                                                          & $46\pm12\pm4\pm2\pm1$
                                                          & $35.8_{-12.8-4.0-1.1}^{+17.6+5.4+1.1}$
                                                          & \multicolumn{2}{c}{$40.3\pm2.6$\;\;[\rm HFAG]}
                                                           \\\hline
   $\bar B^0\to \bar K^{*0}\gamma$
                                                          &
                                                          $54.2\pm13.2\pm6.7$
                                                          & $43\pm11\pm4\pm2\pm1$
                                                          & $38.1^{+17.3+5.5+1.1}_{-12.7-3.8-1.1}$
                                                          & \multicolumn{2}{c}{$40.1\pm2.0$ \;\;[\rm HFAG]}
                                                          \\\hline
   $\bar B_s^0\to \phi\gamma$
                                                          & $39.4\pm10.7\pm5.3$
                                                          & $43\pm11\pm3\pm3\pm1$
                                                          & $35.8_{-10.3-3.5-1.1}^{+13.7+4.9+1.1}$
                                                          & \multicolumn{2}{c}{$57^{+18+12}_{-15-11}$ [Belle]}
                                                          \\
                                                          \hline\hline\hline
        {Modes}     &  QCDF &   This work & \multicolumn{2}{c}{Exp.}\\ \hline
   $B^-\to \rho^-\gamma$
                                                          & $1.16\pm 0.22\pm0.13$
                                                          & $1.15^{+0.57+0.18+0.17}_{-0.39-0.11-0.09}$
                                                          &
                                                          $1.10^{+0.37}_{-0.33}\pm0.09$
                                                          [BaBar]
                                                          &
                                                          $0.55^{+0.42+0.09}_{-0.36-0.08}$
                                                          [Belle]
                                                           \\\hline
   $\bar B^0\to \rho^0\gamma$
                                                          & $0.55\pm0.11\pm0.07$
                                                          & $0.57^{+0.26+0.09+0.08}_{-0.19-0.06-0.04}$
                                                          &
                                                          $0.79^{+0.22}_{-0.20}\pm0.06$ [BaBar]
                                                          &
                                                          $1.25^{+0.37+0.07}_{-0.33-0.06}$ [Belle]
                                                           \\\hline
   $\bar B^0\to \omega\gamma$
                                                          & $0.44\pm0.09\pm0.05$
                                                          & $0.51^{+0.23+0.08+0.08}_{-0.17-0.05-0.03}$
                                                          &
                                                          $0.40^{+0.24}_{-0.20}\pm0.05$ [BaBar]
                                                          & $0.56^{+0.34+0.05}_{-0.27-0.10}$ [Belle]
                                                          \\\hline
   $\bar B_s^0\to  K^{*0}\gamma$
                                                          & $1.26\pm0.25\pm0.18$
                                                          & $1.11_{-0.32-0.12-0.07}^{+0.42+0.15+0.16}$
                                                          & \multicolumn{2}{c}{---}
                                                          \\
                                                          \hline
                                                          \hline
   $\bar B^0\to \phi\gamma$
                                                          & ---
                                                          & $(7.5_{-2.1-0.9-0.5}^{+2.8+2.1+1.1})\times 10^{-6}$
                                                          & \multicolumn{2}{c}{$<0.85$\;\;[\rm HFAG]}
                                                          \\\hline
   $\bar B_s^0\to \rho^0\gamma$
                                                          & ---
                                                          & $(1.7_{-0.4-0.1-0.1}^{+0.4+0.1+0.0})\times10^{-3}$
                                                          & \multicolumn{2}{c}{---}
                                                          \\\hline
   $\bar B_s^0\to \omega\gamma$
                                                          & ---
                                                          & $(1.8_{-0.4-0.2-0.1}^{+0.4+0.1+0.1})\times 10^{-4}$
                                                          & \multicolumn{2}{c}{---}
                                                          \\
 \hline\hline\end{tabular}
\end{center}
 \end{table}

$CP$-averaged branching ratios  of $B\to V\gamma$ decays are listed
in table~\ref{Tab:BRVgamma}. The first error in these entries arises
from the input hadronic parameters, which is dominated by
$B(B_s)$-meson decay constants (taken as $f_{B}=(0.19 \pm 0.02)$ GeV
and $f_{B_s}=(0.23 \pm 0.02)$ GeV) and $B$ ($B_s$) meson wave
function shape parameters (taken as $\omega_B= (0.40\pm0.05)$ GeV
and $\omega_{B_s}= (0.50\pm0.05)$ GeV). The second error is from the
hard scale $t$, defined in Eqs. (A1) -- (A8) in Appendix A, which we
vary from $0.75t$ to $1.25t$ (not changing $1/b_i$), and from
$\Lambda^{(5)}_{QCD}=0.25\pm 0.05$ GeV. This scale-dependence
characterize  the size of next-to-leading order contributions in
pQCD approach. A part of this perturbative improvement coming from
next-to-leading order Wilson coefficients is already available
\cite{Buchalla:1995vs}. However, the complete next-to-leading order
corrections to hard spectator kernels are still missing. The third
error is the combined uncertainties in CKM matrix elements and
angles of the unitarity triangle. It is clear that the largest
uncertainty here is the first one from the input hadronic
parameters.

These ten $B\to V\gamma$ decay channels can be divided into three
different types: $b\to s$ transitions, $b\to d$ transitions and
purely annihilation decays. The first type contains $\bar B^0\to
\bar K^{*0}\gamma$, $B^-\to K^{*-}\gamma$ and $\bar
B_s\to\phi\gamma$. Among these decays, the dominant contribution
from $O_{7\gamma}$, is proportional to
$V_{tb}V_{ts}^*\sim\lambda^2$. This contribution can be related to
the form factor $T_1^{B\to V}$. In the flavor SU(3) symmetry limit,
form factors for these three channels should be equal which could
also relates the three decays. We do obtain similar branching ratios
for this kind decays only with small deviations. The penguin
contribution in $b\to d\gamma$ processes is proportional to
$V_{tb}V_{td}^*\sim\lambda^3$, which is expected to be suppressed by
one order magnitude relative to the $b\to s$ transitions. In our
calculation, branching ratio results for pure annihilation processes
are mainly from two-photon diagrams in Fig. \ref{Feyn:twophoton}.
Thus ${\cal BR}(B\to\phi\gamma)$ is surely consistent with Ref.
\cite{Lu:2006nz}. This feature can also certainly interpret the
large differences between $\bar B_s\to\rho^0\gamma$ and $\bar
B_s\to\omega\gamma$. This contribution is proportional to the charge
of the constitute quark. This factor is $\frac{2}{3}-\frac{-1}{3}$
for $\rho^0$ while $\frac{2}{3}+\frac{-1}{3}$ for $\omega$. Thus the
branching ratio of $B_s\to\rho^0\gamma$ is one order in magnitude
larger than that of $ (B_s\to\omega\gamma)$.

In the literature, there are many studies concentrating on $B\to
V\gamma$ \cite{Li:1998tk,Li:2003kz,Keum:2004is,Lu:2005yz,Li:2006xe,
Ali:2001ez, Beneke:2004dp, Bosch:2001gv,
Bosch:2002bw,Becher:2005fg,Chay:2003kb,
Bosch:2004nd,Ali:2004hn,Ali:2006fa,Ali:2007sj}. Recently, a
comprehensive study \cite{Ball:2006eu} using QCDF method and QCD sum
rules appears. In that paper, the authors used QCDF approach to
calculate all $B\to V\gamma$ approach and included some power
corrections: weak annihilation contributions, the soft-gluon
emission from quark (charm and light-quark) loops and long distance
photon emission from the soft quarks. In Table \ref{Tab:BRVgamma},
we quote them to make a comparison. Their uncertainties come from
form factors, the renormalization scale, the soft-gluon terms, the
CKM parameters, decay constants, Gegenbauer moments, the first
inverse moments of $B$($B_s$) mesons and quark masses, etc. Their
results for branching ratios of $B^-\to K^{*-}\gamma$ and $\bar
B^0\to \bar K^{*0}\gamma$ are about $(20$$-50)\%$ larger than our
predictions. The main reason is differences in form factors $T_1$:
they used $T_1^{B\to K^{*}}=0.31\pm0.04$ while our calculation gives
a smaller $T_1=0.23^{+0.05}_{-0.04}$. The smaller form factor is
also preferred by recent Lattice QCD result: $T_1^{B\to
K^{*}}=0.24\pm0.03^{+0.04}_{-0.01}$. Although the difference is not
too large, it can already induce a sizable difference to branching
ratios. $b\to d$ transitions and $\bar B_s\to\phi\gamma$ are well
consistent with each other, as the effective form factors used in
Ref.~\cite{Ball:2006eu} are smaller and almost equal to our results.
Very recently, the authors in Ref.~\cite{Ali:2007sj} used the more
theoretical approach SCET to investigate the three $b\to s$ decays
channels: $B^-\to K^{*-}\gamma$, $\bar B^0\to \bar K^{*0}\gamma$ and
$\bar B_s\to\phi\gamma$. After integrating out the hard scale $m_b$
which results in SCET$_I$, contributions to $B\to V\gamma$ decay
amplitudes can be divided into two different groups: contributions
from operators $J_A$ and $J_B$. The $B$-type operator is power
suppressed in SCET$_I$ but it can give leading power contributions
when matching onto SCET$_{II}$ as $A$-type operator receives power
suppressions. When performing the matching from SCET$_I$ to
SCET$_{II}$, we have to be cautious about the $A$-type operator as
this term suffers from the end-point singularities. Thus one has to
leave it as a non-perturbative free parameter determined from
experiments or some non-perturbative QCD approaches, but recent
studies using zero-bin subtractions show that this term can also
factorized in rapidity space \cite{Manohar:2006nz}. In
Ref.~\cite{Ali:2007sj}, the authors calculated the two-loop
corrections ($\alpha_s^2$) to short-distance coefficients of the
$A$-type operator (called vertex functions) determined from QCD to
SCET$_I$ matching and utilized the physical $B\to V$ form factor
$T_1$ to extract the soft form factor in SCET. While for the
$B$-type operator which can been factorized into convolutions of
LCDAs and hard kernels, both of the Wilson coefficient and jet
function have been calculated up to one loop order. Short-distance
Wilson coefficients for the $B$-type operator are extracted at the
scale $m_b$ and have been evolved down the intermediate scale
$\mu\sim 1.5$ GeV using renormalization group equations. With these
results, the authors find: compared with the leading order
contribution, the order $\alpha_s$ corrections from vertex
corrections and hard spectator scattering can be as large as 10$\%$
and both of them also provide imaginary amplitudes about $5\%$ in
magnitude; the order $\alpha_s^2$ corrections are not too large. We
quote their final results for branching ratios in
table~\ref{Tab:BRVgamma} and they are also consistent with ours.

The three experimental collaborations, BaBar \cite{Aubert:2004te},
Belle \cite{Nakao:2004th} and CLEO \cite{Coan:1999kh}, have reported
their measurements on ${\cal BR}(B\to K^*\gamma)$. Since all of
these results are well consistent with each other, we quote the
averaging results from Heavy Flavor Averaging Group (HFAG)
\cite{Barberio:2006bi}  in table~\ref{Tab:BRVgamma}. We also include
the very recent result on ${\cal {BR}}(\bar B_s^0\to \phi\gamma)$
\cite{Drutskoy:2007aj}. All of them agree with our calculations. On
the $b\to d$ transition, branching ratios of some channels have been
given by the BaBar \cite{Aubert:2006pu} and Belle \cite{Abe:2005rj}
collaboration. We find that our results agree well with BaBar's
results but not with Belle's central value results. In flavor SU(3)
symmetry limit, the relation ${\cal BR}(B^-\to\rho^-\gamma)=2{\cal
BR}(\bar B^0\to\rho^0\gamma)={\cal BR}(\bar B^0\to\omega\gamma)$
should be held. The small deviation in our calculation is caused by
the SU(3) symmetry breaking effect. Since the electro-magnetic
penguin operator $O_{7\gamma}$ gives the dominant contribution, it
is difficult to understand the results from Belle collaboration: why
is the branching ratio of $\bar B^0\to\rho^0\gamma$ larger than
$B^-\to\rho^-\gamma$. But before we conclude it is the signal for
non-standard model scenarios, it is necessary to re-examine this
channel on the experimental side. All other decay modes, including
$\bar B_s$ decays and annihilation type decays, have not been
experimentally measured.

\subsection{CP asymmetry studies}

The direct CP-asymmetry in $\bar B\to V\gamma$ is defined by:
\begin{eqnarray}
A^{dir}_{\mathrm{CP}}\equiv\frac{{\cal BR}(\bar B\to \bar
V\gamma)-{\cal BR}( B\to V\gamma)}{{\cal BR}(\bar B\to \bar
V\gamma)+{\cal BR}( B\to V\gamma)}=\frac{ |{\cal A}(\bar B\to
 \bar V\gamma)|^2-|{\cal A}( B\to V\gamma)|^2}{ |{\cal A}(\bar B\to \bar
V\gamma)|^2+|{\cal A}( B\to V\gamma)|^2}.\label{cpf}
\end{eqnarray}
In order to give a non-zero direct CP asymmetry, we need two kinds
of contributions with different strong phases and different weak
phases. The magnitude of the CP asymmetry also depends on relative
sizes of the two different amplitudes: if one amplitude is much
larger than the other one, we can only get a small CP asymmetry.
Since there is only penguin contribution in $\bar B^0\to \phi\gamma$
process, the direct CP asymmetry is zero. In other $B\to V \gamma$
decays, there are contributions from the penguin operator and two
kinds of tree operators (proportional to $V_{cb}V^*_{cd,cs}$ and
$V_{ub}V^*_{ud,us}$). Taking these amplitudes into account, we
obtain the numerical results for direct CP asymmetries (in $\%$) in
other $ B\to V\gamma$ decays as:
\begin{eqnarray}
A^{dir}_{\rm{CP}}(B^-\to \rho^-\gamma)&=&12.8^{+0.8+2.9+0.8}_{-0.3-1.8-0.8},\\
A^{dir}_{\mathrm{CP}}(\bar B^0\to \rho^0\gamma)&=&12.4^{+0.2+1.8+0.5}_{-0.4-2.4-0.9},\\
A^{dir}_{\mathrm{CP}}(B^-\to K^{*-}\gamma)&=&-0.4\pm0.0\pm0.1\pm0.0,\label{the1}\\
A^{dir}_{\mathrm{CP}}(\bar B^0\to \bar K^{*0}\gamma)&=&-0.3_{-0.0-0.0-0.0}^{+0.0+0.0+0.0},\label{the2}\\
A^{dir}_{\mathrm{CP}}(\bar B^0\to \omega\gamma)&=&12.1^{+0.0+1.8+0.5}_{-0.2-2.4-0.8},\\
A^{dir}_{\mathrm{CP}}(\bar B^0_s\to \rho^0\gamma)&=&-0.1_{-0.0-0.1-0.0}^{+0.0+0.3+0.0},\\
A^{dir}_{\mathrm{CP}}(\bar B^0_s\to K^{*0}\gamma)&=&12.7_{-0.5-2.3-0.9}^{+0.1+1.6+0.5},\\
A^{dir}_{\mathrm{CP}}(\bar B^0_s\to \omega\gamma)&=&-0.3_{-0.0-0.5-0.0}^{+0.0+0.9+0.0},\\
A^{dir}_{\mathrm{CP}}(\bar B^0_s\to
\phi\gamma)&=&-0.3_{-0.0-0.0-0.0}^{+0.0+0.0+0.0},
\end{eqnarray}
where the three kinds of errors are given as that of the branching
ratios case. It is easy to see that theoretical uncertainties here
are much smaller than branching ratios  in table~\ref{Tab:BRVgamma},
especially the first one from hadronic input parameters, since they
are mostly canceled in eq.(\ref{cpf}). In the three $b\to s$
channels $B^-\to K^{*-}\gamma$, $\bar B^0\to \bar K^{*0}\gamma$ and
$\bar B_s\to \phi\gamma$, CKM matrix element for the magnetic
penguin operator is $V_{tb}V_{ts}^*\sim\lambda^2$, while the tree
operator is either proportional to $V_{cb}V_{cs}^*\sim\lambda^2$ or
$V_{ub}V_{us}^*\sim\lambda^4$. The CKM matrix element in the first
tree operator is almost parallel to the penguin operator. This kind
of contribution has a same weak phase with the penguin contribution.
The second tree operator is small in magnitude. Thus we expect small
CP asymmetries in these channels. Experimentally, both BaBar and
Belle collaboration give their combined  measurements of the two
$B\to K^* \gamma$ channels \cite{exp}:
\begin{eqnarray}
A^{dir}_{\mathrm{CP}}(B \to K^{*}\gamma)&=&\left\{
\begin{array}{ll}
-1.5\pm 4.4 \pm 1.2  , & [{\rm Belle}], \\
-1.3\pm 3.6 \pm 1.0  , & [{\rm BaBar}].
\end{array}\right.
\end{eqnarray}
Although the central value of CP asymmetry is larger than the one in
our calculation in eq.(\ref{the1},\ref{the2}), it is still
consistent with zero.

In annihilation-type decays $B_s\to\rho^0(\omega)\gamma$, the tree
amplitude is also suppressed by CKM matrix elements, thus the CP
asymmetry is small too. In $b\to d$ transitions $B^-\to
\rho^{*-}\gamma$, $\bar B^0\to   \rho^{0}(\omega)\gamma$ and $\bar
B_s\to K^{*0}\gamma$, the CKM matrix element for the magnetic
penguin operator is $V_{tb}V_{td}^*\sim\lambda^3$, while the tree
operator is either proportional to $V_{cb}V_{cd}^*\sim\lambda^3$ or
$V_{ub}V_{ud}^*\sim\lambda^3$. Then tree contribution is not
suppressed and  can be comparative with the penguin contribution.
Thus we expect relatively large CP asymmetries in these four
processes, which are also shown in the above.

Restricting the final vector state $V$ to have definite CP-parity,
the time-dependent decay width for the $B^0 \to f$ decay is:
\begin{eqnarray}
\Gamma(B^0(t)\to f)&=& e^{-\Gamma t}\; \overline \Gamma(B\to f)
\Big[\cosh \Big(\frac{\Delta \Gamma t}{2}\Big)
+H_f \sinh \Big(\frac{\Delta \Gamma t}{2}\Big)\nonumber\\
&&\;\;\;\;-A_{\mathrm{CP}}^{dir}\cos( \Delta m t)- S_f \sin
(\Delta m t) \Big],
\end{eqnarray}
where $\Delta m=m_H-m_L>0$, $\overline{\Gamma}$ is the average decay
width, and  $\Delta \Gamma=\Gamma_H-\Gamma_L$ is the difference of
decay widths for the heavier and lighter $B^0$ mass eigenstates. The
time dependent decay width $\Gamma(\bar B^0(t)\to f)$ is obtained
from the above expression by flipping the signs of the $\cos(\Delta
m t)$ and $\sin(\Delta m t)$ terms. In the $B_d$ system, $\Delta
\Gamma$ is small and can be neglected. In the $B_s$ system, we
expect a much larger decay width difference
$(\Delta\Gamma/\Gamma)_{B_s}=-0.127 \pm 0.024$ \cite{Beneke:1998sy}
within the standard model, while experimentally $(\Delta
\Gamma/\Gamma)_{B_s}=-0.33^{+0.09}_{-0.11}$ \cite{Barberio:2006bi},
so that both $S_f$ and $H_f$,  can be extracted from the time
dependent decays of $B_s$ mesons. The definition of the various
quantities in the above equation are as follows:
\begin{eqnarray}
 S_f(V\gamma)
 &=&  \frac{2 \,{\rm Im}\,\left(\frac{q}{p}({\cal A}_L^* \bar{\cal A}_L +
                                       {\cal A}_R^* \bar{\cal A}_R)\right)}{
        |{\cal A}_L|^2 + |{\cal A}_R|^2 + |\bar{\cal A}_L|^2 + |\bar{\cal
                                       A}_R|^2},\\
 H_f(V\gamma)
 &=&  \frac{2 \,{\rm Re}\,\left(\frac{q}{p}({\cal A}_L^* \bar{\cal A}_L +
                                       {\cal A}_R^* \bar{\cal A}_R)\right)}{
        |{\cal A}_L|^2 + |{\cal A}_R|^2 + |\bar{\cal A}_L|^2 + |\bar{\cal
                                       A}_R|^2}\,,
\end{eqnarray}
where $\bar A$ and $A$ denote the amplitudes for the $\bar B$ and
$B$ meson decays.  $q/p$ is given in terms of the $B^0_q$-$\bar
B^0_q$ mixing matrix $M_{12}$,
\begin{eqnarray}
\frac{q}{p} =\sqrt\frac{M_{12}^*}{M_{12}} =e^{i\phi_q}
\end{eqnarray}
with
\begin{equation}
\phi_d \equiv -{\rm arg}[(V_{td}^* V_{tb})^2] = -2 \beta\,,\qquad
\phi_s \equiv -{\rm arg}[(V_{ts}^* V_{tb})^2] = 2\epsilon.
\end{equation}
where the convention $\mbox{arg}[V_{cb}]=\mbox{arg}[V_{cs}]=0$ is
adopted.

In $b\to D\gamma(D=d,s) $ processes, the dominant contribution to
decay amplitudes comes from the chiral-odd dipole operator $O_7$. As
only left-handed quarks participate in the weak interaction, an
effective operator of this type necessitates, a helicity flip on one
of the external quark lines, which results in a factor $m_b$ (and a
left-handed photon) in $b_R\to D_L\gamma_L$ and a factor $m_D$ (and
a right-handed photon) in $b_L\to D_R\gamma_R$. Hence, the emission
of right-handed photon is suppressed by a factor $m_D/m_b$. In the
$b\to D\gamma$ process, the emitted photon is predominantly
left-handed, and right-handed in $\bar b$ decays. This leads to very
small predictions of $S_f$ and $H_f$. The mixing-induced CP
asymmetry variables are calculated and summarized in table
\ref{Tab:CPMixingVgamma}. $\bar B^0 \to \bar K^{*0} \gamma$ has been
treated as an effective flavor eigenstate. Apparently, the numerical
results agree with our expectations. On the experimental side, the
mixing-induced CP asymmetries have been measured by  Belle and BaBar
as follows \cite{Ushiroda:2005sb,Aubert:2007qj,:2007jf}:
\begin{eqnarray}
 S_f(B\to K^*\gamma\to K_S\pi^0\gamma)&=&\begin{cases}
-0.79^{+0.63}_{-0.50}\pm 0.10,  \;\;\;\, [\mbox{Belle}],\\
-0.08\pm0.31\pm0.05,  [\mbox{BaBar}],
\end{cases}\\
 S_f(\bar B^0\to \rho^0\gamma) &=&-0.83 \pm 0.65\pm 0.18.
\end{eqnarray}
They are consistent with zero since there are large uncertainties in
these results. The theoretical results agree with the experimental
data taking the experimental uncertainty into account. But as this
parameter could be a good probe to detect the non-standard
scenarios, more studies, including both of the precise experimental
studies and the theoretical studies, are strongly deserving.

\begin{table}[tb]
\caption{Mixing-induced CP-asymmetry parameters (in percentage) of
$B\to V\gamma$ decays obtained in pQCD approach. The errors are the
same with table \ref{Tab:BRVgamma}. The $H_f$ parameter in $B_d^0$
decays could hardly be measured as the decay width difference is
small.}
 \label{Tab:CPMixingVgamma}
\begin{center}
 \begin{tabular}{c|c|c}
  \hline\hline
        {Modes}     &  $S_f$ &   $H_f$
          \\  \hline
   $\bar B^0\to \bar K^{*0}\gamma$
                                                          &$-4.0_{-0.0-0.2-0.1}^{+0.1+0.3+0.1}$
                                                          &$4.1_{-0.1-0.3-0.1}^{+0.0+0.2+0.1}$
                                                          \\\hline
   $\bar B_s^0\to \phi\gamma$
                                                          & $0.2_{-0.0-0.0-0.0}^{+0.0+0.0+0.0}$
                                                          & $5.5_{-0.2-0.4-0.0}^{+0.1+0.4+0.0}$
                                                          \\ \hline
   $\bar B^0\to \rho^0\gamma$
                                                          & $0.8_{-0.1-0.2-0.1}^{+0.1+0.0+0.0}$
                                                          & $0.4_{-0.1-0.3-0.1}^{+0.3+0.5+0.1}$
                                                           \\\hline
   $\bar B^0\to \omega\gamma$
                                                          & $0.4_{-0.1-0.2-0.0}^{+0.0+0.2+0.0}$
                                                          & $0.5_{-0.0-0.3-0.0}^{+0.1+0.4+0.1}$
                                                          \\\hline
   $\bar B_s^0\to  K^{*0}\gamma$
                                                          & $0.7_{-0.1-0.2-0.1}^{+0.0+0.4+0.1}$
                                                          & $-0.3_{-0.2-0.3-0.0}^{+0.1+0.3+0.1}$
                                                          \\
 \hline\hline
 \end{tabular}
\end{center}
\end{table}

\subsection{Isospin asymmetry and U-spin asymmetry}

Apart from branching ratios and CP asymmetry variables, we will also
consider some ratios of branching fractions defined below. In the
evaluations for branching fractions, there are many uncertainties,
especially from hadronic input parameters, which can blur our
predictions, but we can improve the accuracy of our predictions by
using ratios of branching fractions. Many uncertainties, such as
those from decay constants, will cancel in these parameters. The
most important ratios are the parameters characterizing isospin
asymmetries which are defined by:
\begin{eqnarray}
A(\rho,\omega) & = &  \frac{\overline{\Gamma}(B^0\to \omega
  \gamma)}{\overline{\Gamma}(B^0\to \rho^0
  \gamma)}-1\,, \label{Eq:AIrhoomega} \\
A_{I}(\rho) & = & \frac{2\overline{\Gamma}(\bar B^0\to
  \rho^0 \gamma)}{\overline{\Gamma}(\bar B^\pm\to
  \rho^\pm \gamma)} - 1\,,\label{Eq:AIrho}\\
A_{I}(K^*) & = & \frac{\overline{\Gamma}(\bar B^0\to
  K^{*0} \gamma) - \overline{\Gamma}(B^\pm\to
  K^{*\pm} \gamma)}{\overline{\Gamma}(\bar B^0\to
  K^{*0} \gamma) + \overline{\Gamma}(B^\pm\to
  K^{*\pm} \gamma)},\label{Eq:AIK}
\end{eqnarray}
where the partial decay rates are CP-averaged.

In the flavor SU(3) symmetry limit and if we neglect diagrams which
are proportional to the quark charge, all of these three parameters
should be equal to 0. The $\omega$ meson decay constant is smaller
than that for $\rho^0$ meson, the $\bar uu$ component contributes
with a different sign and the electomagnetic diagrams with two
photons are different in the charge factor. These differences make
$A(\rho,\omega)$ deviate from 0 (smaller than 0). The origins for
deviations for $A_I(\rho)$ and $A_I(K^*)$ from 0 are similar: the
spectator quarks are different and annihilation diagrams are also
different. Taking on all those power suppressed contributions, our
predictions are
\begin{eqnarray}
A(\rho,\omega) & = &-0.11_{-0.00-0.00-0.00}^{+0.01+0.01+0.00} ,\\
A_{I}(\rho) & = &0.06_{-0.03-0.01-0.02}^{+0.03+0.01+0.04} ,\\
A_{I}(K^*)&=&0.06_{-0.01-0.00-0.00}^{+0.02+0.01+0.00}.
\end{eqnarray}
As we expected, theoretical uncertainties due to the hadronic
parameters are indeed smaller due to cancelations. Using
experimental results listed in table \ref{Tab:BRVgamma}, we can
calculate the isospin asymmetry parameters from experiments
\begin{eqnarray}
 A(\rho,\omega)&=&\begin{cases}
  -0.49^{+0.34}_{-0.30} & [\rm{BaBar}],\\
   -0.55^{+0.27}_{-0.27} & [\rm{Belle}],
  \end{cases}\\
 A_I(\rho) &=&\begin{cases}
   -0.54^{+0.65}_{-0.67}& [\rm{BaBar}],\\
   3.89^{+3.60}_{-4.04} &[\rm{Belle}],
   \end{cases}\\
A_{I}(K^*)&=& 0.03\pm0.04,
\end{eqnarray}
where we have assumed all the uncertainties are not correlated by
adding them quadratically. From the experimental results, except
$A_I(K^*)$, we find that there are large differences between the
results from the two collaborations. Our results are consistent with
them, since the error bars in the experiments are too large. We wish
a more precise measurement on these parameters in the future.

Apart from the isospin symmetry, U-spin symmetry is another kind
symmetry which is well held in strong interactions. U-spin can
connect two different kinds of weak decays
\cite{Soares:1991te,Gronau:2000zy,Hurth:2001yb}: $b\to s$ ($\Delta
S=1$) and $b\to d$ by exchange of $d\leftrightarrow s$. The decay
amplitudes of $b\to s$ process can be expressed as:
\begin{eqnarray}
{\cal A}(B\to f)=V_{ub}^*V_{us}{\cal A}_u+V_{cb}^*V_{cs}{\cal A}_c,\\
{\cal A}(\bar B\to \bar f)=V_{ub}V_{us}^*{\cal
A}_u+V_{cb}V_{cs}^*{\cal A}_c,
\end{eqnarray}
while the decay amplitudes of $UB\to Uf$ are
\begin{eqnarray}
 {\cal A}(UB\to Uf)=V_{ub}^*V_{ud}U{\cal A}_u+V_{cb}^*V_{cd}U{\cal A}_c,\\
 {\cal A}(U\bar B\to U\bar f)=V_{ub}V_{ud}^*U{\cal A}_u+V_{cb}V_{cd}^*U{\cal
 A}_c.
\end{eqnarray}
Using the relation ${\cal A}_u=U{\cal A}_u$ and ${\cal A}_c=U{\cal
A}_c$ in U-spin symmetry limit and the CKM unitarity relation
\begin{eqnarray}
{\mbox Im}(V_{ub}^*V_{us}V_{cd}V_{cs}^*)=-{\mbox
Im}(V_{ub}^*V_{ud}V_{cd}V_{cd}^*),
\end{eqnarray}
we obtain
\begin{eqnarray}
|{\cal A}(B\to f)|^2-|{\cal A}(\bar B\to \bar f)|^2=-|{\cal
A}(UB\to Uf)|^2+|{\cal A}(U\bar B\to U\bar f)|^2.
\end{eqnarray}
This equation relates the differences of partial decay widths. The
following radiative $B$ decays can be related to each other by this
symmetry: $B^-\to\rho^-\gamma$ and $B^-\to K^{*-}\gamma$; $\bar
B^0\to \bar K^{*0}\gamma$ and $\bar B^0_s\to K^{*0}\gamma$. As an
example, we define the following parameter to test U-spin symmetry
breaking:
\begin{eqnarray}
\Delta\equiv{A_{CP}(B^-\to K^{*-}\gamma)}-A_{CP}(B^-\to
\rho^-\gamma)\times\frac{{\cal BR}(B^-\to \rho^-\gamma)}{{\cal
BR}(B^-\to K^{*-}\gamma)}.
\end{eqnarray}
In our pQCD calculation, we find
\begin{eqnarray}
\Delta=(-8.4_{-0.8-2.3-0.6}^{+0.4+1.3+0.3})\times10^{-3}.
\end{eqnarray}
This result is close to 0. The U spin symmetry seems quite good
here. But the most important reason is the small CP asymmetry in
$B^-\to K^{*-}\gamma$ and the small ratio $\frac{{\cal
BR}(B^-\to\rho^-\gamma)}{{\cal BR}(B^-\to K^{*-}\gamma)}$. The
absolute value for the $b\to s$ channel's CP asymmetry is small.
This direct CP asymmetry may be dramatically enhanced by some new
physics with a different weak phase. The precise measurement of CP
asymmetries can at least give a constraint on the non-standard model
scenario parameters.

\section{Calculation of $B\to A(1^3P_1)\gamma$ and $A(1^1P_1)\gamma$ decays} \label{section:BtoA}


The factorization formulae for $B\to A\gamma$ is more complicated
than $B\to V\gamma$ because of the mixing between different mesons:
$K_{1A}$ and $K_{1B}$; $f_1$ and $f_8$; $h_1$ and $h_8$. The real
physical states $K_1(1270)$ and $K_1(1400)$ are mixtures of the
$K_{1A}$ and $K_{1B}$ states with the mixing angle $\theta_K$:
\begin{eqnarray}
|K_1(1270)\rangle&=&|K_{1A}\rangle
{\rm{sin}}\theta_K+|K_{1B}\rangle{\rm{cos}}\theta_K,\\
|K_1(1400)\rangle&=&|K_{1A}\rangle
{\rm{cos}}\theta_K-|K_{1B}\rangle{\rm{sin}}\theta_K.
\end{eqnarray}
In flavor SU(3) symmetry limit, these mesons can not mix with each
other; but since  $s$ quark is heavier than the $u,d$ quarks,
$K_1(1270)$ and $K_1(1400)$ are not purely $1^3P_1$ or $1^1P_1$
states. In general, the mixing angle can be determined by
experimental data. The partial decay rate for $\tau^-\to
K_1\nu_\tau$ is given by:
\begin{eqnarray}
{\Gamma}(\tau^-\to
K_1\nu_\tau)=\frac{m_\tau^3}{16\pi}G_F^2|V_{us}|^2f_A^2\left(1-\frac{m_A^2}{m_\tau^2}
\right)^2\left(1+\frac{2m_A^2}{m_\tau^2}\right),
\end{eqnarray}
with the measured results for branching fractions \cite{Yao:2006px}:
\begin{eqnarray}
{\cal BR}(\tau^-\to K_1(1270)\nu_\tau)=(4.7\pm1.1)\times 10^{-3},\;
{\cal BR}(\tau^-\to K_1(1400)\nu_\tau)=(1.7\pm2.6)\times
10^{-3}.\label{eq:BRintau}
\end{eqnarray}
We can straightforward obtain the longitudinal decay constant (in
MeV):
\begin{eqnarray}
|f_{K_1(1270)}|=169_{-21}^{+19};\;\;\;
|f_{K_1(1400)}|=125_{-125}^{+~74}.\label{eq:decayconstantK1}
\end{eqnarray}
In principle, one can combine the decay constants for $K_{1A}$,
$K_{1B}$ evaluated in QCD sum rules with the above results to
determine the mixing angle $\theta_K$. But since there are large
uncertainties in Eq.~(\ref{eq:decayconstantK1}), the constraint on
the mixing angle is expected to be rather smooth:
\begin{eqnarray}
 -143^\circ<\theta_K<-120^\circ,\;\;\;{\rm or}\;\;
 -49^\circ<\theta_K<-27^\circ,\;\;\;{\rm or}\;\;
 37^\circ<\theta_K<60^\circ,\;\;\;{\rm or}\;\;
 131^\circ<\theta_K<153^\circ,
\end{eqnarray}
where we have taken the uncertainties from the branching ratios in
Eq.(\ref{eq:BRintau}) and the first Gegenbauer moment $a_1^K$ into
account but neglected the mass differences as usual.  For
simplicity, we use two reference values in Ref.~\cite{Yang:2007zt}
\begin{eqnarray}
\theta_K=\pm 45^\circ.
\end{eqnarray}
Besides, the flavor-octet and the flavor-singlet can also mix with
each other:
\begin{eqnarray}
 |f_1(1285)\rangle&=&|f_{1}\rangle
{\rm{cos}}\theta_{^3P_1}+|f_{8}\rangle{\rm{sin}}\theta_{^3P_1},\;\;\;
 |f_1(1420)\rangle=-|f_{1}\rangle
{\rm{sin}}\theta_{^3P_1}+|f_{8}\rangle{\rm{cos}}\theta_{^3P_1},\\
 |h_1(1170)\rangle&=&|h_{1}\rangle
{\rm{cos}}\theta_{^1P_1}+|h_{8}\rangle{\rm{sin}}\theta_{^1P_1},\;\;\;
 |h_1(1380)\rangle=-|h_{1}\rangle
{\rm{sin}}\theta_{^1P_1}+|h_{8}\rangle{\rm{cos}}\theta_{^1P_1}.
\end{eqnarray}
The references points are chosen as: $\theta_{^3P_1}=38^\circ$ or
$\theta_{^3P_1}=50^\circ$; $\theta_{^1P_1}=10^\circ$ or
$\theta_{^1P_1}=45^\circ$ \cite{Yang:2007zt}. We should point out
that if the mixing angle is $\theta=35.3^\circ$, the mixing is
ideal: $f_1(1285)$ is made up of $\frac{\bar uu+\bar dd}{\sqrt 2}$
while $f_1(1420)$ is composed of $\bar ss$. Thus some of the form
factors are very small which will of course give small production
rates of this meson.

Apart from these differences, the expression for $B\to A \gamma$ is
different from $B\to V\gamma$ in more aspects. Since the twist-2
LCDA $\phi_{||}$ is normalized to $a^{||}_0$, we should replace the
decay constant $f_V$ by $f_Aa_0^{||}$ in the first two annihilation
diagrams. As there is no overlap between an axial-vector meson and a
photon (wrong parity), there is no contribution from the two photon
electromagnetic operator diagrams in Fig.~\ref{Feyn:twophoton}.
Regardless of these differences, the factorization formulae for
$B\to A\gamma$ decays can be obtained from the corresponding $B\to
V\gamma$ ones using the replacement in Eq. (\ref{eq:relationofAVDA})
if the electroweak current is $\sigma_{\mu\nu}(1+\gamma_5)$ or
$\gamma_\mu(1-\gamma_5)$ type. If the current is
$\sigma_{\mu\nu}(1-\gamma_5)$, we should add an additional minus
sign. In annihilation diagrams, if the electroweak current is $LL$
or $SP$ in the lower two diagrams, we need replace the distribution
amplitudes as in Eq.~(\ref{eq:relationofAVDA}); while we add a minus
sign if the current is $LR$. We show the formulas in Appendix
\ref{sec:Agamma}.


\begin{table}[tb]
\caption{$CP$-averaged branching ratios ($\times
 10^{-6}$) of $B\to A\gamma$ decays obtained in pQCD approach
using two different mixing angles; the errors for these entries
correspond to uncertainties in the input hadronic quantities, from
the scale-dependence, and CKM matrix elements, the Gegenbauer
moments of the axial-vector mesons respectively. }
 \label{Tab:BRAgamma}
\begin{center}
 \begin{tabular}{c|c|c|c}
   \hline\hline
        {Modes}     &  $\theta_K=45^\circ$ &  $\theta_K=-45^\circ$& Exp.
          \\  \hline
     $B^-\to K^{-}_{1}(1270)\gamma$
                                                          & $134_{-49-18-4-38}^{+68+21+4+41}$
                                                          & $1.4_{-0.7-0.6-0.0-2.0}^{+1.2+0.3+0.0+5.0}$
                                                          & $42.8 \pm 9.4 \pm 4.3$ \cite{Abe:2004kr}
                                                           \\\hline
   $\bar B^0\to \bar K^{0}_{1}(1270)\gamma$
                                                          & $141_{-48-18-4-41}^{+64+19+4+45}$
                                                          & $1.4_{-0.6-0.5-0.0-1.9}^{+0.9+0.3+0.0+5.4}$
                                                          &
                                                          \\\hline
     $B^-\to K^{-}_{1}(1400)\gamma$
                                                          & $1.4_{-0.7-0.6-0.0-2.0}^{+1.2+0.3+0.0+5.0}$
                                                          & $134_{-49-18-4-38}^{+68+21+4+41}$
                                                          & $<14.4$
                                                           \\\hline
   $\bar B^0\to \bar K^{0}_{1}(1400)\gamma$
                                                          & $1.4_{-0.6-0.5-0.0-1.9}^{+0.9+0.3+0.0+5.4}$
                                                          & $141_{-48-18-4-41}^{+64+19+4+45}$
                                                          &
                                                          \\\hline
     $\bar B_s\to K^{0}_{1}(1270)\gamma$
                                                          & $0.19_{-0.06-0.03-0.01-0.22}^{+0.07+0.02+0.02+0.34}$
                                                          & $0.38_{-0.15-0.07-0.03-0.32}^{+0.24+0.09+0.07+0.44}$
                                                          &
                                                           \\\hline
   $\bar B_s\to  K^{0}_{1}(1400)\gamma$
                                                          & $0.38_{-0.15-0.07-0.03-0.32}^{+0.24+0.09+0.07+0.44}$
                                                          & $0.19_{-0.06-0.03-0.01-0.22}^{+0.07+0.02+0.02+0.34}$
                                                          &
                                                          \\ \hline\hline
   {Modes}     &  $\theta_{^3P_1}=38^\circ$ &  $\theta_{^3P_1}=50^\circ$&
          \\  \hline
   $\bar B^0\to f_{1}(1285)\gamma$
                                                          & $1.7_{-0.6-0.2-0.1-0.4}^{+0.8+0.2+0.2+0.5}$
                                                          & $1.6_{-0.5-0.2-0.1-0.4}^{+0.7+0.2+0.2+0.4}$
                                                          &
                                                          \\\hline
   $\bar B^0\to f_{1}(1420)\gamma$
                                                          & $(4.9_{-1.7-1.1-0.3-2.7}^{+2.3+0.6+0.7+3.9})\times10^{-3}$
                                                          & $0.11_{-0.04-0.02-0.01-0.04}^{+0.05+0.01+0.02+0.04}$
                                                          &
                                                          \\\hline
   $\bar B_s^0\to f_{1}(1285)\gamma$
                                                          & $0.11_{-0.04-0.01-0.00-0.03}^{+0.05+0.01+0.00+0.03}$
                                                          & $3.8_{-1.2-0.4-0.1-0.7}^{+1.6+0.4+0.1+0.7}$
                                                          &
                                                          \\\hline
   $\bar B_s^0\to f_{1}(1420)\gamma$
                                                          & $61.9_{-18.9-6.0-1.8-15.5}^{+24.5+5.5+1.8+17.4}$
                                                          & $58.2_{-17.7-5.6-1.7-14.8}^{+22.9+5.1+1.6+16.7}$
                                                          &
                                                          \\\hline\hline
        {Modes}     &  $\theta_{^1P_1}=10^\circ$ &  $\theta_{^1P_1}=45^\circ$&
          \\  \hline
   $\bar B^0\to h_{1}(1170)\gamma$
                                                          & $0.99_{-0.33-0.13-0.06-0.21}^{+0.43+0.16+0.14+0.24}$
                                                          & $1.24_{-0.41-0.16-0.08-0.27}^{+0.55+0.20+0.18+0.31}$
                                                          &
                                                          \\\hline
   $\bar B^0\to h_{1}(1380)\gamma$
                                                          & $0.28_{-0.09-0.04-0.02-0.06}^{+0.12+0.05+0.04+0.07}$
                                                          & $(2.0_{-0.7-0.3-0.1-0.3}^{+0.8+0.3+0.3+0.3})\times10^{-2}$
                                                          &
                                                          \\\hline
   $\bar B_s^0\to h_{1}(1170)\gamma$
                                                          & $7.9_{-2.2-0.7-0.2-1.6}^{+2.9+1.0+0.2+1.8}$
                                                          & $2.3_{-0.7-0.3-0.1-0.6}^{+0.9+0.3+0.1+0.7}$
                                                          &
                                                          \\\hline
   $\bar B_s^0\to h_{1}(1380)\gamma$
                                                          & $44.4_{-12.8-4.1-1.3-9.7}^{+16.8+5.6+1.3+11.0}$
                                                          & $50.0_{-14.3-4.5-1.5-10.7}^{+18.8+6.3+1.5+12.2}$
                                                          &
                                                          \\\hline
 \end{tabular}
\end{center}
 \end{table}

\begin{table}[tb]
\caption{Direct CP asymmetries of $B\to A\gamma$ decays obtained in
pQCD approach using two different mixing angles; the errors for
these entries correspond to uncertainties in the input hadronic
quantities, from the scale-dependence, and CKM matrix elements,
respectively. }
 \label{Tab:AcpAgamma}
\begin{center}
  \begin{tabular}{c|c|cc}
  \hline\hline
        {Modes}     &  $\theta_K=45^\circ$ &  $\theta_K=-45^\circ$&
          \\  \hline
     $B^-\to K^{-}_{1}(1270)\gamma$
                                                          & $-0.6_{-0.0-0.1-0.0-0.1}^{+0.0+0.2+0.0+0.1}$
                                                          & $-3.8_{-1.3-0.3-0.3-13.0}^{+0.7+1.1+1.8+4.0}$
                                                          &
                                                           \\\hline
   $\bar B^0\to \bar K^{0}_{1}(1270)\gamma$
                                                          & $-0.2\pm0.0\pm0.0\pm0.0\pm0.0$
                                                          & $0.1_{-0.1-0.3-0.0-0.4}^{+0.0+0.1+0.0+2.2}$
                                                          &
                                                          \\\hline
     $B^-\to K^{-}_{1}(1400)\gamma$
                                                          & $-3.8_{-1.3-0.3-0.3-13.0}^{+0.7+1.1+1.8+4.0}$
                                                          & $-0.6_{-0.0-0.1-0.0-0.1}^{+0.0+0.2+0.0+0.1}$
                                                          &
                                                           \\\hline
   $\bar B^0\to \bar K^{0}_{1}(1400)\gamma$
                                                          & $0.1_{-0.1-0.3-0.0-0.4}^{+0.0+0.1+0.0+2.2}$
                                                          & $-0.2\pm0.0\pm0.0\pm0.0\pm0.0$
                                                          &
                                                          \\\hline
     $\bar B_s\to K^{0}_{1}(1270)\gamma$
                                                          & $-8.4_{-2.9-3.4-0.4-11.2}^{+0.0+0.5+0.4+5.1}$
                                                          & $-3.1_{-0.1-3.1-0.6-14.1}^{+3.9+4.0+0.8+8.6}$
                                                          &
                                                           \\\hline
   $\bar B_s\to  K^{0}_{1}(1400)\gamma$
                                                          & $-3.1_{-0.1-3.1-0.6-14.1}^{+3.9+4.0+0.8+8.6}$
                                                          & $-8.4_{-2.9-3.4-0.4-11.2}^{+0.0+0.5+0.4+5.1}$
                                                          &
                                                          \\ \hline\hline
        {Modes}     &  $\theta_{^3P_1}=38^\circ$ &  $\theta_{^3P_1}=50^\circ$&
          \\  \hline
   $\bar B^0\to f_{1}(1285)\gamma$
                                                          & $3.4_{-0.1-0.5-0.2-0.1}^{+0.6+0.8+0.2+0.7}$
                                                          & $3.4_{-0.1-0.4-0.2-0.1}^{+0.7+0.8+0.2+0.8}$
                                                          &
                                                          \\\hline
   $\bar B^0\to f_{1}(1420)\gamma$
                                                          & $7.1_{-3.1-7.7-0.5-2.3}^{+0.0+0.7+0.3+0.0}$
                                                          & $4.1_{-0.3-1.9-0.3-0.4}^{+0.1+0.5+0.2+0.1}$
                                                          &
                                                          \\\hline
   $\bar B_s^0\to f_{1}(1285)\gamma$
                                                          & $-0.1_{-0.0-0.1-0.0-0.2}^{+0.3+0.1+0.0+0.1}$
                                                          & $-0.2_{-0.0-0.1-0.0-0.0}^{+0.1+0.0+0.0+0.0}$
                                                          &
                                                          \\\hline
   $\bar B_s^0\to f_{1}(1420)\gamma$
                                                          & $-0.2\pm0.0\pm0.0\pm0.0\pm0.0$
                                                          & $-0.2\pm0.0\pm0.0\pm0.0\pm0.0$
                                                          &
                                                          \\\hline\hline
        {Modes}     &  $\theta_{^1P_1}=10^\circ$ &  $\theta_{^1P_1}=45^\circ$&
          \\  \hline
   $\bar B^0\to h_{1}(1170)\gamma$
                                                          & $10.2_{-0.9-2.5-0.7-0.4}^{+0.0+1.4+0.4+0.0}$
                                                          & $10.1_{-0.5-2.3-0.7-0.3}^{+0.1+1.7+0.4+0.2}$
                                                          &
                                                          \\\hline
   $\bar B^0\to h_{1}(1380)\gamma$
                                                          & $9.8_{-0.0-2.0-0.7-0.0}^{+0.8+2.4+0.4+1.1}$
                                                          & $11.3_{-5.1-4.2-0.7-3.5}^{+0.0+0.0+0.5+0.0}$
                                                          &
                                                          \\\hline
   $\bar B_s^0\to h_{1}(1170)\gamma$
                                                          & $-0.2\pm0.0\pm0.0\pm0.0\pm0.0$
                                                          & $-0.1_{-0.0-0.1-0.0-0.0}^{+0.0+0.0+0.0+0.0}$
                                                          &
                                                          \\\hline
   $\bar B_s^0\to h_{1}(1380)\gamma$
                                                          & $-0.2\pm0.0\pm0.0\pm0.0\pm0.0$
                                                          & $-0.2\pm0.0\pm0.0\pm0.0\pm0.0$
                                                          &
                                                          \\\hline
 \end{tabular}
\end{center}
 \end{table}

Branching ratios (in unit of $10^{-6}$) and direct CP asymmetries
(in $\%$) for $B\to (a_1,b_1)\gamma$ processes are calculated
straightforward as follows:
\begin{eqnarray}
{\cal BR}(B^-\to a^-_1(1260)\gamma)&=&3.0_{-1.1-0.3-0.2-0.7}^{+1.6+0.4+0.4+0.8},\\
{\cal BR}(\bar B^0\to a^0_1(1260)\gamma)&=&1.5_{-0.5-0.2-0.1-0.4}^{+0.7+0.2+0.2+0.4},\\
{\cal BR}(\bar B_s\to a^0_1(1260)\gamma)&=&(2.1_{-0.5-0.1-0.1-0.0}^{+0.6+0.3+0.1+0.0})\times10^{-4},\\
{\cal BR}(B^-\to b^-_1(1235)\gamma)&=&2.0_{-0.7-0.3-0.1-0.5}^{+1.0+0.4+0.3+0.6},\\
{\cal BR}(\bar B^0\to b^0_1(1235)\gamma)&=&1.1_{-0.3-0.1-0.1-0.2}^{+0.5+0.2+0.2+0.3},\\
{\cal BR}(\bar B_s\to
b^0_1(1235)\gamma)&=&(5.4_{-0.9-2.5-0.2-1.8}^{+1.0+6.4+0.3+2.1})\times10^{-5},
\end{eqnarray}
\begin{eqnarray}
{A_{CP}^{dir}}(B^-\to a^-_1(1260)\gamma)&=&11.2_{-0.5-2.4-0.8-1.3}^{+2.3+3.0+0.9+2.7},\\
{A_{CP}^{dir}}(\bar B^0\to a^0_1(1260)\gamma)&=&3.8_{-0.5-0.5-0.3-0.7}^{+0.3+0.3+0.2+0.4},\\
{A_{CP}^{dir}}(\bar B_s\to a^0_1(1260)\gamma)&=&0.8_{-0.1-1.5-0.0-0.0}^{+0.1+0.8+0.1+0.0},\\
{A_{CP}^{dir}}(B^-\to b^-_1(1235)\gamma)&=&16.0_{-0.5-2.7-1.1-0.7}^{+1.3+4.2+0.7+1.7},\\
{A_{CP}^{dir}}(\bar B^0\to b^0_1(1235)\gamma)&=&11.0_{-0.3-2.5-0.7-0.2}^{+0.2+1.9+0.5+0.3},\\
{A_{CP}^{dir}}(\bar B_s\to
b^0_1(1235)\gamma)&=&-0.5_{-0.0-1.5-0.0-0.0}^{+0.0+2.7+0.0+0.0},
\end{eqnarray}
while we give branching ratios and CP asymmetries for $B\to
K_1(f_1,h_1)\gamma$ in table \ref{Tab:BRAgamma} and
\ref{Tab:AcpAgamma}, respectively. The errors for these entries
correspond to uncertainties in the input hadronic quantities, the
scale-dependence, CKM matrix elements, and the Gegenbauer moments of
the axial-vector mesons. It is noted that  theoretical uncertainties
for branching ratios are quite large. The branching fractions of
$B\to a_1(1260)(b_1(1235))\gamma$ are larger than that of
$B\to\rho\gamma$, as we have shown that $B\to A$ form factors are
larger. As we have mentioned in the above, there are some
ambiguities in the quark content of $B\to K_1(f_1,h_1)\gamma$: these
mesons are mixtures but the mixing angles are not uniquely
determined. The reference points for the mixing angles are two-fold,
thus we give two different kinds of results collected in these two
tables. The branching ratios and CP asymmetries are very sensitive
to the mixing angles, which is not quite constrained.

\begin{figure}
\vspace{-0.0cm}
\begin{center}
\psfig{file=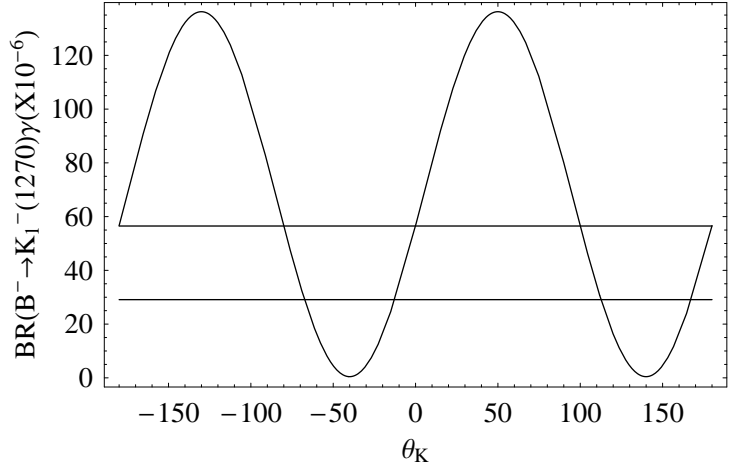,width=10.0cm,angle=0}
\end{center}
 \vspace{-1.0cm}
\caption{The $\theta_K$ dependence of the $B^-\to K_1^-(1270)\gamma$
branching ratio. The region between the two horizontal lines are
allowed by the experimental $1\sigma$ bound, where we add the
statistic and the systematic uncertainties linearly: $29.1<{\cal
BR}<56.5$. } \label{Feyn:thetadependence}
\end{figure}

Experimentalist gave results for $B^-\to K_1^-(1270)\gamma$
\cite{Abe:2004kr} shown also in table~\ref{Tab:BRAgamma}. Compared
with it, our result for $B^-\to K_1^-(1270)\gamma$, is about 3 times
larger, when $\theta_K=45^\circ$; or very smaller than the
experimental results, when $\theta_K=-45^\circ$.  In
Fig.~\ref{Feyn:thetadependence}, we show the  strong dependence of
the branching ratio on the mixing angle. At $\theta_K=45^\circ$, the
$B^-\to K_1^-(1270)\gamma$ receives almost a maximal branching
ratio. The current $B^-\to K_1^-(1270)\gamma$ experiment implies our
chosen two reference points of the mixing angle are not favored.
From this Fig.~\ref{Feyn:thetadependence}, we could read out the
experimental constrained mixing angle value, which are also two
fold.  However large hadronic uncertainties and the missing
next-to-leading order corrections \cite{Li:2006jv} plus the still
large experimental error bars make this constraint not very
effective. Except for these two processes, other decay modes have
not been measured. Here we refrain from a direct comparison with the
previous studies on $B\to A\gamma$
\cite{Cheng:2004yj,Kwon:2004ri,Jamil Aslam:2005mc,Aslam:2006vh,Jamil
Aslam:2006bw,Lee:2006qj}, as the analysis is similar in $B\to A$
form factors which has been performed in section
\ref{section:formfactor}.

\section{Summary}

pQCD approach is based on $k_T$ factorization where we keep the
transverse momentum of valence quarks in the meson, to smear the
endpoint singularity. $k_T$ resummation of double logarithms results
in the Sudakov factor. Resummation of double logarithms from the
threshold region leads to the jet function. Sudakov factor and jet
function can suppress the contribution from the large $b$ region and
small $x$ region, respectively. This makes the pQCD approach
self-consistent. Inspired by the success of pQCD approach in
non-leptonic B decays \cite{pqcd}, we give a comprehensive study on
the charmless $B_{(s)}\to V(A)\gamma$ decays in pQCD approach.

Semi-leptonic and radiative decays are somewhat simpler than
non-leptonic decays as only one hadronic meson involved in the final
state. In this case, the dominant amplitude can be parameterized in
form factors. In order to make precise prediction and extract CKM
matrix elements, we have to know the behavior of form factors. In
pQCD approach, the final state meson moves nearly on the light-cone
and a hard-gluon-exchange is required. Thus the dominant
contribution is from the hard region which can be factorized. In
section \ref{section:formfactor}, we have used the same input
hadronic parameters with Ref. \cite{Ali:2007ff} and updated all the
$B\to V$ decay form factors in pQCD approach. Compared with the
results evaluated from other approaches, we find: despite of a
number of theoretical differences in different approaches, all the
numerical results of the form factors are surprisingly consistent
with each other.

The 10 $B\to V\gamma$ decay channels can be divided into three
categories based on their dominant quark transition $b\to s$, $b\to
d$ and the annihilation topology. Our prediction on the first
category of decays ${\cal BR}(B\to K^*\gamma)$ is consistent with
the averaged value from experiments. On the $b\to d$ transition,
branching ratios have been given by BaBar and Belle collaborations
with still large error bars. We find our results are well consistent
with BaBar's results but a little far from the Belle's central value
results in some channels. We also give our predictions on the purely
annihilation type decays with very small branching ratios in SM. In
three $b\to s$ transitions $B^-\to K^{*-}\gamma$, $\bar B^0\to \bar
K^{*0}\gamma$ and $\bar B_s\to \phi\gamma$, the direct CP asymmetry
is small, since the tree contribution is suppressed by the CKM
matrix element. In the $b\to d$ transitions $B^-\to \rho^{-}\gamma$,
$\bar B^0\to \bar \rho^{0}(\omega)\gamma$ and $\bar B_s\to
K^{*0}\gamma$, the tree contribution can be comparable with the
penguin contribution. Thus we obtain large CP asymmetries in these
four processes. In SM, the two quantities $S_f$ and $H_f$ in
time-dependent decay are expected to be rather small. This is due to
the fact that the dominant contribution to decay amplitudes comes
from the chiral-odd dipole operator $O_7$. Except for a few decays
discussed in the above, all other decay modes including $\bar B_s$
decays and annihilation type decays, remain essentially unexplored.
We wish a wealth of measurements at the $B$ factories and other
experiments in the future.

In section~\ref{section:formfactor}, we also study $B\to A$ form
factors. As the quark contents (to be precise the mixing angle) for
the axial-vectors have not been uniquely determined, we give two
different kinds of results for the form factors according to
different mixing angles. For the axial-vector mesons $f_1$, we have
used the mixing angle between the octet and singlet:
$\theta=38^\circ(50^\circ)$ which is close to the ideal mixing angle
$\theta=35.3^\circ$. With this mixing angle, one can easily check
that the lighter meson $f_1(1285)$ is made almost up of $\frac{\bar
uu+\bar dd}{\sqrt 2}$ while the heavier meson $f_1(1420)$ is
composed of $\bar ss$. Thus partial decay widths of $B\to
f_1(1420)\gamma$ and $B_s\to f_1(1285)\gamma$ are suppressed by the
flavor structure. In Fig.~\ref{Feyn:thetadependence}, we show the
strong dependence of the $B^-\to K_1^-(1270)\gamma$ decay branching
ratio on the mixing angle $\theta_K$.  Our calculation can be used
to constrain this mixing angle using experimental measurements
provided with well understood hadronic inputs. The study of higher
resonance production in $B$ decays can help us to uncover the
mysterious structure of these excited states.

\section*{Acknowledgements}

This work is partly supported by National Science Foundation of
China under the Grant No.~10475085 and 10625525. We would like to
acknowledge S.-Y. Li, Y. Li, Y.-L. Shen, X.-X. Wang, Y.-M. Wang,
K.-C. Yang, M.-Z. Yang and H. Zou for valuable discussions.

\appendix

\section{pQCD functions}\label{PQCDfunctions}

In this appendix, we group the functions which appear in the
factorization formulae. The hard scales are chosen as
\begin{eqnarray}
 t_a&=&\mbox{max}\{{\sqrt {x_2}m_{B},1/b_1,1/b_2}\},\;\;\;
 t_a^\prime=\mbox{max}\{{\sqrt{x_1}m_{B},1/b_1,1/b_2}\},\\
 t_b&=&\mbox{max}\{\sqrt {x_1x_2}m_{B},\sqrt{(1-x_2)}m_{B},1/b_1,1/b_2\},\\
 t_b^\prime&=&\mbox{max}\{\sqrt{x_1x_2}m_{B},\sqrt{1+x_1}m_{B},1/b_1,1/b_2\},\\
 t_c&=&\mbox{max}\{\sqrt{|x_1-x_2|}m_{B},\sqrt {x_2} m_B,1/b_1,1/b_2\},\\
 t_c^\prime&=&\mbox{max}\{\sqrt{|x_1-x_2|}m_{B},\sqrt {x_1} m_B,1/b_1,1/b_2\},\\
 t_d&=&\mbox{max}\{\sqrt{x_1x_2}m_B,\sqrt{|B^2|},1/b_1\},\\
 t_e&=&\mbox{max}\{\sqrt {1+x_1}m_{B},1/b_1\},\;\;\;
 t_e^\prime=\mbox{max}\{\sqrt {x_1}m_{B},1/b_1\},\\
 t_f&=&\mbox{max}\{\sqrt {1-x_2}m_{B},1/b_2\},\;\;\;
 t_f^\prime=\mbox{max}\{\sqrt {x_2}m_{B},1/b_2\}.
\end{eqnarray}

The functions $h_i$ in decay amplitudes are from the propagators of
virtual quark and gluon and are defined by:
\begin{eqnarray}
h_e(A,B,b_1,b_2)&=&\Big[\theta(A)K_0(\sqrt
{A}m_Bb_1)+\theta(-A)i\frac{\pi}{2}H_0(\sqrt
{-A}m_{B}b_1)\Big]\nonumber\\
&&\times \bigg\{\theta(b_1-b_2)\Big[\theta(B)K_0(\sqrt B
m_{B}b_1)I_0(\sqrt Bm_{B}b_2)\nonumber\\
&&\;\;\;+\theta(-B)i\frac{\pi}{2}H_0^{(1)}(\sqrt {-B}
m_{B}b_1)J_0(\sqrt {-B}m_{B}b_2)\Big]+ (b_1\leftrightarrow
b_2)\bigg\},
\end{eqnarray}
where $H_0^{(1)}(z) = \mathrm{J}_0(z) + i\, \mathrm{Y}_0(z)$.

The Sudakov factor from threshold resummation is universal,
independent of flavors of internal quarks, twists,  and the specific
processes. To simplify the analysis, the following parametrization
has been used  \cite{Li:2002mi}:
\begin{eqnarray}
S_t(x)=\frac{2^{1+2c}\Gamma(3/2+c)}{\sqrt{\pi}\Gamma(1+c)}
[x(1-x)]^c\;, \label{str}
\end{eqnarray}
with $c=0.4$. This parametrization, symmetric under the
interchange of $x$ and $1-x$, is convenient for evaluation of the
amplitudes. It is obvious that the threshold resummation modifies
the end-point behavior of the meson distribution amplitudes,
rendering them vanish faster at $x\to 0$.

The evolution factors $E^{(\prime)}_e$ and $E^{(\prime)}_a$ are
given by
\begin{eqnarray}
E_e(t)&=&\alpha_s(t) S_t(x_2)\exp[-S_B(t)-S_2(t)],
 \ \ \ \
 E'_e(t)=\alpha_s(t)S_t(x_1)
 \exp[-S_B(t)-S_2(t)],\\
E_a(t)&=&S_t(x_1)
 \exp[-S_B(t)],\
 \ \ \
E'_a(t)= S_t(x_2)\exp[-S_2(t)],
\end{eqnarray}
in which the Sudakov exponents are defined as
\begin{eqnarray}
S_B(t)&=&s\left(x_1\frac{m_{B}}{\sqrt
2},b_1\right)+\frac{5}{3}\int^t_{1/b_1}\frac{d\bar \mu}{\bar
\mu}\gamma_q(\alpha_s(\bar \mu)),\\
S_2(t)&=&s\left(x_2\frac{m_{B}}{\sqrt
2},b_2\right)+s\left((1-x_2)\frac{m_{B}}{\sqrt
2},b_2\right)+2\int^t_{1/b_2}\frac{d\bar \mu}{\bar
\mu}\gamma_q(\alpha_s(\bar \mu)),
\end{eqnarray}
with the quark anomalous dimension $\gamma_q=-\alpha_s/\pi$. The
explicit form for the  function $s(Q,b)$ is:
\begin{eqnarray}
s(Q,b)&=&~~\frac{A^{(1)}}{2\beta_{1}}\hat{q}\ln\left(\frac{\hat{q}}
{\hat{b}}\right)-
\frac{A^{(1)}}{2\beta_{1}}\left(\hat{q}-\hat{b}\right)+
\frac{A^{(2)}}{4\beta_{1}^{2}}\left(\frac{\hat{q}}{\hat{b}}-1\right)
-\left[\frac{A^{(2)}}{4\beta_{1}^{2}}-\frac{A^{(1)}}{4\beta_{1}}
\ln\left(\frac{e^{2\gamma_E-1}}{2}\right)\right]
\ln\left(\frac{\hat{q}}{\hat{b}}\right)
\nonumber \\
&&+\frac{A^{(1)}\beta_{2}}{4\beta_{1}^{3}}\hat{q}\left[
\frac{\ln(2\hat{q})+1}{\hat{q}}-\frac{\ln(2\hat{b})+1}{\hat{b}}\right]
+\frac{A^{(1)}\beta_{2}}{8\beta_{1}^{3}}\left[
\ln^{2}(2\hat{q})-\ln^{2}(2\hat{b})\right],
\end{eqnarray} where the variables are defined by
\begin{eqnarray}
\hat q\equiv \mbox{ln}[Q/(\sqrt 2\Lambda)],~~~ \hat b\equiv
\mbox{ln}[1/(b\Lambda)], \end{eqnarray} and the coefficients
$A^{(i)}$ and $\beta_i$ are \begin{eqnarray}
\beta_1=\frac{33-2n_f}{12},~~\beta_2=\frac{153-19n_f}{24},\nonumber\\
A^{(1)}=\frac{4}{3},~~A^{(2)}=\frac{67}{9}
-\frac{\pi^2}{3}-\frac{10}{27}n_f+\frac{8}{3}\beta_1\mbox{ln}(\frac{1}{2}e^{\gamma_E}),
\end{eqnarray}
$n_f$ is the number of the quark flavors and $\gamma_E$ is the Euler
constant. We will use the one-loop running coupling constant, i.e.
we pick up only the four terms in the first line of the expression
for the function $s(Q,b)$.

\section{Analytic Formulae for the $B \to V\gamma$ decay
amplitudes} \label{section:BtoVgammaformulae}

 The analytic formulae for $B^-\to\rho^-\gamma$ is:
\begin{eqnarray}
 {\cal A}^i(B^-\to\rho^{-}\gamma)
&=&  \frac{G_F}{\sqrt{2}}  V_{ub}V_{ud}^{*} \Big\{ {\cal
M}_{1u}^{i(a)} +{\cal M}_{1u}^{i(b)}(Q_u)+{\cal M}_{2u}^{i}+ {\cal
M}_{ann}^{i(a,LL)}(a_1,Q_{u})+ {\cal
M}_{ann}^{i(b,LL)}(a_1,Q_{d},Q_{u})\Big \}\nonumber
   \\
  &&+\frac{G_F}{\sqrt{2}}  V_{cb}V_{cd}^{*} \Big\{ {\cal
M}_{1c}^{i(a)} +{\cal M}_{1c}^{i(b)}(Q_u)+{\cal M}_{2c}^{i}
 \Big\}\nonumber
   \\
  &&- \frac{G_F}{\sqrt{2}} V_{tb}V_{td}^{*} \Big\{{\cal
  M}_{7\gamma}^i+{\cal  M}_{8g}^{i(a)}+{\cal   M}_{8g}^{i(b)}(Q_u)+ {\cal
M}_{ann}^{i(a,LL)}(a_4+a_{10},Q_{u})\nonumber
   \\
  &&\;\;\;+ {\cal
M}_{ann}^{i(b,LL)}(a_4+a_{10},Q_{d},Q_{u})+ {\cal
M}_{ann}^{i(SP)}(a_6+a_8,Q_{d},Q_u) \Big \},
\end{eqnarray}
while the expression for $B^-\to K^{*-}\gamma$ is basically the same
except with the only difference in the CKM matrix elements:
$V_{qd}\to V_{qs}$. The formulas for other channels are
\begin{eqnarray}
 \sqrt 2{\cal A}^i(\bar B^0\to\rho^{0}\gamma)
&=&  \frac{G_F}{\sqrt{2}}  V_{ub}V_{ud}^{*} \Big\{ {\cal
M}_{ann}^{i(a,LL)}(a_2,Q_{d})+{\cal
M}_{ann}^{i(b,LL)}(a_2,Q_u,Q_{u})-{\cal M}_{1u}^{i(a)} -{\cal
M}_{1u}^{i(b)}(Q_d)-{\cal M}_{2u}^{i} \Big \}\nonumber
   \\
  &&+\frac{G_F}{\sqrt{2}}  V_{cb}V_{cd}^{*} \Big\{ -{\cal
M}_{1c}^{i(a)} -{\cal M}_{1c}^{i(b)}(Q_d)-{\cal M}_{2c}^{i}
 \Big\}- \frac{G_F}{\sqrt{2}} V_{tb}V_{td}^{*} \Big\{-{\cal
  M}_{7\gamma}^i-{\cal  M}_{8g}^{i(a)}\nonumber
   \\
  &&-{\cal   M}_{8g}^{i(b)}(Q_d)+ {\cal
M}_{ann}^{i(a,LL)}(-a_4+\frac{3}{2}a_7+\frac{3}{2}a_9+\frac{1}{2}a_{10},Q_{d})\nonumber
   \\
  &&+ {\cal
M}_{ann}^{i(b,LL)}(a_3+a_9,Q_{u},Q_{u})+ {\cal
M}_{ann}^{i(b,LR)}(a_5+a_7,Q_{u},Q_{u})\nonumber\\
&&+ {\cal
M}_{ann}^{i(b,LL)}(-a_3-a_4+\frac{1}{2}a_9+\frac{1}{2}a_{10},Q_{d},Q_{d})+
{\cal M}_{ann}^{i(b,LR)}(-a_5+\frac{1}{2}a_7,Q_{d},Q_{d})\nonumber
   \\
  &&+ {\cal
M}_{ann}^{i(SP)}(-a_6+\frac{1}{2}a_8,Q_{d},Q_d)+ {\cal
M}_{en}^{i}(Q_u-Q_d) \Big \},
\end{eqnarray}
\begin{eqnarray}
 \sqrt 2{\cal A}^i(\bar B^0\to\omega\gamma)
&=&  \frac{G_F}{\sqrt{2}}  V_{ub}V_{ud}^{*} \Big\{ {\cal
M}_{ann}^{i(a,LL)}(a_2,Q_{d})+{\cal
M}_{ann}^{i(b,LL)}(a_2,Q_u,Q_{u})+{\cal M}_{1u}^{i(a)} +{\cal
M}_{1u}^{i(b)}(Q_d)+{\cal M}_{2u}^{i} \Big \}\nonumber
   \\
  &&+\frac{G_F}{\sqrt{2}}  V_{cb}V_{cd}^{*} \Big\{ {\cal
M}_{1c}^{i(a)} +{\cal M}_{1c}^{i(b)}(Q_d)+{\cal M}_{2c}^{i}
 \Big\}- \frac{G_F}{\sqrt{2}} V_{tb}V_{td}^{*} \Big\{{\cal
  M}_{7\gamma}^i+{\cal  M}_{8g}^{i(a)}\nonumber
   \\
  &&+{\cal   M}_{8g}^{i(b)}(Q_d)+ {\cal
M}_{ann}^{i(a,LL)}(2a_3+a_4+2a_5+\frac{1}{2}a_7+\frac{1}{2}a_9-\frac{1}{2}a_{10},Q_{d})\nonumber
   \\
  &&+ {\cal
M}_{ann}^{i(b,LL)}(a_3+a_9,Q_{u},Q_{u})+ {\cal
M}_{ann}^{i(b,LR)}(a_5+a_7,Q_{u},Q_{u})\nonumber\\
&&+ {\cal
M}_{ann}^{i(b,LL)}(a_3+a_4-\frac{1}{2}a_9-\frac{1}{2}a_{10},Q_{d},Q_{d})+
{\cal M}_{ann}^{i(b,LR)}(a_5-\frac{1}{2}a_7,Q_{d},Q_{d})\nonumber
   \\
  &&
+ {\cal M}_{ann}^{i(SP)}(a_6-\frac{1}{2}a_8,Q_{d},Q_d) + {\cal
M}_{en}^{i}(Q_u+Q_d)\Big \},
\end{eqnarray}
\begin{eqnarray}
 {\cal A}^i(\bar B^0\to\bar K^{*0}\gamma)
&=&  \frac{G_F}{\sqrt{2}}  V_{ub}V_{us}^{*} \Big\{ {\cal
M}_{1u}^{i(a)} +{\cal M}_{1u}^{i(b)}(Q_d)+{\cal M}_{2u}^{i}\Big
\}\nonumber
   \\
  &&+\frac{G_F}{\sqrt{2}}  V_{cb}V_{cs}^{*} \Big\{ {\cal
M}_{1c}^{i(a)} +{\cal M}_{1c}^{i(b)}(Q_d)+{\cal M}_{2c}^{i}
 \Big\}\nonumber
   \\
  &&- \frac{G_F}{\sqrt{2}} V_{tb}V_{ts}^{*} \Big\{{\cal
  M}_{7\gamma}^i+{\cal  M}_{8g}^{i(a)}+{\cal   M}_{8g}^{i(b)}(Q_d)+ {\cal
M}_{ann}^{i(a,LL)}(a_4-\frac{1}{2}a_{10},Q_{d})\nonumber
   \\
  &&+ {\cal
M}_{ann}^{i(b,LL)}(a_4-\frac{1}{2}a_{10},Q_{s},Q_{d})+ {\cal
M}_{ann}^{i(SP)}(a_6-\frac{1}{2}a_8,Q_{s},Q_d) \Big \}.
\end{eqnarray}
The expression for $\bar B_s\to K^{*0}\gamma$ can be obtained by
replacing $V_{qs}$ by $V_{qd}$ from $\bar B^0\to\bar K^{*0}\gamma$.

The formulas for the $B_s \to \phi \gamma$ decay are
\begin{eqnarray}
 {\cal A}^i(\bar B_s\to\phi\gamma)
&=&  \frac{G_F}{\sqrt{2}}  V_{ub}V_{us}^{*} \Big\{ {\cal
M}_{1u}^{i(a)} +{\cal M}_{1u}^{i(b)}(Q_s)+{\cal M}_{2u}^{i}\Big
\}\nonumber
   \\
  &&+\frac{G_F}{\sqrt{2}}  V_{cb}V_{cs}^{*} \Big\{ {\cal
M}_{1c}^{i(a)} +{\cal M}_{1c}^{i(b)}(Q_s)+{\cal M}_{2c}^{i}
 \Big\}\nonumber
   \\
  &&- \frac{G_F}{\sqrt{2}} V_{tb}V_{ts}^{*} \Big\{{\cal
  M}_{7\gamma}^i+{\cal  M}_{8g}^{i(a)}+{\cal
  M}_{8g}^{i(b)}(Q_s)\nonumber
   \\
  &&+ {\cal
M}_{ann}^{i(a,LL)}(a_3+a_4+a_5-\frac{1}{2}a_7-\frac{1}{2}a_9-\frac{1}{2}a_{10},Q_{s})\nonumber\\
  &&+ {\cal
M}_{ann}^{i(b,LL)}(a_3+a_4-\frac{1}{2}a_9-\frac{1}{2}a_{10},Q_{s},Q_{s})+
{\cal M}_{ann}^{i(b,LR)}(a_5-\frac{1}{2}a_7,Q_{s},Q_{s})\nonumber
   \\&&+
{\cal M}_{ann}^{i(SP)}(a_6-\frac{1}{2}a_8,Q_{s},Q_s)+ {\cal
M}_{en}^{i}(Q_s) \Big \}.
\end{eqnarray}
For the annihilation type decays, we have
\begin{eqnarray}
 {\cal A}^i(\bar B^0\to\phi\gamma)
&=& - \frac{G_F}{\sqrt{2}} V_{tb}V_{td}^{*} \Big\{{\cal
M}_{ann}^{i(a,LL)}(a_3+a_5-\frac{1}{2}a_7-\frac{1}{2}a_9,Q_{s})  +
{\cal
M}_{en}^{i}(Q_s)\nonumber\\
&&+ {\cal M}_{ann}^{i(b,LL)}(a_3-\frac{1}{2}a_9,Q_{s},Q_{s})+
{\cal M}_{ann}^{i(b,LR)}(a_5-\frac{1}{2}a_7,Q_{s},Q_{s})\Big \},
\end{eqnarray}
\begin{eqnarray}
 \sqrt 2{\cal A}^i(\bar B_s\to\rho^0\gamma)
&=&  \frac{G_F}{\sqrt{2}}  V_{ub}V_{us}^{*} \Big\{{\cal
M}_{ann}^{i(a,LL)}(a_2,Q_{s})+{\cal
M}_{ann}^{i(b,LL)}(a_2,Q_{u},Q_{u})\Big \}\nonumber
   \\
  &&- \frac{G_F}{\sqrt{2}} V_{tb}V_{ts}^{*} \Big\{{\cal
M}_{ann}^{i(a,LL)}(\frac{3}{2}a_7+\frac{3}{2}a_9,Q_{s})+ {\cal
M}_{en}^{i}(Q_u-Q_d)\nonumber\\
  &&+ {\cal
M}_{ann}^{i(b,LL)}(a_3+a_9,Q_{u},Q_{u})+ {\cal
M}_{ann}^{i(b,LR)}(a_5+a_7,Q_{u},Q_{u})\nonumber\\
&&+ {\cal M}_{ann}^{i(b,LL)}(-a_3+\frac{1}{2}a_9,Q_{d},Q_{d})+
{\cal M}_{ann}^{i(b,LR)}(-a_5+\frac{1}{2}a_7,Q_{d},Q_{d})\Big \},
\end{eqnarray}
\begin{eqnarray}
 \sqrt 2{\cal A}^i(\bar B_s\to\omega\gamma)
&=&  \frac{G_F}{\sqrt{2}}  V_{ub}V_{us}^{*} \Big\{{\cal
M}_{ann}^{i(a,LL)}(a_2,Q_{s})+{\cal
M}_{ann}^{i(b,LL)}(a_2,Q_{u},Q_{u})\Big \}\nonumber
   \\
  &&- \frac{G_F}{\sqrt{2}} V_{tb}V_{ts}^{*} \Big\{{\cal
M}_{ann}^{i(a,LL)}(2a_3+2a_5+\frac{1}{2}a_7+\frac{1}{2}a_9,Q_{s})+
{\cal
M}_{en}^{i}(Q_u+Q_d) \nonumber\\
  &&+ {\cal
M}_{ann}^{i(b,LL)}(a_3+a_9,Q_{u},Q_{u})+ {\cal
M}_{ann}^{i(b,LR)}(a_5+a_7,Q_{u},Q_{u})\nonumber\\
&&+ {\cal M}_{ann}^{i(b,LL)}(a_3-\frac{1}{2}a_9,Q_{d},Q_{d})+ {\cal
M}_{ann}^{i(b,LR)}(a_5-\frac{1}{2}a_7,Q_{d},Q_{d})\Big \}.
\end{eqnarray}

\section{Analytic Formulae for the $B\to A \gamma$  decay
amplitudes}\label{sec:Agamma}

The expression for $B\to ^1P_1 \gamma$ is different from $B\to
V\gamma$ in various aspects. The first two annihilation diagrams
vanish for neutral axial vector mesons (not including $K_{1B}$).
There is not any two-photon diagram contribution in $B\to A\gamma$
decays. The explicit formula for the $B^-\to b_1^{-}(1235)\gamma$
decay amplitude is
\begin{eqnarray}
 {\cal A}^i(B^-\to b_1^{-}(1235)\gamma)
&=&  \frac{G_F}{\sqrt{2}}  V_{ub}V_{ud}^{*} \Big\{ {\cal
M}_{1u}^{i(a)} +{\cal M}_{1u}^{i(b)}(Q_u)+{\cal M}_{2u}^{i}+ {\cal
M}_{ann}^{i(b,LL)}(a_1,Q_{d},Q_{u})\Big \}\nonumber
   \\
  &&+\frac{G_F}{\sqrt{2}}  V_{cb}V_{cd}^{*} \Big\{ {\cal
M}_{1c}^{i(a)} +{\cal M}_{1c}^{i(b)}(Q_u)+{\cal M}_{2c}^{i}
 \Big\}\nonumber
   \\
  &&- \frac{G_F}{\sqrt{2}} V_{tb}V_{td}^{*} \Big\{{\cal
  M}_{7\gamma}^i+{\cal  M}_{8g}^{i(a)}+{\cal   M}_{8g}^{i(b)}(Q_u)\nonumber
   \\
  &&\;\;\;+ {\cal
M}_{ann}^{i(b,LL)}(a_4+a_{10},Q_{d},Q_{u})+ {\cal
M}_{ann}^{i(SP)}(a_6+a_8,Q_{d},Q_u) \Big \},
\end{eqnarray}
while the expression for U-spin related process $B^-\to
K^{-}_{1B}\gamma$ is basically the same except with the only
difference in the CKM¡¡matrix elements: $V_{qd}\to V_{qs}$. The
formulas for the neutral decays modes are
\begin{eqnarray}
 \sqrt 2{\cal A}^i(\bar B^0\to b_1^{0}(1235)\gamma)
&=&  \frac{G_F}{\sqrt{2}}  V_{ub}V_{ud}^{*} \Big\{{\cal
M}_{ann}^{i(b,LL)}(a_2,Q_u,Q_{u})-{\cal M}_{1u}^{i(a)} -{\cal
M}_{1u}^{i(b)}(Q_d)-{\cal M}_{2u}^{i} \Big \}\nonumber
   \\
  &&+\frac{G_F}{\sqrt{2}}  V_{cb}V_{cd}^{*} \Big\{ -{\cal
M}_{1c}^{i(a)} -{\cal M}_{1c}^{i(b)}(Q_d)-{\cal M}_{2c}^{i}
 \Big\}- \frac{G_F}{\sqrt{2}} V_{tb}V_{td}^{*} \Big\{-{\cal
  M}_{7\gamma}^i\nonumber
   \\
  &&-{\cal   M}_{8g}^{i(b)}(Q_d)+ {\cal
M}_{ann}^{i(b,LL)}(a_3+a_9,Q_{u},Q_{u})+ {\cal
M}_{ann}^{i(b,LR)}(a_5+a_7,Q_{u},Q_{u})\nonumber\\
&&+ {\cal
M}_{ann}^{i(b,LL)}(-a_3-a_4+\frac{1}{2}a_9+\frac{1}{2}a_{10},Q_{d},Q_{d})-{\cal
M}_{8g}^{i(a)}\nonumber
   \\
  &&+
{\cal M}_{ann}^{i(b,LR)}(-a_5+\frac{1}{2}a_7,Q_{d},Q_{d})+ {\cal
M}_{ann}^{i(SP)}(-a_6+\frac{1}{2}a_8,Q_{d},Q_d)\Big \},
\end{eqnarray}
\begin{eqnarray}
 {\cal A}^i(\bar B^0\to\bar K^{0}_{1B}\gamma)
&=&  \frac{G_F}{\sqrt{2}}  V_{ub}V_{us}^{*} \Big\{ {\cal
M}_{1u}^{i(a)} +{\cal M}_{1u}^{i(b)}(Q_d)+{\cal M}_{2u}^{i}\Big
\}\nonumber
   \\
  &&+\frac{G_F}{\sqrt{2}}  V_{cb}V_{cs}^{*} \Big\{ {\cal
M}_{1c}^{i(a)} +{\cal M}_{1c}^{i(b)}(Q_d)+{\cal M}_{2c}^{i}
 \Big\}\nonumber
   \\
  &&- \frac{G_F}{\sqrt{2}} V_{tb}V_{ts}^{*} \Big\{{\cal
  M}_{7\gamma}^i+{\cal  M}_{8g}^{i(a)}+{\cal   M}_{8g}^{i(b)}(Q_d)+ {\cal
M}_{ann}^{i(a,LL)}(a_4-\frac{1}{2}a_{10},Q_{d})\nonumber
   \\
  &&+ {\cal
M}_{ann}^{i(b,LL)}(a_4-\frac{1}{2}a_{10},Q_{s},Q_{d})+ {\cal
M}_{ann}^{i(SP)}(a_6-\frac{1}{2}a_8,Q_{s},Q_d) \Big \},
\end{eqnarray}
and the expression for $\bar B_s\to K^{0}_{1B}\gamma$ can be
obtained by replacing $V_{qs}$ by $V_{qd}$.

The decay amplitudes involving $h_1(1170)$ and $h_1(1380)$ can
obtained as
\begin{eqnarray}
 A^i(B\to h_1(1170)\gamma)&=&A^i(B\to h_1\gamma) {\rm cos}\theta
_{^1P_1}+A^i(B\to h_8\gamma) {\rm sin}\theta _{^1P_1},\\
 A^i(B\to h_1(1380)\gamma)&=&-A^i(B\to h_1\gamma) {\rm sin}\theta
_{^1P_1}+A^i(B\to h_8\gamma) {\rm cos}\theta _{^1P_1},
\end{eqnarray}
where $B$ denotes $\bar B^0$ or $\bar B_s^0$ with
\begin{eqnarray}
 \sqrt 2{\cal A}^i(\bar B^0\to h_8\gamma)
&=&  \frac{G_F}{\sqrt{2}}  V_{ub}V_{ud}^{*} \frac{1}{\sqrt 6}\Big\{
{\cal M}_{ann}^{i(b,LL)}(a_2,Q_u,Q_{u})+{\cal M}_{1u}^{i(a)} +{\cal
M}_{1u}^{i(b)}(Q_d)+{\cal M}_{2u}^{i} \Big \}\nonumber
   \\
  &&+\frac{G_F}{\sqrt{2}}  V_{cb}V_{cd}^{*} \frac{1}{\sqrt 6}\Big\{ {\cal
M}_{1c}^{i(a)} +{\cal M}_{1c}^{i(b)}(Q_d)+{\cal M}_{2c}^{i}
 \Big\}\nonumber
   \\
  &&- \frac{G_F}{\sqrt{2}} V_{tb}V_{td}^{*}\frac{1}{\sqrt 6}\Big\{{\cal
  M}_{7\gamma}^i+{\cal  M}_{8g}^{i(a)}
+ {\cal M}_{ann}^{i(SP)}(a_6-\frac{1}{2}a_8,Q_{d},Q_d)\Big
\}\nonumber
   \\
  &&+{\cal   M}_{8g}^{i(b)}(Q_d)+ {\cal
M}_{ann}^{i(b,LL)}(a_3+a_9,Q_{u},Q_{u})+ {\cal
M}_{ann}^{i(b,LR)}(a_5+a_7,Q_{u},Q_{u})\nonumber\\
&&+ {\cal
M}_{ann}^{i(b,LL)}(a_3+a_4-\frac{1}{2}a_9-\frac{1}{2}a_{10},Q_{d},Q_{d})+
{\cal M}_{ann}^{i(b,LR)}(a_5-\frac{1}{2}a_7,Q_{d},Q_{d})\nonumber\\
&& -2{\cal M}_{ann}^{i(b,LL)}(a_3-\frac{1}{2}a_9,Q_{s},Q_{s})-2
{\cal M}_{ann}^{i(b,LR)}(a_5-\frac{1}{2}a_7,Q_{s},Q_{s})\Big \},
\end{eqnarray}
\begin{eqnarray}
 \sqrt 2{\cal A}^i(\bar B^0\to h_1\gamma)
&=&  \frac{G_F}{\sqrt{2}}  V_{ub}V_{ud}^{*} \frac{1}{\sqrt 3}\Big\{
{\cal M}_{ann}^{i(b,LL)}(a_2,Q_u,Q_{u})+{\cal M}_{1u}^{i(a)} +{\cal
M}_{1u}^{i(b)}(Q_d)+{\cal M}_{2u}^{i} \Big \}\nonumber
   \\
  &&+\frac{G_F}{\sqrt{2}}  V_{cb}V_{cd}^{*} \frac{1}{\sqrt 3}\Big\{ {\cal
M}_{1c}^{i(a)} +{\cal M}_{1c}^{i(b)}(Q_d)+{\cal M}_{2c}^{i}
 \Big\}\nonumber
   \\
  &&- \frac{G_F}{\sqrt{2}} V_{tb}V_{td}^{*}\frac{1}{\sqrt 3}\Big\{{\cal
  M}_{7\gamma}^i+{\cal  M}_{8g}^{i(a)}
+ {\cal M}_{ann}^{i(SP)}(a_6-\frac{1}{2}a_8,Q_{d},Q_d)\Big
\}\nonumber
   \\
  &&+{\cal   M}_{8g}^{i(b)}(Q_d)+ {\cal
M}_{ann}^{i(b,LL)}(a_3+a_9,Q_{u},Q_{u})+ {\cal
M}_{ann}^{i(b,LR)}(a_5+a_7,Q_{u},Q_{u})
  \nonumber\\
  &&+ {\cal
M}_{ann}^{i(b,LL)}(a_3+a_4-\frac{1}{2}a_9-\frac{1}{2}a_{10},Q_{d},Q_{d})+
{\cal M}_{ann}^{i(b,LR)}(a_5-\frac{1}{2}a_7,Q_{d},Q_{d})\nonumber\\
&& +{\cal M}_{ann}^{i(b,LL)}(a_3-\frac{1}{2}a_9,Q_{s},Q_{s})+ {\cal
M}_{ann}^{i(b,LR)}(a_5-\frac{1}{2}a_7,Q_{s},Q_{s})\Big \},
\end{eqnarray}
\begin{eqnarray}
 {\cal A}^i(\bar B_s\to h_8\gamma)
&=&  \frac{G_F}{\sqrt{2}}  V_{ub}V_{us}^{*}  \frac{-2}{\sqrt
6}\Big\{ {\cal M}_{1u}^{i(a)} +{\cal M}_{1u}^{i(b)}(Q_s)+{\cal
M}_{2u}^{i}\Big \}\nonumber
   \\
  &&+\frac{G_F}{\sqrt{2}}  V_{cb}V_{cs}^{*}  \frac{-2}{\sqrt
6}\Big\{ {\cal M}_{1c}^{i(a)} +{\cal M}_{1c}^{i(b)}(Q_s)+{\cal
M}_{2c}^{i}
 \Big\}\nonumber
   \\
  &&- \frac{G_F}{\sqrt{2}} V_{tb}V_{ts}^{*}  \frac{-2}{\sqrt
6}\Big\{{\cal
  M}_{7\gamma}^i+{\cal  M}_{8g}^{i(a)}+{\cal
  M}_{8g}^{i(b)}(Q_s)+
{\cal M}_{ann}^{i(SP)}(a_6-\frac{1}{2}a_8,Q_{s},Q_s)\nonumber
   \\
  &&+ {\cal
M}_{ann}^{i(b,LL)}(a_3+a_4-\frac{1}{2}a_9-\frac{1}{2}a_{10},Q_{s},Q_{s})+
{\cal M}_{ann}^{i(b,LR)}(a_5-\frac{1}{2}a_7,Q_{s},Q_{s})\Big
\},\nonumber\\
 &&+\frac{G_F}{\sqrt{2}}  V_{ub}V_{us}^{*} \frac{1}{\sqrt
6}\Big\{{\cal M}_{ann}^{i(b,LL)}(a_2,Q_{u},Q_{u})\Big \}\nonumber
   \\
  &&- \frac{G_F}{\sqrt{2}} V_{tb}V_{ts}^{*} \frac{1}{\sqrt
6}\Big\{{\cal M}_{ann}^{i(b,LL)}(a_3+a_9,Q_{u},Q_{u})+ {\cal
M}_{ann}^{i(b,LR)}(a_5+a_7,Q_{u},Q_{u})\nonumber\\
&&+ {\cal M}_{ann}^{i(b,LL)}(a_3-\frac{1}{2}a_9,Q_{d},Q_{d})+ {\cal
M}_{ann}^{i(b,LR)}(a_5-\frac{1}{2}a_7,Q_{d},Q_{d})\Big \},
\end{eqnarray}
\begin{eqnarray}
 {\cal A}^i(\bar B_s\to h_1\gamma)
&=&  \frac{G_F}{\sqrt{2}}  V_{ub}V_{us}^{*}  \frac{1}{\sqrt 3}\Big\{
{\cal M}_{1u}^{i(a)} +{\cal M}_{1u}^{i(b)}(Q_s)+{\cal
M}_{2u}^{i}\Big \}\nonumber
   \\
  &&+\frac{G_F}{\sqrt{2}}  V_{cb}V_{cs}^{*}  \frac{1}{\sqrt
3}\Big\{ {\cal M}_{1c}^{i(a)} +{\cal M}_{1c}^{i(b)}(Q_s)+{\cal
M}_{2c}^{i}
 \Big\}\nonumber
   \\
  &&- \frac{G_F}{\sqrt{2}} V_{tb}V_{ts}^{*}  \frac{1}{\sqrt
3}\Big\{{\cal
  M}_{7\gamma}^i+{\cal  M}_{8g}^{i(a)}+{\cal
  M}_{8g}^{i(b)}(Q_s)+
{\cal M}_{ann}^{i(SP)}(a_6-\frac{1}{2}a_8,Q_{s},Q_s)\nonumber
   \\
  &&+ {\cal
M}_{ann}^{i(b,LL)}(a_3+a_4-\frac{1}{2}a_9-\frac{1}{2}a_{10},Q_{s},Q_{s})+
{\cal M}_{ann}^{i(b,LR)}(a_5-\frac{1}{2}a_7,Q_{s},Q_{s})\Big
\},\nonumber\\
 &&+\frac{G_F}{\sqrt{2}}  V_{ub}V_{us}^{*} \frac{1}{\sqrt
3}\Big\{{\cal M}_{ann}^{i(b,LL)}(a_2,Q_{u},Q_{u})\Big \}\nonumber
   \\
  &&- \frac{G_F}{\sqrt{2}} V_{tb}V_{ts}^{*} \frac{1}{\sqrt
3}\Big\{{\cal M}_{ann}^{i(b,LL)}(a_3+a_9,Q_{u},Q_{u})+ {\cal
M}_{ann}^{i(b,LR)}(a_5+a_7,Q_{u},Q_{u})\nonumber\\
&&+ {\cal M}_{ann}^{i(b,LL)}(a_3-\frac{1}{2}a_9,Q_{d},Q_{d})+ {\cal
M}_{ann}^{i(b,LR)}(a_5-\frac{1}{2}a_7,Q_{d},Q_{d})\Big \}.
\end{eqnarray}

The annihilation type decay amplitude is:
\begin{eqnarray}
 \sqrt 2{\cal A}^i(\bar B_s\to b_1^0(1235)\gamma)
&=&  \frac{G_F}{\sqrt{2}}  V_{ub}V_{us}^{*} \Big\{{\cal
M}_{ann}^{i(b,LL)}(a_2,Q_{u},Q_{u})\Big \}\nonumber
   \\
  &&- \frac{G_F}{\sqrt{2}} V_{tb}V_{ts}^{*} \Big\{{\cal
M}_{ann}^{i(b,LL)}(a_3+a_9,Q_{u},Q_{u})+ {\cal
M}_{ann}^{i(b,LR)}(a_5+a_7,Q_{u},Q_{u})\nonumber\\
&&+ {\cal M}_{ann}^{i(b,LL)}(-a_3+\frac{1}{2}a_9,Q_{d},Q_{d})+ {\cal
M}_{ann}^{i(b,LR)}(-a_5+\frac{1}{2}a_7,Q_{d},Q_{d})\Big \}.
\end{eqnarray}

In the following, we will give the analytic factorization formulae
for $B\to ^3P_1\gamma$ which is similar with $B\to V\gamma$ except
for some differences in the flavor structure and zero contribution
from the two-photon diagrams,
\begin{eqnarray}
 {\cal A}^i(B^-\to a_1^{-}(1235)\gamma)
&=&  \frac{G_F}{\sqrt{2}}  V_{ub}V_{ud}^{*} \Big\{ {\cal
M}_{1u}^{i(a)} +{\cal M}_{1u}^{i(b)}(Q_u)+{\cal M}_{2u}^{i}+ {\cal
M}_{ann}^{i(a,LL)}(a_1,Q_{u})\nonumber
   \\
  &&+ {\cal
M}_{ann}^{i(b,LL)}(a_1,Q_{d},Q_{u})\Big \}+\frac{G_F}{\sqrt{2}}
V_{cb}V_{cd}^{*} \Big\{ {\cal M}_{1c}^{i(a)} +{\cal
M}_{1c}^{i(b)}(Q_u)+{\cal M}_{2c}^{i}
 \Big\}\nonumber
   \\
  &&- \frac{G_F}{\sqrt{2}} V_{tb}V_{td}^{*} \Big\{{\cal
  M}_{7\gamma}^i+{\cal  M}_{8g}^{i(a)}+{\cal   M}_{8g}^{i(b)}(Q_u)+ {\cal
M}_{ann}^{i(a,LL)}(a_4+a_{10},Q_{u})\nonumber
   \\
  &&\;\;\;+ {\cal
M}_{ann}^{i(b,LL)}(a_4+a_{10},Q_{d},Q_{u})+ {\cal
M}_{ann}^{i(SP)}(a_6+a_8,Q_{d},Q_u) \Big \},
\end{eqnarray}
while the expression for $B^-\to K^{-}_{1A}\gamma$ is   the same
except with the only difference in the CKM matrix elements:
$V_{qd}\to V_{qs}$. The formulas for other channels are
\begin{eqnarray}
 \sqrt 2{\cal A}^i(\bar B^0\to a_1^{0}(1235)\gamma)
&=&  \frac{G_F}{\sqrt{2}}  V_{ub}V_{ud}^{*} \Big\{ {\cal
M}_{ann}^{i(a,LL)}(a_2,Q_{d})+{\cal
M}_{ann}^{i(b,LL)}(a_2,Q_u,Q_{u})-{\cal M}_{1u}^{i(a)}\nonumber
   \\
  && -{\cal
M}_{1u}^{i(b)}(Q_d)-{\cal M}_{2u}^{i} \Big \}+\frac{G_F}{\sqrt{2}}
V_{cb}V_{cd}^{*} \Big\{ -{\cal M}_{1c}^{i(a)} -{\cal
M}_{1c}^{i(b)}(Q_d)-{\cal M}_{2c}^{i}
 \Big\}\nonumber
   \\
  &&- \frac{G_F}{\sqrt{2}} V_{tb}V_{td}^{*} \Big\{-{\cal
  M}_{7\gamma}^i-{\cal  M}_{8g}^{i(a)}\nonumber
   \\
  &&-{\cal   M}_{8g}^{i(b)}(Q_d)+ {\cal
M}_{ann}^{i(a,LL)}(-a_4+\frac{3}{2}a_7+\frac{3}{2}a_9+\frac{1}{2}a_{10},Q_{d})\nonumber
   \\
  &&+ {\cal
M}_{ann}^{i(b,LL)}(a_3+a_9,Q_{u},Q_{u})+ {\cal
M}_{ann}^{i(b,LR)}(a_5+a_7,Q_{u},Q_{u})\nonumber\\
&&+ {\cal
M}_{ann}^{i(b,LL)}(-a_3-a_4+\frac{1}{2}a_9+\frac{1}{2}a_{10},Q_{d},Q_{d})
   \\
  &&+
{\cal M}_{ann}^{i(b,LR)}(-a_5+\frac{1}{2}a_7,Q_{d},Q_{d})+ {\cal
M}_{ann}^{i(SP)}(-a_6+\frac{1}{2}a_8,Q_{d},Q_d) \Big \},\nonumber
\end{eqnarray}
\begin{eqnarray}
 {\cal A}^i(\bar B^0\to\bar K^{0}_{1A}\gamma)
&=&  \frac{G_F}{\sqrt{2}}  V_{ub}V_{us}^{*} \Big\{ {\cal
M}_{1u}^{i(a)} +{\cal M}_{1u}^{i(b)}(Q_d)+{\cal M}_{2u}^{i}\Big
\}\nonumber
   \\
  &&+\frac{G_F}{\sqrt{2}}  V_{cb}V_{cs}^{*} \Big\{ {\cal
M}_{1c}^{i(a)} +{\cal M}_{1c}^{i(b)}(Q_d)+{\cal M}_{2c}^{i}
 \Big\}\nonumber
   \\
  &&- \frac{G_F}{\sqrt{2}} V_{tb}V_{ts}^{*} \Big\{{\cal
  M}_{7\gamma}^i+{\cal  M}_{8g}^{i(a)}+{\cal   M}_{8g}^{i(b)}(Q_d)+ {\cal
M}_{ann}^{i(a,LL)}(a_4-\frac{1}{2}a_{10},Q_{d})\nonumber
   \\
  &&+ {\cal
M}_{ann}^{i(b,LL)}(a_4-\frac{1}{2}a_{10},Q_{s},Q_{d})+ {\cal
M}_{ann}^{i(SP)}(a_6-\frac{1}{2}a_8,Q_{s},Q_d) \Big \},
\end{eqnarray}
and the expression for $\bar B_s\to K^{0}_{1A}\gamma$ can be
obtained by replacing $V_{qs}$ by $V_{qd}$.

The decay amplitudes involving $f_1(1285)$ and $f_1(1420)$ can
obtained as
\begin{eqnarray}
 A^i(B\to f_1(1285)\gamma)&=&A^i(B\to f_1\gamma) {\rm cos}\theta
_{^3P_1}+A^i(B\to f_8\gamma) {\rm sin}\theta _{^3P_1},\\
 A^i(B\to h_1(1420)\gamma)&=&-A^i(B\to f_1\gamma) {\rm sin}\theta
_{^3P_1}+A^i(B\to f_8\gamma) {\rm cos}\theta _{^3P_1},
\end{eqnarray}
where $B$ denotes $\bar B^0$ or $\bar B_s^0$ with
\begin{eqnarray}
 \sqrt 2{\cal A}^i(\bar B^0\to f_8\gamma)
&=&  \frac{G_F}{\sqrt{2}}  V_{ub}V_{ud}^{*} \frac{1}{\sqrt 6}\Big\{
{\cal M}_{ann}^{i(a,LL)}(a_2,Q_{d})+{\cal
M}_{ann}^{i(b,LL)}(a_2,Q_u,Q_{u})+{\cal M}_{1u}^{i(a)}\nonumber
   \\
  && +{\cal
M}_{1u}^{i(b)}(Q_d)+{\cal M}_{2u}^{i} \Big \}+\frac{G_F}{\sqrt{2}}
V_{cb}V_{cd}^{*} \frac{1}{\sqrt 6}\Big\{ {\cal M}_{1c}^{i(a)}+{\cal
M}_{1c}^{i(b)}(Q_d)+{\cal M}_{2c}^{i}
 \Big\}\nonumber
   \\
  && - \frac{G_F}{\sqrt{2}} V_{tb}V_{td}^{*} \frac{1}{\sqrt 6}\Big\{{\cal
  M}_{7\gamma}^i+{\cal  M}_{8g}^{i(a)}+{\cal   M}_{8g}^{i(b)}(Q_d)\nonumber
   \\
  &&+ {\cal
M}_{ann}^{i(a,LL)}(2a_3+a_4+2a_5+\frac{1}{2}a_7+\frac{1}{2}a_9-\frac{1}{2}a_{10},Q_{d})\nonumber
   \\
  &&+ {\cal
M}_{ann}^{i(b,LL)}(a_3+a_9,Q_{u},Q_{u})+ {\cal
M}_{ann}^{i(b,LR)}(a_5+a_7,Q_{u},Q_{u})\nonumber\\
&&+ {\cal
M}_{ann}^{i(b,LL)}(a_3+a_4-\frac{1}{2}a_9-\frac{1}{2}a_{10},Q_{d},Q_{d})+
{\cal M}_{ann}^{i(b,LR)}(a_5-\frac{1}{2}a_7,Q_{d},Q_{d})\nonumber
   \\
  &&
+ {\cal M}_{ann}^{i(SP)}(a_6-\frac{1}{2}a_8,Q_{d},Q_d)-2 {\cal
M}_{ann}^{i(a,LL)}(a_3+a_5-\frac{1}{2}a_7-\frac{1}{2}a_9,Q_{s})\nonumber\\
&&-2 {\cal M}_{ann}^{i(b,LL)}(a_3-\frac{1}{2}a_9,Q_{s},Q_{s})-2
{\cal M}_{ann}^{i(b,LR)}(a_5-\frac{1}{2}a_7,Q_{s},Q_{s})\Big \},
\end{eqnarray}

\begin{eqnarray}
 \sqrt 2{\cal A}^i(\bar B^0\to f_1\gamma)
&=&  \frac{G_F}{\sqrt{2}}  V_{ub}V_{ud}^{*} \frac{1}{\sqrt 3}\Big\{
{\cal M}_{ann}^{i(a,LL)}(a_2,Q_{d})+{\cal
M}_{ann}^{i(b,LL)}(a_2,Q_u,Q_{u})+{\cal M}_{1u}^{i(a)}\nonumber
   \\
  && +{\cal
M}_{1u}^{i(b)}(Q_d)+{\cal M}_{2u}^{i} \Big \}+\frac{G_F}{\sqrt{2}}
V_{cb}V_{cd}^{*} \frac{1}{\sqrt 3}\Big\{ {\cal M}_{1c}^{i(a)} +{\cal
M}_{1c}^{i(b)}(Q_d)+{\cal M}_{2c}^{i}
 \Big\}\nonumber
   \\
  &&- \frac{G_F}{\sqrt{2}} V_{tb}V_{td}^{*} \frac{1}{\sqrt 3}\Big\{{\cal
  M}_{7\gamma}^i+{\cal  M}_{8g}^{i(a)}\nonumber
   \\
  &&+{\cal   M}_{8g}^{i(b)}(Q_d)+ {\cal
M}_{ann}^{i(a,LL)}(2a_3+a_4+2a_5+\frac{1}{2}a_7+\frac{1}{2}a_9-\frac{1}{2}a_{10},Q_{d})\nonumber
   \\
  &&+ {\cal
M}_{ann}^{i(b,LL)}(a_3+a_9,Q_{u},Q_{u})+ {\cal
M}_{ann}^{i(b,LR)}(a_5+a_7,Q_{u},Q_{u})\nonumber\\
&&+ {\cal
M}_{ann}^{i(b,LL)}(a_3+a_4-\frac{1}{2}a_9-\frac{1}{2}a_{10},Q_{d},Q_{d})+
{\cal M}_{ann}^{i(b,LR)}(a_5-\frac{1}{2}a_7,Q_{d},Q_{d})\nonumber
   \\
  &&
+ {\cal M}_{ann}^{i(SP)}(a_6-\frac{1}{2}a_8,Q_{d},Q_d)+ {\cal
M}_{ann}^{i(a,LL)}(a_3+a_5-\frac{1}{2}a_7-\frac{1}{2}a_9,Q_{s})\nonumber\\
&&+ {\cal M}_{ann}^{i(b,LL)}(a_3-\frac{1}{2}a_9,Q_{s},Q_{s})+ {\cal
M}_{ann}^{i(b,LR)}(a_5-\frac{1}{2}a_7,Q_{s},Q_{s})\Big \},
\end{eqnarray}
\begin{eqnarray}
 \sqrt 2{\cal A}^i(\bar B_s\to f_8\gamma)
&=&  \frac{G_F}{\sqrt{2}}  V_{ub}V_{us}^{*} \frac{1}{\sqrt
6}\Big\{{\cal M}_{ann}^{i(a,LL)}(a_2,Q_{s})+{\cal
M}_{ann}^{i(b,LL)}(a_2,Q_{u},Q_{u})\Big \}\nonumber
   \\
  &&- \frac{G_F}{\sqrt{2}} V_{tb}V_{ts}^{*} \frac{1}{\sqrt
6} \Big\{{\cal
M}_{ann}^{i(a,LL)}(2a_3+2a_5+\frac{1}{2}a_7+\frac{1}{2}a_9,Q_{s}) \nonumber\\
  &&+ {\cal
M}_{ann}^{i(b,LL)}(a_3+a_9,Q_{u},Q_{u})+ {\cal
M}_{ann}^{i(b,LR)}(a_5+a_7,Q_{u},Q_{u})\nonumber\\
&&+ {\cal M}_{ann}^{i(b,LL)}(a_3-\frac{1}{2}a_9,Q_{d},Q_{d})+ {\cal
M}_{ann}^{i(b,LR)}(a_5-\frac{1}{2}a_7,Q_{d},Q_{d})\Big \}\nonumber\\
 && +\frac{G_F}{\sqrt{2}}  V_{ub}V_{us}^{*} \frac{-2}{\sqrt
6}\Big\{ {\cal M}_{1u}^{i(a)} +{\cal M}_{1u}^{i(b)}(Q_s)+{\cal
M}_{2u}^{i}\Big \}\nonumber
   \\
  &&+\frac{G_F}{\sqrt{2}}  V_{cb}V_{cs}^{*}\frac{-2}{\sqrt
6} \Big\{ {\cal M}_{1c}^{i(a)} +{\cal M}_{1c}^{i(b)}(Q_s)+{\cal
M}_{2c}^{i}
 \Big\}\nonumber
   \\
  &&- \frac{G_F}{\sqrt{2}} V_{tb}V_{ts}^{*} \frac{-2}{\sqrt
6}\Big\{{\cal
  M}_{7\gamma}^i+{\cal  M}_{8g}^{i(a)}+{\cal
  M}_{8g}^{i(b)}(Q_s)+
{\cal M}_{ann}^{i(SP)}(a_6-\frac{1}{2}a_8,Q_{s},Q_s)\nonumber
   \\
  &&+ {\cal
M}_{ann}^{i(a,LL)}(a_3+a_4+a_5-\frac{1}{2}a_7-\frac{1}{2}a_9-\frac{1}{2}a_{10},Q_{s})\\
  &&+ {\cal
M}_{ann}^{i(b,LL)}(a_3+a_4-\frac{1}{2}a_9-\frac{1}{2}a_{10},Q_{s},Q_{s})+
{\cal M}_{ann}^{i(b,LR)}(a_5-\frac{1}{2}a_7,Q_{s},Q_{s}) \Big
\},\nonumber
\end{eqnarray}
\begin{eqnarray}
 \sqrt 2{\cal A}^i(\bar B_s\to f_1\gamma)
&=&  \frac{G_F}{\sqrt{2}}  V_{ub}V_{us}^{*} \frac{1}{\sqrt
3}\Big\{{\cal M}_{ann}^{i(a,LL)}(a_2,Q_{s})+{\cal
M}_{ann}^{i(b,LL)}(a_2,Q_{u},Q_{u})\Big \}\nonumber
   \\
  &&- \frac{G_F}{\sqrt{2}} V_{tb}V_{ts}^{*} \frac{1}{\sqrt
3} \Big\{{\cal
M}_{ann}^{i(a,LL)}(2a_3+2a_5+\frac{1}{2}a_7+\frac{1}{2}a_9,Q_{s}) \nonumber\\
  &&+ {\cal
M}_{ann}^{i(b,LL)}(a_3+a_9,Q_{u},Q_{u})+ {\cal
M}_{ann}^{i(b,LR)}(a_5+a_7,Q_{u},Q_{u})\nonumber\\
&&+ {\cal M}_{ann}^{i(b,LL)}(a_3-\frac{1}{2}a_9,Q_{d},Q_{d})+ {\cal
M}_{ann}^{i(b,LR)}(a_5-\frac{1}{2}a_7,Q_{d},Q_{d})\Big \}\nonumber\\
 && +\frac{G_F}{\sqrt{2}}  V_{ub}V_{us}^{*} \frac{1}{\sqrt
3}\Big\{ {\cal M}_{1u}^{i(a)} +{\cal M}_{1u}^{i(b)}(Q_s)+{\cal
M}_{2u}^{i}\Big \}\nonumber
   \\
  &&+\frac{G_F}{\sqrt{2}}  V_{cb}V_{cs}^{*}\frac{1}{\sqrt
3} \Big\{ {\cal M}_{1c}^{i(a)} +{\cal M}_{1c}^{i(b)}(Q_s)+{\cal
M}_{2c}^{i}
 \Big\}\nonumber
   \\
  &&- \frac{G_F}{\sqrt{2}} V_{tb}V_{ts}^{*} \frac{1}{\sqrt
3}\Big\{{\cal
  M}_{7\gamma}^i+{\cal  M}_{8g}^{i(a)}+{\cal
  M}_{8g}^{i(b)}(Q_s)+
{\cal M}_{ann}^{i(SP)}(a_6-\frac{1}{2}a_8,Q_{s},Q_s)\nonumber
   \\
  &&+ {\cal
M}_{ann}^{i(a,LL)}(a_3+a_4+a_5-\frac{1}{2}a_7-\frac{1}{2}a_9-\frac{1}{2}a_{10},Q_{s})\\
  &&+ {\cal
M}_{ann}^{i(b,LL)}(a_3+a_4-\frac{1}{2}a_9-\frac{1}{2}a_{10},Q_{s},Q_{s})+
{\cal M}_{ann}^{i(b,LR)}(a_5-\frac{1}{2}a_7,Q_{s},Q_{s}) \Big
\}.\nonumber
\end{eqnarray}

For annihilation type decays, we have
\begin{eqnarray}
 \sqrt 2{\cal A}^i(\bar B_s\to a_1^0(1235)\gamma)
&=&  \frac{G_F}{\sqrt{2}}  V_{ub}V_{us}^{*} \Big\{{\cal
M}_{ann}^{i(a,LL)}(a_2,Q_{s})+{\cal
M}_{ann}^{i(b,LL)}(a_2,Q_{u},Q_{u})\Big \}\nonumber
   \\
  &&- \frac{G_F}{\sqrt{2}} V_{tb}V_{ts}^{*} \Big\{{\cal
M}_{ann}^{i(a,LL)}(\frac{3}{2}a_7+\frac{3}{2}a_9,Q_{s})\nonumber\\
  &&+ {\cal
M}_{ann}^{i(b,LL)}(a_3+a_9,Q_{u},Q_{u})+ {\cal
M}_{ann}^{i(b,LR)}(a_5+a_7,Q_{u},Q_{u})\\
&&+ {\cal M}_{ann}^{i(b,LL)}(-a_3+\frac{1}{2}a_9,Q_{d},Q_{d})+ {\cal
M}_{ann}^{i(b,LR)}(-a_5+\frac{1}{2}a_7,Q_{d},Q_{d})\Big \}.\nonumber
\end{eqnarray}
Decay amplitudes of $A^i(B\to K_1(1270)\gamma)$ and $A^i(B\to
K_1(1400)\gamma)$ (here $B$ denotes $\bar B_{u,d,s}$ and $K_1$
denotes $K_1^{-}(\bar K_1^0)$) can be obtained by:
\begin{eqnarray}
  A^i(B\to K_1(1270)\gamma)&=&{\rm sin}(\theta_K) A^i(B\to K_{1A}\gamma)
   + {\rm cos}(\theta_K) A^i(B\to K_{1B}\gamma),\\
     A^i(B\to K_1(1400)\gamma)&=&- {\rm sin}(\theta_K) A^i(B\to K_{1B}\gamma)
     +{\rm cos}(\theta_K) A^i(B\to K_{1A}\gamma)
   .
\end{eqnarray}


\end{document}